 \documentclass[aps,prd,preprint,floatfix,nofootinbib,superscriptaddress,showpacs]{revtex4}
 \usepackage{graphicx}

\def\Z0{${\em Z^0\/}$}

\def\r#1 {$^{#1}$}

\newcommand{\gevc} { {\rm GeV/c}}
\newcommand{\gevcc}{ {\rm GeV/c^2}}

\def\gepsfcentered#1{
  \def\testit{#1}
  \def\lbracket{[}
  \ifx\testit\lbracket
    \let\dofilecmd=\gepsfwithopt
  \else
    \let\dofilecmd=\gepsfnoopt
  \fi
  \dofilecmd}

\def\gepsfnoopt#1{
  \begin{center}
  \leavevmode
  \epsffile{#1}
  \end{center}}

\def\gepsfwithopt#1 #2 #3 #4]#5{
  \begin{center}
  \leavevmode
  \gepsfmaxx=0.94\textwidth
  \epsffile[#1 #2 #3 #4]{#5}
  \end{center}}

\newdimen\gepsfmaxx
\def\epsfsize#1#2{
  \ifnum \epsfxsize=0
    \ifnum \epsfysize=0
      \ifnum #1 > \gepsfmaxx
        \gepsfmaxx
      \else
        #1
      \fi
    \else
      \epsfxsize
    \fi
  \else
    \epsfxsize
  \fi
}

\begin{document}

 \bibliographystyle{apsrev}

%%%%%%%%%%%%%%%%%%%%%%%%%%%%%%%%%%%%%%%%%%%%%%%%%%%%%
 \title {Measurement of correlated {\boldmath $b \bar{b}$} production
         in {\boldmath $p \bar{p}$} collisions at {\boldmath $\sqrt{s}=$}
         1960 GeV}
 \affiliation{Institute of Physics, Academia Sinica, Taipei, Taiwan 11529, Republic of China} 
\affiliation{Argonne National Laboratory, Argonne, Illinois 60439} 
\affiliation{Institut de Fisica d'Altes Energies, Universitat Autonoma de Barcelona, E-08193, Bellaterra (Barcelona), Spain} 
\affiliation{Baylor University, Waco, Texas  76798} 
\affiliation{Istituto Nazionale di Fisica Nucleare, University of Bologna, I-40127 Bologna, Italy} 
\affiliation{Brandeis University, Waltham, Massachusetts 02254} 
\affiliation{University of California, Davis, Davis, California  95616} 
\affiliation{University of California, Los Angeles, Los Angeles, California  90024} 
\affiliation{University of California, San Diego, La Jolla, California  92093} 
\affiliation{University of California, Santa Barbara, Santa Barbara, California 93106} 
\affiliation{Instituto de Fisica de Cantabria, CSIC-University of Cantabria, 39005 Santander, Spain} 
\affiliation{Carnegie Mellon University, Pittsburgh, PA  15213} 
\affiliation{Enrico Fermi Institute, University of Chicago, Chicago, Illinois 60637} 
\affiliation{Comenius University, 842 48 Bratislava, Slovakia; Institute of Experimental Physics, 040 01 Kosice, Slovakia} 
\affiliation{Joint Institute for Nuclear Research, RU-141980 Dubna, Russia} 
\affiliation{Duke University, Durham, North Carolina  27708} 
\affiliation{Fermi National Accelerator Laboratory, Batavia, Illinois 60510} 
\affiliation{University of Florida, Gainesville, Florida  32611} 
\affiliation{Laboratori Nazionali di Frascati, Istituto Nazionale di Fisica Nucleare, I-00044 Frascati, Italy} 
\affiliation{University of Geneva, CH-1211 Geneva 4, Switzerland} 
\affiliation{Glasgow University, Glasgow G12 8QQ, United Kingdom} 
\affiliation{Harvard University, Cambridge, Massachusetts 02138} 
\affiliation{Division of High Energy Physics, Department of Physics, University of Helsinki and Helsinki Institute of Physics, FIN-00014, Helsinki, Finland} 
\affiliation{University of Illinois, Urbana, Illinois 61801} 
\affiliation{The Johns Hopkins University, Baltimore, Maryland 21218} 
\affiliation{Institut f\"{u}r Experimentelle Kernphysik, Universit\"{a}t Karlsruhe, 76128 Karlsruhe, Germany} 
\affiliation{High Energy Accelerator Research Organization (KEK), Tsukuba, Ibaraki 305, Japan} 
\affiliation{Center for High Energy Physics: Kyungpook National University, Taegu 702-701, Korea; Seoul National University, Seoul 151-742, Korea; SungKyunKwan University, Suwon 440-746, Korea} 
\affiliation{Ernest Orlando Lawrence Berkeley National Laboratory, Berkeley, California 94720} 
\affiliation{University of Liverpool, Liverpool L69 7ZE, United Kingdom} 
\affiliation{University College London, London WC1E 6BT, United Kingdom} 
\affiliation{Centro de Investigaciones Energeticas Medioambientales y Tecnologicas, E-28040 Madrid, Spain} 
\affiliation{Massachusetts Institute of Technology, Cambridge, Massachusetts  02139} 
%\affiliation{Institute of Particle Physics: McGill University, Montr\'{e}al, Canada H3A~2T8; and University of Toronto, Toronto, Canada M5S~1A7} 
\affiliation{University of Michigan, Ann Arbor, Michigan 48109} 
\affiliation{Michigan State University, East Lansing, Michigan  48824} 
\affiliation{University of New Mexico, Albuquerque, New Mexico 87131} 
\affiliation{Northwestern University, Evanston, Illinois  60208} 
\affiliation{The Ohio State University, Columbus, Ohio  43210} 
\affiliation{Okayama University, Okayama 700-8530, Japan} 
\affiliation{Osaka City University, Osaka 588, Japan} 
\affiliation{University of Oxford, Oxford OX1 3RH, United Kingdom} 
\affiliation{University of Padova, Istituto Nazionale di Fisica Nucleare, Sezione di Padova-Trento, I-35131 Padova, Italy} 
\affiliation{LPNHE, Universite Pierre et Marie Curie/IN2P3-CNRS, UMR7585, Paris, F-75252 France} 
\affiliation{University of Pennsylvania, Philadelphia, Pennsylvania 19104} 
\affiliation{Istituto Nazionale di Fisica Nucleare Pisa, Universities of Pisa, Siena and Scuola Normale Superiore, I-56127 Pisa, Italy} 
\affiliation{University of Pittsburgh, Pittsburgh, Pennsylvania 15260} 
\affiliation{Purdue University, West Lafayette, Indiana 47907} 
\affiliation{University of Rochester, Rochester, New York 14627} 
\affiliation{The Rockefeller University, New York, New York 10021} 
\affiliation{Istituto Nazionale di Fisica Nucleare, Sezione di Roma 1, University of Rome ``La Sapienza," I-00185 Roma, Italy} 
\affiliation{Rutgers University, Piscataway, New Jersey 08855} 
\affiliation{Texas A\&M University, College Station, Texas 77843} 
\affiliation{Istituto Nazionale di Fisica Nucleare, University of Trieste/\ Udine, Italy} 
\affiliation{University of Tsukuba, Tsukuba, Ibaraki 305, Japan} 
\affiliation{Tufts University, Medford, Massachusetts 02155} 
\affiliation{Waseda University, Tokyo 169, Japan} 
\affiliation{Wayne State University, Detroit, Michigan  48201} 
\affiliation{University of Wisconsin, Madison, Wisconsin 53706} 
\affiliation{Yale University, New Haven, Connecticut 06520} 
\author{T.~Aaltonen}
\affiliation{Division of High Energy Physics, Department of Physics, University of Helsinki and Helsinki Institute of Physics, FIN-00014, Helsinki, Finland}
\author{A.~Abulencia}
\affiliation{University of Illinois, Urbana, Illinois 61801}
\author{J.~Adelman}
\affiliation{Enrico Fermi Institute, University of Chicago, Chicago, Illinois 60637}
\author{T.~Affolder}
\affiliation{University of California, Santa Barbara, Santa Barbara, California 93106}
\author{T.~Akimoto}
\affiliation{University of Tsukuba, Tsukuba, Ibaraki 305, Japan}
\author{M.G.~Albrow}
\affiliation{Fermi National Accelerator Laboratory, Batavia, Illinois 60510}
\author{S.~Amerio}
\affiliation{University of Padova, Istituto Nazionale di Fisica Nucleare, Sezione di Padova-Trento, I-35131 Padova, Italy}
\author{D.~Amidei}
\affiliation{University of Michigan, Ann Arbor, Michigan 48109}
\author{A.~Anastassov}
\affiliation{Rutgers University, Piscataway, New Jersey 08855}
\author{K.~Anikeev}
\affiliation{Fermi National Accelerator Laboratory, Batavia, Illinois 60510}
% ...
\collaboration{CDF Collaboration\footnote{With visitors from $^a$University of Athens, 15784 Athens, Greece, 
$^b$University of Bristol, Bristol BS8 1TL, United Kingdom, 
$^c$University Libre de Bruxelles, B-1050 Brussels, Belgium, 
$^d$Cornell University, Ithaca, NY  14853, 
$^e$University of Cyprus, Nicosia CY-1678, Cyprus, 
$^f$University College Dublin, Dublin 4, Ireland, 
$^g$University of Edinburgh, Edinburgh EH9 3JZ, United Kingdom, 
$^h$University of Heidelberg, D-69120 Heidelberg, Germany, 
$^i$Universidad Iberoamericana, Mexico D.F., Mexico, 
$^j$University of Manchester, Manchester M13 9PL, England, 
$^k$Nagasaki Institute of Applied Science, Nagasaki, Japan, 
$^l$University de Oviedo, E-33007 Oviedo, Spain, 
$^m$University of London, Queen Mary College, London, E1 4NS, England, 
$^n$University of California Santa Cruz, Santa Cruz, CA  95064, 
$^o$Texas Tech University, Lubbock, TX  79409, 
$^p$University of California, Irvine, Irvine, CA  92697, 
$^q$IFIC(CSIC-Universitat de Valencia), 46071 Valencia, Spain 
}}
\noaffiliation

%%%%%%%%%%%%%%%%%%%%%%%%%%%%%%%%%%%%%%%%%%%%%%%%%%%%%
 \noaffiliation

%%%%%%%%%%%%%%%%%%%%
 \begin{abstract}
%%%%%%%%%%%%%%%%%%%%
 We present a measurement of the correlated $b\bar{b}$
 production cross section. The data used in this analysis were taken
 with the  upgraded CDF detector (CDF II) at the Fermilab Tevatron collider,
 and correspond to an integrated luminosity of 742 pb$^{-1}$. 
 We utilize muon pairs with invariant mass $5 \leq m_{\mu\mu} \leq 80 \;\gevcc$
 produced by $b\bar{b}$ double semileptonic decays. For muons with 
 $p_T \geq 3 \; \gevc$ and $|\eta| \leq 0.7$, that are produced by
 $b$ and $\bar{b}$ quarks with  $p_T \geq 2\; \gevc$ and $|y| \leq 1.3$,
 we measure $\sigma_{b\rightarrow\mu,\bar{b}\rightarrow \mu}= 1549 \pm 133$ pb.
 We compare this result with theoretical predictions and previous measurements.
 We also report the  measurement of 
 $\sigma_{c\rightarrow\mu,\bar{c}\rightarrow \mu}$, 
 a by-product of the study of the background to $b\bar{b}$ production.
 \end{abstract} 
%%%%%%%%%%%%%%%%%%%%
 \preprint{FERMILAB-PUB-07-500-E}
 \maketitle
%%%%%%%%%%%%%%%%%%%%%%%%%%%%%%%%%%%%%%%%%%%%%%%%%%
 \section {Introduction}  \label{sec:ss-intro}
%%%%%%%%%%%%%%%%%%%%%%%%%%%%%%%%%%%%%%%%%%%%%%%%%%
 Measurements of the cross section for producing, in hadronic collisions, both
 $b$ and $\bar{b}$ quarks centrally and above a given transverse momentum  
 threshold (typically $p_T \geq 5-20 \; \gevc$), referred to as   
 $\sigma_{b\bar{b}}$ or $b \bar{b}$ correlations, provide an important test
 of the predictive power of quantum chromodynamics (QCD).
 Experimentally, $b \bar{b}$ correlations at the Tevatron are inferred
 from the production rate above a given $p_T$ threshold of some of the
 decay products  (leptons or tracks consistent with a secondary displaced 
 vertex) of both $b$ and $\bar{b}$ hadrons. In QCD calculations, the long- 
 and short-distance dynamics of the hadronic hard-scattering cross section 
 are factorized into nonperturbative parton distribution functions (PDF) 
 and fragmentation functions, and perturbatively calculable hard-scattering
 functions. At the perturbative level, the hard-scattering function can be
 evaluated at leading-order (LO) and next-to-leading order (NLO) with the 
 {\sc mnr} Monte Carlo program~\cite{mnr}.
 In contrast with the exact NLO prediction of the single $b$ quark production
 cross section~\footnote{
 The single $b$ quark cross section can be evaluated at exact NLO accuracy 
 with the {\sc nde} Monte Carlo generator~\cite{nde}. The calculation
 is affected by an uncertainty as large as 50\% due to the choice of 
 renormalization and factorization scales and by additional, but smaller,
 uncertainties due to choice of the PDF fits or the $b$-quark 
 mass~\cite{mlmri}.
 At perturbative level, the large scale dependence of the NLO calculation
 is interpreted as a symptom of large higher-order 
 contributions~\cite{qcdan}.}, the exact NLO calculation of 
 $\sigma_{b\bar{b}}$ appears to be a robust perturbative QCD prediction. 
 As noted in Ref.~\cite{ajets}, the exact LO and NLO prediction of 
 $\sigma_{b\bar{b}}$ are equal within a few percent, and the NLO result 
 does not change by more than 15\% when varying the renormalization and 
 factorization scales by a factor of two and the $b$ quark pole mass 
 ($m_b=4.75 \; \gevcc$) by $0.25 \; \gevcc$. The exact NLO prediction of
 $\sigma_{b\bar{b}}$ is quite insensitive to the choice of PDF 
 fits when they include HERA data and yield a value of the QCD coupling
 strength consistent with LEP data 
 ($\alpha_s(m_Z) \simeq 0.118$)~\cite{bstatus, mrst,cteq}.
 However, when comparing to the data, the apparent 
 robustness of the $\sigma_{b\bar{b}}$ calculation could be spoiled by the
 inclusion of nonperturbative fragmentation functions that connect 
 $b$-quark and $b$-hadron distributions. Traditionally, data to theory 
 comparisons use a fragmentation model based on the Peterson 
 function~\cite{pet} with the $\epsilon$ parameter set to 0.006 according 
 to fits to $e^+e^-$ data~\cite{chrin}. However, as noted in 
 Ref.~\cite{cana}, the Peterson fragmentation function has been tuned to 
 the data in conjunction with  LO parton-level cross sections evaluated 
 with parton-shower event generators, and cannot
 be consistently convoluted with the exact NLO calculation. As an example,
 the FONLL calculation~\cite{fonll} implements the exact NLO prediction 
 of the single $b$-quark cross section with the resummation of ($p_T/m_b$)
 logarithms with next-to-leading accuracy (NLL). A calculation with the
 same level of accuracy, available for the production of $b$ quarks at 
 $e^+e^-$ colliders~\cite{f1}, has been used to extract consistent 
 nonperturbative fragmentation functions from LEP and SLC data~\cite{f2}.
 These fragmentation functions appear to be harder than the Peterson 
 fragmentation~\cite{cana}. Unfortunately, they also cannot be consistently
 convoluted with the exact NLO calculation of  $\sigma_{b\bar{b}}$ for 
 which NLL logarithmic corrections have yet to be evaluated.

 Alternatively, the production of pairs of $b$ and $\bar{b}$ hadrons can 
 be estimated with event generators that are based on the LO calculation
 combined with a leading-logarithmic (LL) treatment of higher orders via
 the parton shower approximation, such as the {\sc herwig}~\cite{herwig}
 and {\sc pythia}~\cite{pythia} Monte Carlo programs. The {\sc mc@nlo} 
 event generator~\cite{mcnlo} merges the exact NLO matrix element with 
 the LL shower evolution and hadronization performed by the {\sc herwig}
 parton-shower Monte Carlo. In some cases, event generators that combine
 exact LO or NLO calculations with LL parton-shower simulations return
 parton-level cross sections that are quite different from the exact NLO
 calculation~\cite{mnr}. The {\sc mc@nlo} method suffers the additional
 problem that the {\sc herwig} model of the $b$ quark hadronization has 
 been tuned to $e^+e^-$ data using LO parton-level cross sections.
 The benefits and pitfalls of each theoretical approach are discussed in
 more detail in Refs.~\cite{mcnlo,pnason,bdis}.

 Precise measurements of the pair production of $b$ and $\bar{b}$ hadrons 
 at the Tevatron could contribute to improve the modeling of fragmentation
 functions consistent with the exact NLO calculation. Unfortunately, as 
 noted in Ref.~\cite{bstatus}, the status of the $\sigma_{b\bar{b}}$
 measurements at the Tevatron is quite disconcerting. Five measurements of
 $\sigma_{b\bar{b}}$ have been performed by the CDF and D${\not\! {\rm O}}$
 collaborations. Reference~\cite{bstatus} compares the results of
 different experiments using $R_{2b}$, the ratio of the measured 
 $\sigma_{b\bar{b}}$  to the exact NLO prediction (the $b$-quark and 
 $b$-hadron distributions are connected via the LL {\sc herwig} 
 fragmentation model or the Peterson fragmentation function).

 The study in Ref.~\cite{ajets} (CDF) uses two central jets with 
 $E_T \geq 15$ GeV, each containing a secondary vertex due to $b$- or
 $\bar{b}$-quark decays. The measurement yields $R_{2b}=1.2 \pm 0.3$.

 The study in Ref.~\cite{shears} (CDF) uses events containing two central jets
 with $E_T\geq 30$ and 20 GeV, respectively; pairs of $b$ jets are also 
 identified by requiring the presence of displaced secondary vertices. 
 This study yields~\footnote{
 Ref.~\cite{shears} compares the data to the {\sc mc@nlo}
 prediction, which is 12\% smaller than the exact NLO prediction.  
 This difference was not appreciated in Ref.~\cite{bstatus}.}
 $R_{2b}=1.1 \pm 0.3$.

 The study in Ref.~\cite{derw} (CDF) uses events containing  muons 
 from $b$-quark semileptonic decays that recoil against a jet that
 contains tracks with large impact parameter ($b$ jet). This study 
 yields $R_{2b}=1.5 \pm 0.2$ for $b$ and $\bar{b}$ quarks produced 
 centrally with $p_T \geq 12\; \gevc$.

 References~\cite{2mucdf} (CDF) and~\cite{d0b2} (D${\not\!{\rm O}})$
 report measurements that use two central muons arising from $b$-quark 
 semileptonic decays. The measurements yield $R_{2b}=3.0 \pm 0.6$ and
 $R_{2b}=2.3 \pm 0.7 $ for central $b$ and $\bar{b}$ quarks with
 $p_T \geq 6$ and $7\;  \gevc$, respectively.

 The five measurements yield $<R_{2b}>=1.8$ with a 0.8 RMS 
 deviation~\cite{bstatus}. Such a large RMS deviation is a likely indication
 of experimental difficulties~\footnote{ 
 This includes the possibility that in some cases the NLO prediction has 
 been evaluated incorrectly.}.
 This type of discrepancy could result from an underestimate of  the 
 kinematic and detector acceptance for semileptonic $b$ decays or of the 
 underlying background. However, measurements of the single $b$-quark 
 production cross section based upon detection of semileptonic $b$-quark 
 decays suggest otherwise 
 because they are approximately 35\% smaller than those based on detection
 of $J/\psi$ mesons from $b$-quark decays~\cite{bstatus,bjk}.
 The present discrepancy could also be explained by postulating the 
 production of additional objects with a 100\% semileptonic branching 
 ratio and a cross section of the order of 1/10 of the $b$ cross section
 as investigated in Ref.~\cite{ajets}. Therefore, it is of interest to 
 clarify the experimental situation. This paper reports a new measurement
 of $\sigma_{b\bar{b}}$ that uses dimuons arising from $b \bar{b}$ 
 production. At the Tevatron, dimuon events result from decays of heavy
 quark pairs ($b\bar{b}$ and $c\bar{c}$), the Drell-Yan process, charmonium
 and bottomonium decays, and decays of $\pi$ and $K$ mesons. Background to
 dimuon events also comes from the misidentification of $\pi$ or $K$ mesons.
 As in previous studies~\cite{2mucdf,bmix}, we make use of the precision
 tracking provided by the CDF silicon microvertex detector to evaluate the
 fractions of muons due to long-lived $b$- and $c$-hadron decays, and to 
 the other background contributions.

 Sections~\ref{sec:ss-det} and~\ref{sec:ss-anal} describe the detector
 systems relevant to this analysis and the data selection, respectively.
 The analysis method is discussed in Sec.~\ref{sec:ss-meth}, while
 the heavy flavor composition of  the dimuon sample is determined in 
 Sec.~\ref{sec:ss-comp}. The kinematic and detector acceptance is 
 evaluated in Sec.~\ref{sec:ss-acc}. The dimuon cross section is derived
 and compared to theoretical expectation and previous measurements in
 Sec.~\ref{sec:ss-disc}. 
 Our conclusions are summarized in Sec.~\ref{sec:ss-concl}. 
%%%%%%%%%%%%%%%%%%%%%%%%%%%%%%%%%%%%%%
 \section{CDF II detector and trigger}
 \label{sec:ss-det}
%%%%%%%%%%%%%%%%%%%%%%%%%%%%%%%
 CDF II is a multipurpose detector, equipped with a charged particle 
 spectrometer and a finely segmented calorimeter. In this section, we 
 describe the detector components that are relevant to this analysis. 
 The description of these subsystems can be found in  
 Refs.~\cite{det1,det2,det3_0,det3,det4_0,det4,det5,det6,det7,det8}.
 Two devices inside the 1.4 T solenoid are used for measuring the momentum
 of charged particles: the silicon vertex detector (SVXII and ISL) and the
 central tracking chamber (COT). The SVXII detector consists of 
 microstrip sensors arranged in six cylindrical shells with radii between 1.5
 and 10.6 cm, and with a total $z$ coverage~\footnote{
 In the CDF coordinate system, $\theta$ and $\phi$ are the polar and azimuthal
 angles of a track, respectively, defined with respect to the proton beam
 direction, $z$. The pseudorapidity $\eta$ is defined as 
 $-\log \;\tan (\theta/2)$. The transverse momentum of a particle is 
 $p_T= p \; \sin (\theta)$. The rapidity is defined as 
 $y=1/2 \cdot \log ( (E+p_z)/(E-p_z) )$, where $E$ and $p_z$ are the energy
 and longitudinal momentum of the particle associated with the track.} 
 of 90 cm. The first SVXII layer, also referred to as L00 detector,
  is made of single-sided sensors mounted on the beryllium 
 beam pipe. The remaining five SVXII layers are made of double-sided sensors
 and are divided
 into three contiguous five-layer sections along the beam direction $z$.
 The vertex $z$-distribution for $p\bar{p}$ collisions is approximately 
 described by a Gaussian function with a sigma of 28 cm. The transverse 
 profile of the Tevatron beam is circular and has an RMS spread of 
 $\simeq 25\; \mu$m in the horizontal and vertical directions. The SVXII 
 single-hit resolution is approximately $11\; \mu$m and allows a track impact
 parameter~\footnote{
 The impact parameter $d$ is the distance of closest approach of a track to
 the primary event vertex in the transverse plane.}
 resolution of approximately $35\; \mu$m, when also including the effect of
 the beam transverse size. The two additional silicon layers of the ISL help
 to link tracks in the COT to hits in the SVXII. 
 The COT is a cylindrical drift chamber containing 96 sense wire layers grouped
 into eight alternating superlayers of axial and stereo wires. Its active 
 volume covers $|z| \leq 155$ cm and 40 to 140 cm in radius. The transverse 
 momentum resolution of tracks reconstructed using COT hits
 is $\sigma(p_T)/p_T^2 \simeq 0.0017\; [\gevc]^{-1}$. COT tracks are 
 extrapolated into the SVXII detector and refitted adding hits consistent
 with the track extrapolation.

 The central muon detector (CMU) is located around the central electromagnetic
 and hadronic calorimeters, which have a thickness of 5.5 interaction lengths
 at normal incidence. The CMU detector covers a nominal pseudorapidity range
 $|\eta| \leq 0.63$ relative to the center of the detector, and  is segmented
 into two barrels of 24 modules, each covering 15$^\circ$ in $\phi$. Every
 module is further segmented  into three  submodules, each covering
 4.2$^\circ$ in $\phi$ and consisting of four layers of drift chambers.
 The smallest drift unit, called a stack, covers a 1.2$^\circ$ angle in $\phi$.
 Adjacent pairs of stacks are combined together into a tower. A track segment
 (hits in two out of four layers of a stack) detected in a tower is referred
 to as a CMU stub. A second set of muon drift chambers (CMP) is located behind
 an additional steel absorber of 3.3 interaction lengths. The chambers are
 640 cm long and are arranged axially to form a box around the central 
 detector. The CMP detector covers a nominal pseudorapidity range 
 $|\eta| \leq 0.54$ relative to the center of the detector. Muons which
 produce a stub in both CMU and CMP systems are called CMUP muons.

 The luminosity is measured using gaseous Cherenkov counters (CLC) that monitor
 the rate of inelastic $p\bar{p}$ collisions. The inelastic $p\bar{p}$ cross
 section at $\sqrt{s}=1960$ GeV is scaled from measurements at $\sqrt{s}=1800$
 GeV using the calculations in Ref.~\cite{sigmatot}. The integrated luminosity
 is determined with a 6\% systematic uncertainty~\cite{klimen}.
 
 CDF uses a three-level trigger system. At Level 1 (L1), data from every beam 
 crossing are stored in a pipeline capable of buffering data from 42 beam 
 crossings. The L1 trigger either rejects events or copies them into one of 
 the four Level 2 (L2) buffers. Events that pass the L1 and L2 selection 
 criteria are sent to the Level 3 (L3) trigger, a cluster of computers 
 running  speed-optimized reconstruction code.  

 For this study, we select events with two muon candidates identified by the L1
 and L2 triggers. The L1 trigger uses tracks with $p_T \geq 1.5 \; \gevc$ found
 by a fast track processor (XFT). The XFT examines COT hits from the four
 axial superlayers and provides $r-\phi$ information. The XFT finds tracks
 with $p_T \geq 1.5 \; \gevc$ in azimuthal sections of 1.25$^\circ$. The XFT
 passes the tracks to a set of extrapolation units that determine the CMU
 towers in which a CMU stub  should be found if the track is a muon. If a 
 stub is found, a L1 CMU primitive is generated.
 The L1 dimuon trigger requires at least two CMU primitives, separated by at
 least two CMU towers. The L2 trigger additionally requires that at least one
 of the muons has a CMUP stub matched to an XFT track with 
 $p_T \geq 3 \;\gevc$. All these trigger requirements are emulated by the
 detector simulation on a run-by-run basis.
 The L3 trigger requires a pair of CMUP muons with  invariant mass larger than
 $5 \; \gevcc$, and $|\delta z_0| \leq 5$ cm, where $z_0$ is the $z$ 
 coordinate of the muon track at its point of closest approach to the beam 
 line in the $r-\phi$ plane. These requirements define the dimuon trigger
 used in this analysis.

 We use additional triggers in order to measure detection efficiencies and
 verify the detector simulation. The first trigger (CMUP$p_T$4) selects 
 events with at least one CMUP primitive with
 $p_T \geq 4 \; \gevc$ identified by both the L1 and L2 triggers,
 and an additional muon found by the L3 algorithms.
 Events collected with this trigger are used to measure the muon trigger 
 efficiency. The second trigger requires a L1 CMUP primitive
 with $p_T \geq 4 \; \gevc$ accompanied by a L2 requirement of an additional
 track with $p_T \geq 2\; \gevc$ and impact parameter $0.12 \leq d \leq1$ mm
 as measured by the Silicon Vertex Trigger (SVT)~\cite{svt}. 
 The SVT  calculates the impact parameter of each XFT track,
 with respect to the beam line, with a 50 $\mu$m resolution that includes the 
 25 $\mu$m contribution of the beam transverse width. Events selected
 with this trigger ($\mu-$SVT) are used to verify the muon detector acceptance
 and the muon reconstruction efficiency. The last trigger ({\sc charm}) 
 acquires events with two SVT tracks with $p_T \geq 2 \; \gevc$ and with
 impact parameter $0.12 \leq d \leq 1$ mm. In this data sample,
 we reconstruct $D^0 \rightarrow K \pi$ decays to measure the 
 probability that a charged hadron mimics the signal of a CMUP muon. 
 We also use $J/\psi \rightarrow \mu^+\mu^-$ events acquired with the 
 $J/\psi$ trigger. At L1 and L2, this trigger requires two CMU primitives 
 corresponding to tracks with $p_T \geq 1.5 \; \gevc$. At L3, muons are 
 required to have opposite charges and an invariant mass in the window 
 $2.7-4.0 \; \gevcc$. These events are used to calibrate the efficiency of
 the SVXII detector and of stricter requirements used for selecting CMUP muons.
%%%%%%%%%%%%%%%%%%%%%%%%%%%%%
%%%%%%%%%%%%%%%%%%%%%%%%%%%%%%%
 \section{Data selection} \label{sec:ss-anal}
%%%%%%%%%%%%%%%%%%%%%%%%%%%%%%%
 In this analysis, we select events acquired with the dimuon trigger and 
 which contain two and only two CMUP muons with same or opposite charge.
 Events are reconstructed offline 
 taking advantage of more refined calibration constants and reconstruction 
 algorithms. COT tracks are extrapolated into the SVXII detector, and refitted
 adding  hits  consistent  with the track extrapolation. Stubs reconstructed 
 in the CMU and CMP detectors are matched to tracks with $p_T \geq 3 \; \gevc$.
 A track is identified as a CMUP muon if $\Delta r\phi$, the distance in the 
 $r-\phi$ plane between the track projected to the CMU (CMP) chambers and a
 CMU (CMP) stub, is less than 20 (40) cm. We require that muon-candidate 
 stubs correspond to a L1 CMU primitive, and correct the muon momentum
 for energy losses in the detector.

 To ensure an accurate impact parameter measurement, each muon track is
 required to be reconstructed in the SVXII detector with hits in the two inner
 layers and in at least two of the remaining four external layers. We evaluate
 the impact parameter of each muon track with respect to the  primary vertex.
 We reconstruct  primary vertices using all tracks with SVXII hits
 that are consistent with originating from a common vertex.
 In events in which more than one interaction vertex has been reconstructed 
 we use the one closest in $z$ to the average of the muon track $z_0$-positions
 and within a 6 cm distance. The primary vertex coordinates transverse to the
 beam direction have RMS uncertainties of approximately $3\; \mu$m, depending
 on the number of SVXII tracks associated with the primary vertex and the 
 event topology.

 Muon pairs arising from  cascade decays of a single $b$ quark are removed 
 by selecting dimuon candidates with invariant mass greater than 
 5 GeV/c$^2$. We also reject muon pairs with invariant mass larger than
 80 GeV/c$^2$ that are mostly contributed by $Z^0$ decays.
%%%%%%%%%%%%%%%%%%%%%%%%%%%%%%%%%%%%%
 \section{Method of analysis}\label{sec:ss-meth}
%%%%%%%%%%%%%%%%%%%%%%%%%%%%%%%%%%%%%
 For muons originating from the decay of long lived particles, the impact
 parameter is $d=|\beta \gamma c t \sin(\delta)|$, where $t$ is the proper
 decay time of the parent particle from which the muon track originates,
 $\delta$ is the decay angle of the muon track with respect to the
 direction of the parent particle, and $\beta \gamma$ is the Lorentz boost
 factor. The impact parameter distribution of  muon tracks is proportional 
 to the lifetime of the parent particle. The markedly different  
 distributions for muons from $b$ decays, $c$ decays, and other sources
 allow the determination of the parent fractions.
  
 We determine the $b\bar{b}$ and $c\bar{c}$ content of the data following 
 the method already used in Refs.~\cite{2mucdf,bmix}. The procedure is to 
 fit the observed impact parameter distribution of the muon pairs with the
 expected impact parameter distributions of leptons from various sources.
 After data selection, the main sources of reconstructed muons are
 semileptonic decays of bottom and charmed hadrons, prompt decays of quarkonia,
 and Drell-Yan production.

\begin{figure}
 \begin{center}
 \vspace{-0.3in}
 \leavevmode
\includegraphics*[width=\textwidth]{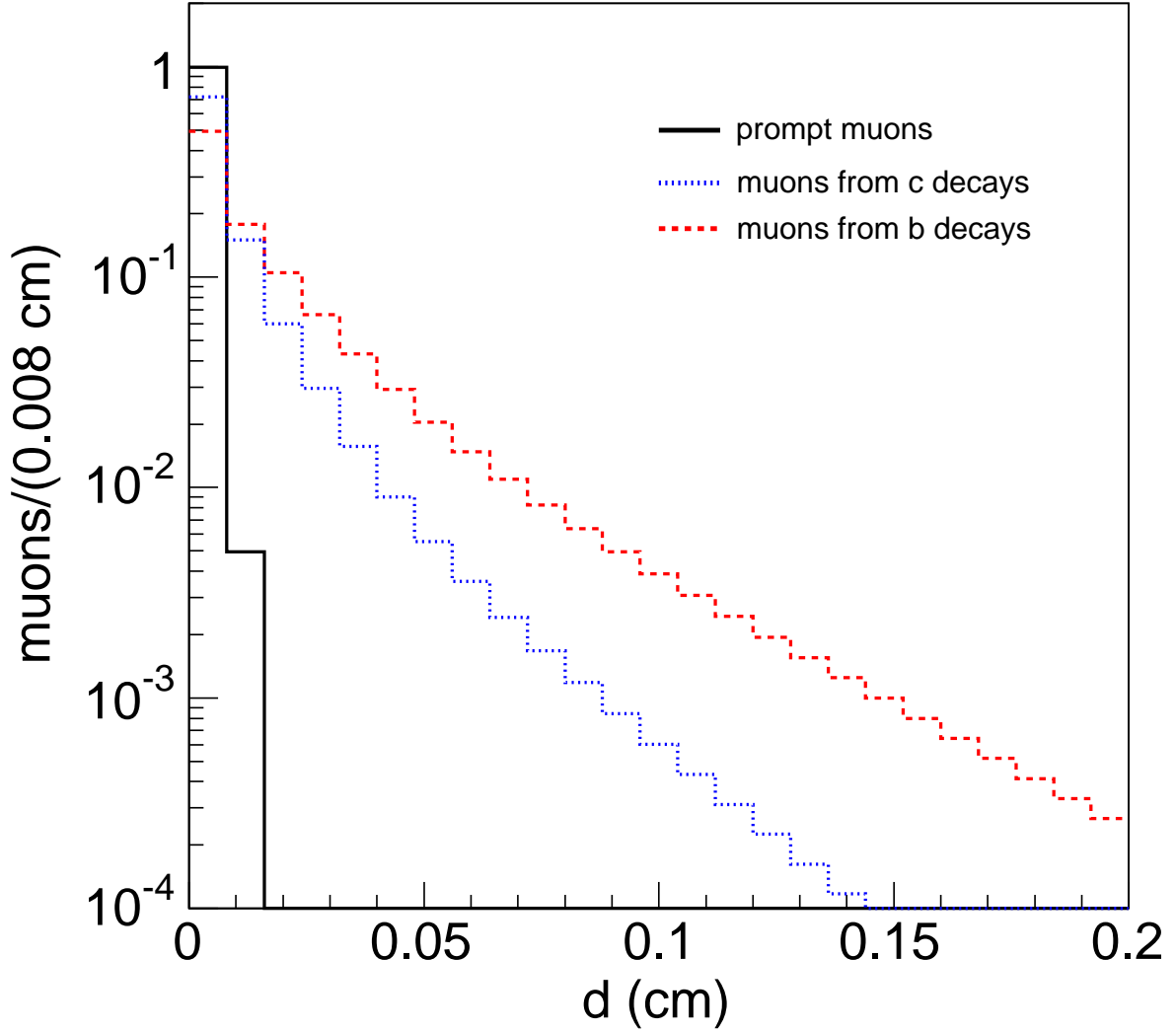}
 \caption[]{Impact parameter distributions of muons coming from 
            $b$- and $c$-hadron decays (simulation) and of prompt muons (data).
	    Distributions are normalized to unit area.}
 \label{fig:fig_1}
 \end{center}
 \end{figure}

\begin{figure}[]
 \begin{center}
 \vspace{-0.2in}
 \leavevmode
 \includegraphics*[width=\textwidth]{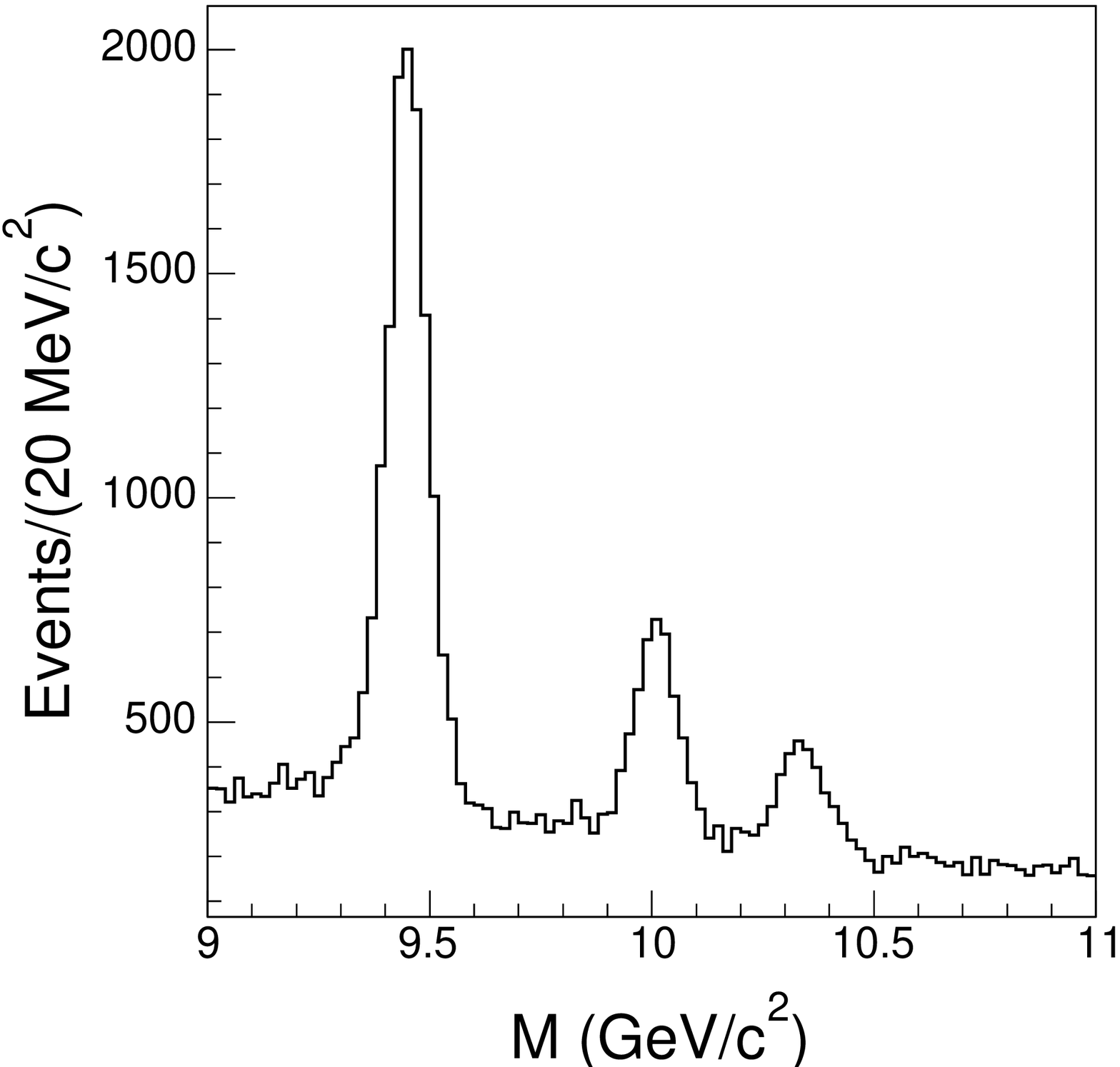}
 \caption[]{Distribution of the invariant mass of muon pairs in the $\Upsilon$
            region. The prompt template  in Fig.~\ref{fig:fig_1} is
	    derived using muons with invariant mass between 9.28 and 9.6
            GeV/$c^2$. The background is sideband subtracted using dimuons 
            with invariant mass between 9.04 and 9.2 GeV/$c^2$ and between 
            9.64 and 9.8 GeV/$c^2$.}
 \label{fig:fig_2}
 \end{center}
 \end{figure}

 Monte Carlo simulations are used to model the impact parameter distributions
 of muons from $b$- and $c$-hadron decays. We use the {\sc herwig}
 Monte Carlo program~\cite{herwig}, the settings of which are described in
 Appendix~A, to generate hadrons with heavy flavors that are subsequently 
 decayed using the {\sc evtgen} Monte Carlo program~\cite{evtgen}. 
 The detector response to particles produced by the above generators
 is modeled with the CDF~II detector simulation that in turn is based on 
 the {\sc geant} Monte Carlo program~\cite{geant}. Impact parameter 
 distributions of muon tracks in simulated $b$- and $c$-hadron decays 
 are shown in Fig.~\ref{fig:fig_1}. Since lifetimes of bottom and charmed 
 hadrons ($c\tau_{B} \simeq 476\; \mu$m and $c\tau_{C} \simeq 213\; \mu$m)
 are much larger than the average SVXII impact parameter resolution
 ($\simeq 28\; \mu$m), the dominant factor determining the impact parameter
 distribution is the kinematics of the semileptonic decays which is well
 modeled by the {\sc evtgen} program.
 The impact parameter distribution of muons from prompt sources, such as
 quarkonia decays and Drell-Yan production, is constructed using muons from
 $\Upsilon(1S)$ decays (see Fig.~\ref{fig:fig_2}). Muons from $\pi$ and $K$
 in-flight decays are also regarded as prompt tracks since the track
 reconstruction algorithm rejects those with appreciable kinks. 
 Tracks associated with $\pi$ and $K$ mesons which mimic the lepton
 signal (fake muons) are mostly prompt. The small contribution to fake 
 muons of pion and kaon tracks arising from the decay of hadrons with
 heavy flavor is evaluated separately in Sec.~\ref{sec:ss-fake}.
%%%%%%%%%%%%%%%%%%%%%%%%%%
%%%%%%%%%%%%%%%%%%%%%%%%%
%%%%%%%%%%%%%%%%%%%%%%%%%%
%%%%%%%%%%%%%%%%%%%%%%%%%
  Since there are two muons in an event, the fit is performed in the 
  two-dimensional space of impact parameters. Each axis represents the impact
  parameter of one of the two muons. In filling the histograms, the 
  muon assignment is randomized. The two-dimensional impact
  parameter technique exploits the fact that muon impact parameters are
  independent uncorrelated variables~\footnote{
  The correlation between the two impact parameters, $\rho=\frac{\int \int
  (d_1-<d_1>)(d_2-<d_2>) \delta d_1 \delta d_2}{\sigma_{d_1} \sigma_{d_2}}$, 
  is approximately 0.01 in the data and the heavy flavor simulation.}. 
  The two-dimensional template distributions for each type of event are
  made by combining the relevant one-dimensional distributions in
  Fig.~\ref{fig:fig_1}.

\begin{figure}[htb]
 \begin{center}
% \vspace{-0.2in}
 \leavevmode
 \includegraphics*[width=\textwidth]{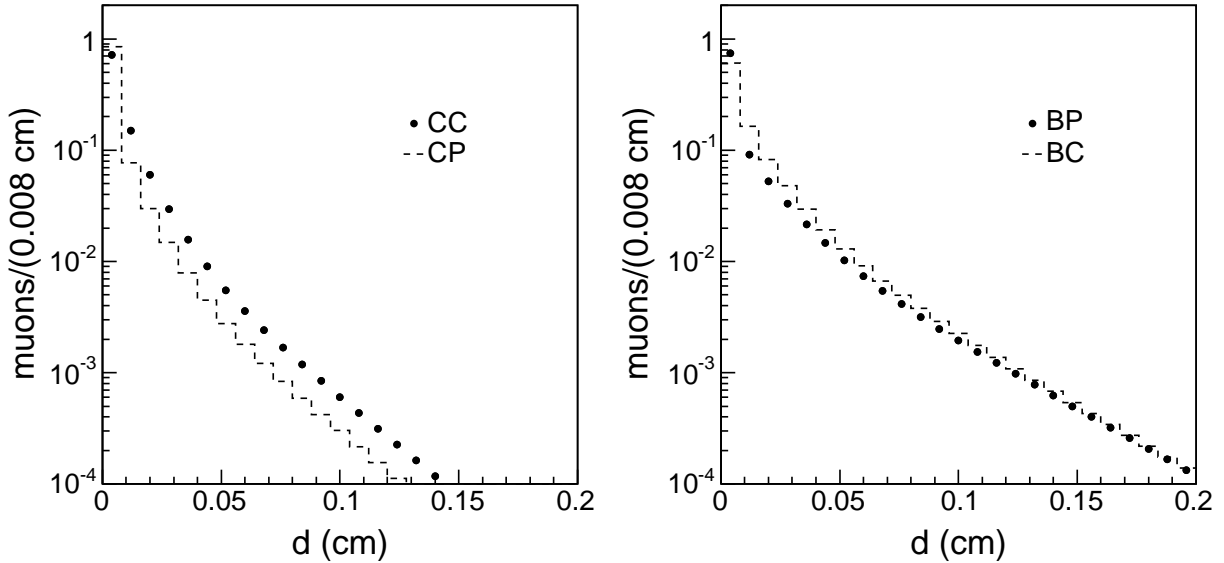}
 \caption[]{Projections of the two-dimensional impact parameter distributions 
            of some components used to fit the dimuon data
           (see text).}
 \label{fig:fig_3}
 \end{center}
 \end{figure}

  We use a binned maximum log likelihood method~\cite{minuit} to fit the
  dimuon impact parameter distribution. The likelihood function $L$
  is defined as
  \begin{eqnarray}
     L = \prod_i \prod_j [ l_{ij}^{n(i,j)} e^{-l_{ij}}/n(i,j)!] 
  \end{eqnarray}
  where $n(i,j)$ is the number of events in the $(i,j)$-th bin. The function 
  $l_{ij}$ is defined as
  \begin{eqnarray}
   l_{ij} &  = & BB \cdot S_b(i)\cdot S_b(j) + CC \cdot S_c(i)\cdot S_c(j) 
               + PP \cdot S_p(i)\cdot S_p(j) +   \\  \nonumber
   & &    0.5 \cdot [ BP \cdot (S_b(i)\cdot S_p(j) + S_p(i)\cdot S_b(j) ) +
        CP \cdot (S_c(i)\cdot S_p(j) + S_p(i)\cdot S_c(j) ) + \\ \nonumber
   & &     BC \cdot (S_b(i)\cdot S_c(j) + S_c(i)\cdot S_b(j) )  ]    
   \end{eqnarray}
  where $S_b$, $S_c$, and $S_p$ are the impact parameter templates shown in
  Fig.~\ref{fig:fig_1}. The fit parameters $BB$, $CC$, and $PP$ represent 
  the $b\bar{b}$, $c\bar{c}$, and prompt dimuon contributions, respectively.
  The fit parameter $BP$ ($CP$) estimates the number of events in which there
  is only one $b$ ($c$) quark in the detector acceptance and the second 
  lepton is produced by the decay or the misidentification of $\pi$ or $K$
  mesons~\footnote{
  According to the simulation, approximately 86\% of the $b\bar{b}$ and 
  $c\bar{c}$ events with an identified muon from heavy flavor decay do not 
  contain a second hadron with heavy flavor in the detector acceptance. 
  Therefore, following the procedure of Ref.~\cite{bmix}, we start by
  ignoring the small  fake muon contribution due to  $\pi$ and $K$ mesons 
  from heavy flavor decays, that is estimated in Sec.~\ref{sec:ss-fake}.}.
  The fit parameter $BC$ estimates the number of events in which
  both bottom and charmed quarks are  final state partons of the hard 
  scattering. According to the simulation, the $BC$ component is 
  $\simeq 4.6$\% of the $BB$ component and the $CP$ component is 
  $\simeq 83$\% of the $BP$ component~\footnote{
  In the simulation, events containing a muon from heavy flavor decay and
  a prompt track are mostly contributed by NLO diagrams, such as flavor 
  excitation and gluon splitting, in which a heavy flavor quark recoils
  against a gluon or a light quark. The cross section for producing at
  least one $c$ quark in the kinematic acceptance of this study is 2.6 times
  larger than that of a $b$ quark, but the contribution of NLO terms
  in $c\bar{c}$ production is approximately 3.6 times larger than for 
  $b\bar{b}$ production. However, the kinematic acceptance for muons 
  from $c$ decays is $\simeq 23$\% of that for $b$ decays.}.
  Figure~\ref{fig:fig_3} shows projections of the two-dimensional 
  distributions for each  type of  mixed contribution. By comparing with 
  Fig.~\ref{fig:fig_1}, one notes that the $BB$ and $PP$ components have
  impact parameter distributions markedly different from any other 
  contribution, whereas the $CC$, $CP$, $BP$, and $BC$ components have 
  quite similar shapes. Using Monte Carlo pseudoexperiments we have 
  verified that, as observed in previous studies~\cite{2mucdf,bmix},
  the likelihood function is not capable of disentangling these four 
  components. Therefore, Eq.~(1) is supplemented with the term
 \begin{eqnarray}
  0.5 \cdot (\frac{(CP-0.83\cdot BP)^2}{(CP+0.83^2 \cdot BP+(0.14 \cdot BP)^2)}
  + \frac{ (BC- 0.046 \cdot BB)^2} {(BC+0.046^2 \cdot BB+(0.013 \cdot BB)^2)})
\end{eqnarray}
 that constrains the ratios $CP/BP$ and $BC/BB$ to the values predicted
 by the simulation within their theoretical uncertainties approximated with 
 Gaussian functions~\footnote{
  Using other PDF fits available in the PDF library~\cite{pdf}, the 
  $c$-to-$b$ ratio of the flavor excitation cross section in the simulation 
  varies up to $\pm30$\%~\cite{topxsec}.
  The ratio of the $c$-to-$b$ gluon splitting cross section also changes 
  by $\pm30$\% when varying the $c$- and $b$-quark pole mass by 
  $0.5\; \gevcc$~\cite{topxsec}.
  We use as Gaussian uncertainty the 60\% variation divided by $\sqrt{12}$.
  The ratio of $bc$ to $b\bar{b}$ production depends on the $c$-quark 
  structure function, and varies up to $\pm 50$\% when using other PDF
  fits~\cite{pdf}.}.
%%%%%%%%%%%%%%%%%%%%%%%%%%
%%%%%%%%%%%%%%%%%%%%%%%%%
\section{Heavy flavor composition of the dimuon sample}\label{sec:ss-comp}
 In this section, we first determine the dimuon sample composition by 
 fitting the impact parameter distribution with the templates described in
 the previous section. We then evaluate and remove the contribution of muons 
 faked by pion or kaon tracks from heavy flavor decays. Lastly, we estimate
 the systematic uncertainty of the result due to the fit likelihood function
 and simulated templates.
%%%
\subsection{Result of the fit to the impact parameter distribution}
\label{sec:ss-res}
\begin{figure}[htb]
 \begin{center}
% \vspace{-0.2in}
 \leavevmode
 \includegraphics*[width=\textwidth]{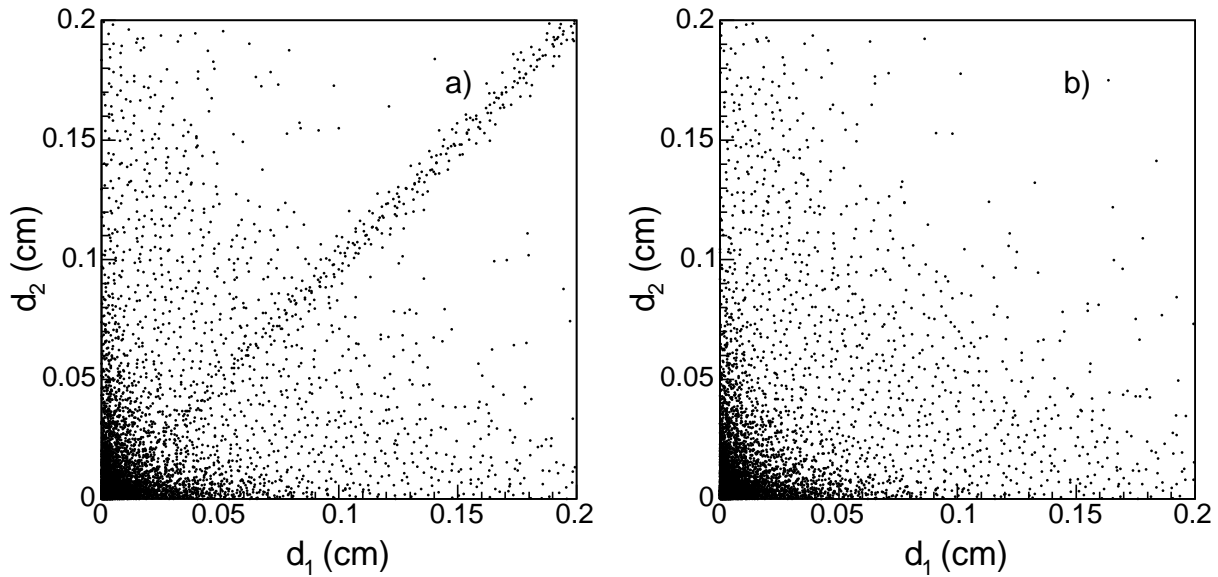}
 \caption[]{Two-dimensional impact parameter distributions of muon pairs 
            (a) before and (b) after cosmic removal.}
 \label{fig:fig_4}
 \end{center}
 \end{figure}

\begin{figure}[htb]
 \begin{center}
 \vspace{-0.2in}
 \leavevmode
 \includegraphics*[width=\textwidth]{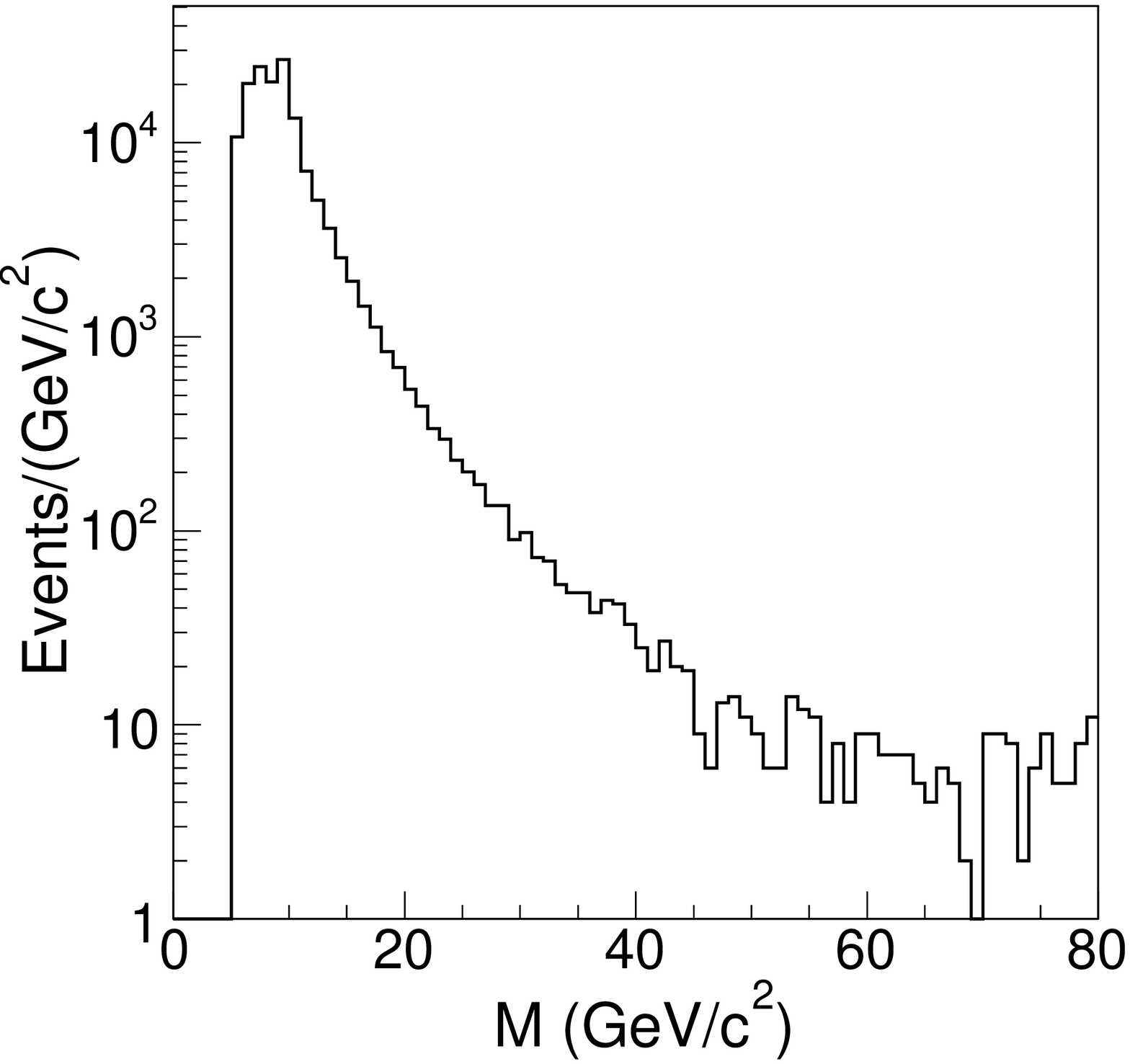}
 \caption[]{Invariant mass spectrum of the muon pairs used in this study.}
 \label{fig:fig_5}
 \end{center}
 \end{figure}

%%%%
  The two-dimensional impact parameter distribution of the 161948 
  muon pairs selected in this analysis is plotted in Fig.~\ref{fig:fig_4}(a).
%%%%%%%%%%%%%%%%%%%%%%%%%%
%%%%%%%%%%%%%%%%%%%%%%%%%
  An appreciable fraction of events cluster along the diagonal line
  $d_1 = d_2$. These events are due to cosmic rays and we remove them by
  requiring the azimuthal angle between muons with opposite charge to be
  smaller than 3.135 radians (see Fig.~\ref{fig:fig_4}(b)).
  In the simulation, the $\delta \phi \leq 3.135$ requirement has a 99.3\%
  efficiency when applied to muon pairs arising from $b\bar{b}$ and 
  $c\bar{c}$ production.
  The invariant mass spectrum of the remaining 143677 events is shown in  
  Fig.~\ref{fig:fig_5}.
 %%%%%%%%%%%%%%%%%%%%%%%%%%
%%%%%%%%%%%%%%%%%%%%%%%%% 

\begin{figure}[htb]
 \begin{center}
 \vspace{-0.2in}
 \leavevmode
 \includegraphics*[width=\textwidth]{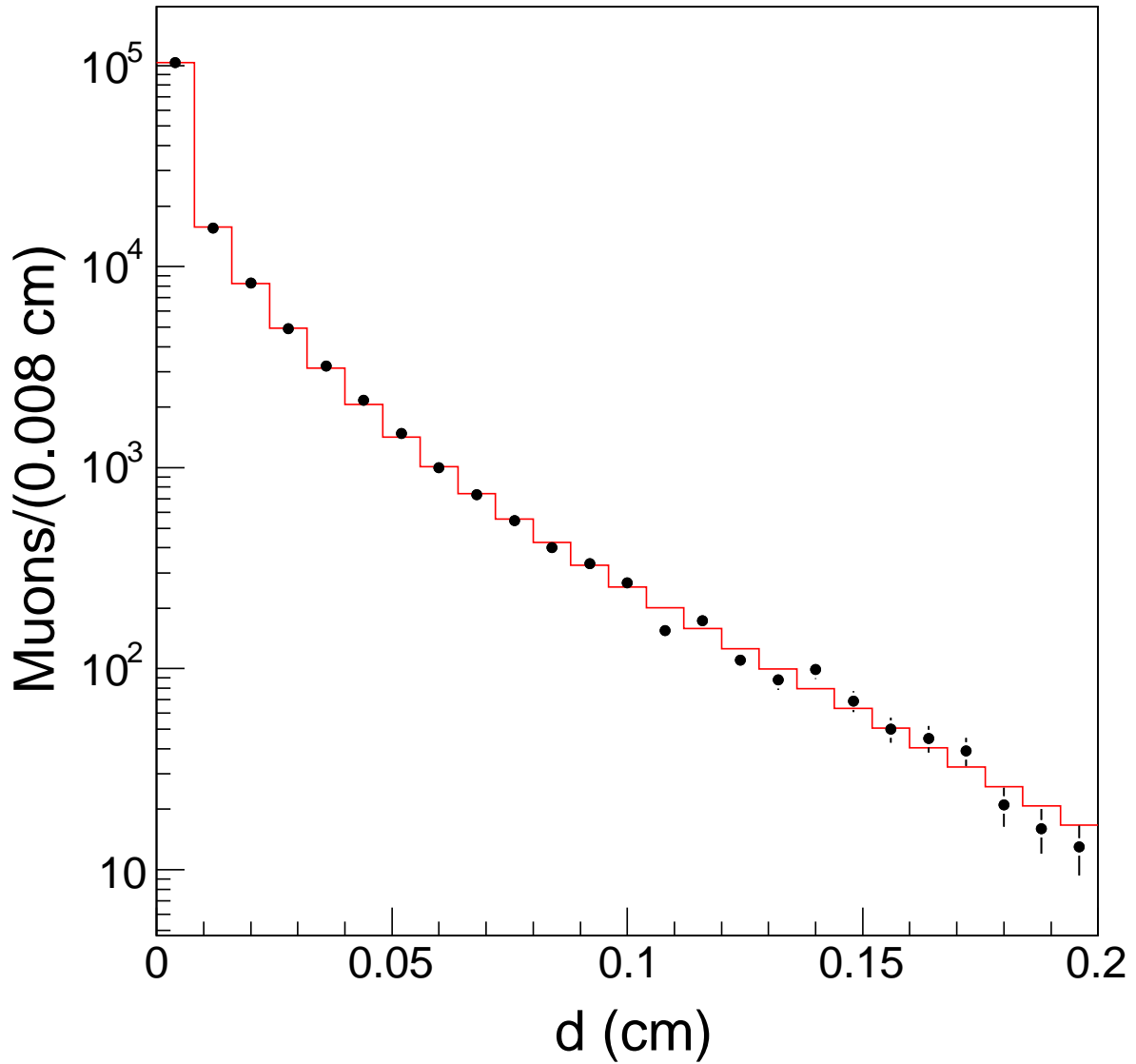}
 \caption[]{The projection of the  two-dimensional impact parameter
            distribution of muon pairs onto one of the two axes is 
            compared to the fit result (histogram).}
 \label{fig:fig_6}
 \end{center}
 \end{figure}

\begin{figure}[htb]
 \begin{center}
 \vspace{-0.2in}
 \leavevmode
 \includegraphics*[width=\textwidth]{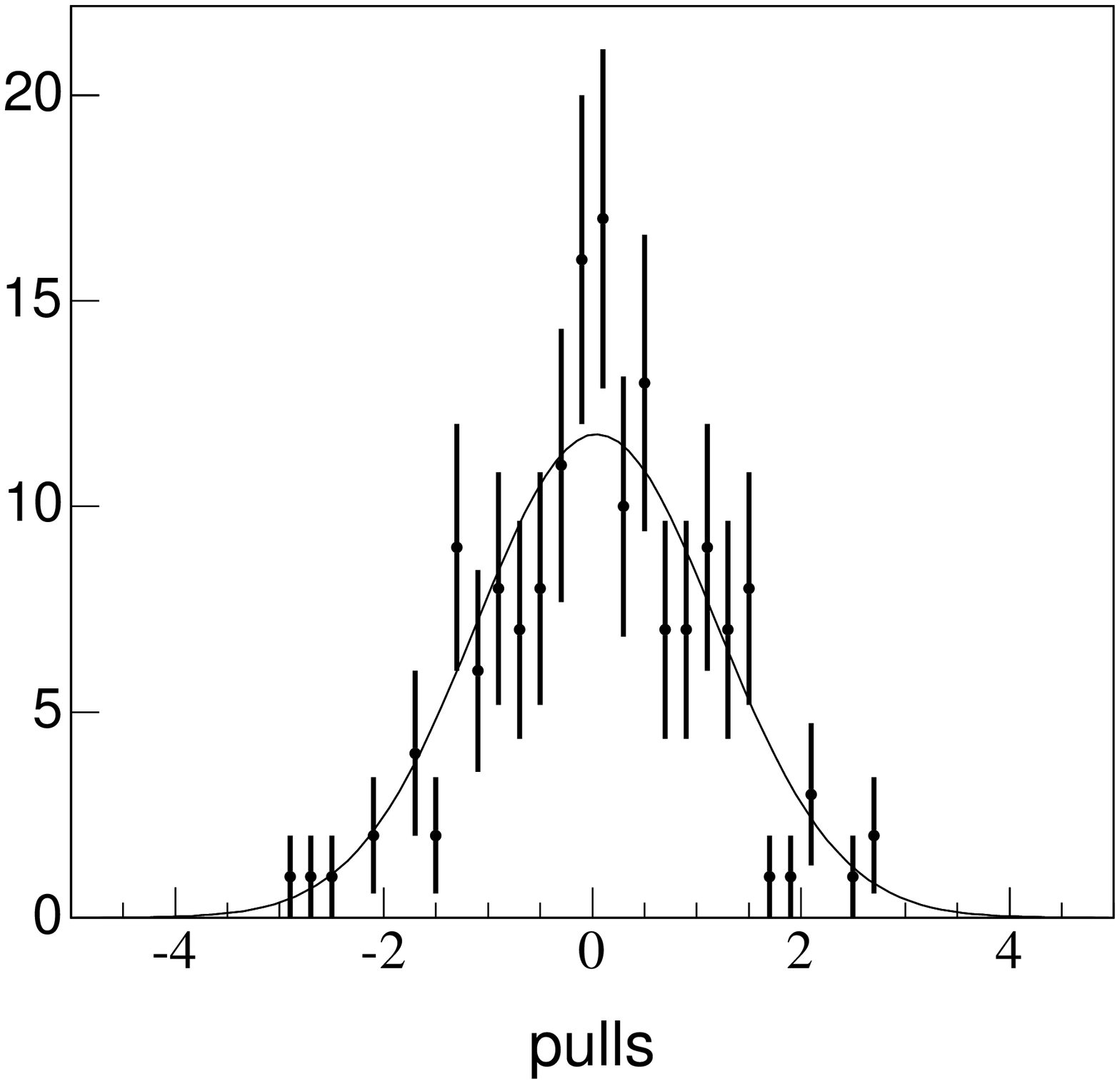}
 \caption[]{Distribution of the pulls - (data-fit)/$\sqrt{\rm fit}$ - of the 
            fit listed in Table~\ref{tab:tab_1} for the impact parameter bins
            with at least 10 entries. The solid line represents a Gaussian
 distribution with unit RMS.}
 \label{fig:fig_6bis}
 \end{center}
 \end{figure}

  The result of the fit to the two-dimensional impact parameter distribution
  of the data using the likelihood function in Eqs.~(1-3) is shown in 
  Table~\ref{tab:tab_1}. The parameter correlation matrix is listed in
  Table~\ref{tab:tab_2}.
%%%%%%%%%%%%%%%%%%%%%%%%%%%%%%%%%%
 \begin{table}
 \caption[]{Number of events attributed to the different dimuon sources  by
           the fit to the impact parameter distribution. 
           The errors correspond to a 0.5 change of $-\ln L$.}
\begin{center}
\begin{ruledtabular}
 \begin{tabular}{lc}
 Component  &  No. of Events      \\
  $BB$      & $54583 \pm 678$     \\ 
  $CC$      & $24458 \pm 1565$    \\
  $PP$      & $41556 \pm ~651$    \\
  $BP$      & $10598 \pm ~744$    \\
  $CP$      & $10024 \pm 1308$    \\
  $BC$      & $~2165 \pm ~693$    \\
 \end{tabular}
 \end{ruledtabular}
\end{center}
 \label{tab:tab_1}
 \end{table}
%%%%%%%%%%%%%%%%%%%%%%%%%%%%%%%%%%%%%%%%%%%
 \begin{table}
 \caption[]{Parameter correlation coefficients returned by the fit listed
           in Table~\ref{tab:tab_1}. }
\begin{center}
\begin{ruledtabular}
 \begin{tabular}{lccccc}
  Component & $BB$ & $CC$ & $PP$ & $BP$ & $CP$                \\
  $CC$  & $-0.46$  &          &         &   &                 \\
  $PP$  & $~0.09$  & $~0.18$  &         &   &                 \\
  $BP$  & $~0.01$  & $-0.43$  & $-0.14$ &   &                 \\
  $CP$  & $~0.27$  & $-0.69$  & $-0.71$ & $~0.13$  &          \\
  $BC$  & $-0.42$  & $-0.19$  & $~0.15$  & $-0.18$ & $-0.06$  \\
 \end{tabular}
 \end{ruledtabular}
 \end{center}
 \label{tab:tab_2}
 \end{table}
%%%%%%%%%%%%%%%%%%%%%%%%%%%%%%%%%%%%%%%%%%%
 The projection of the two-dimensional impact parameter distribution is 
 compared to the fit result in Fig.~\ref{fig:fig_6} and the distribution 
 of the fit residuals is plotted in Fig.~\ref{fig:fig_6bis}.
 The best fit returns  $-\ln L= 1078$. The probability of the fit to the
 data is determined with Monte Carlo pseudoexperiments. In each
 experiment, we randomly generate different components with average size as
 determined by the fit to the data~\footnote{ 
 We have performed 1000 pseudoexperiments starting from the following\
 component sizes: $BB_0=54600$, $CC_0=24500$, $PP_0 =41500$, $BP_0=10200$,
 $CP_0=10000$, and $BC_0=2200$.}
 and allow for Poisson fluctuations; the impact parameter distribution for
 each component is randomly generated from the corresponding templates used
 to fit the data. We find that 16.5\% of the fits to the pseudoexperiments
 return a $-\ln L$ value equal or larger than $1078$. The values of the 
 different components returned by the fits to each pseudoexperiment have 
 Gaussian distributions with sigmas equal to the corresponding errors 
 listed in Table~\ref{tab:tab_1}. The fit result is quite insensitive to 
 the constraint $CP/BB=0.83 \pm 0.14$ in Eq.~(3). If the uncertainty is
 increased from 0.14 to 0.28, the size of the $BB$ and $CC$ components 
 returned by the fit changes by less than half of the corresponding errors
 listed in Table~\ref{tab:tab_1}. However, without this constraint, the fit
 returns a $CC$ component 30\% smaller than the standard fit together with
 a ratio $CP/BP \simeq 3.5$ that would be difficult to account for. 
%%%%%%%%%%%%%%%%%%%%%%%%%%
%%%%%%%%%%%%%%%%%%%%%%%%%
%%%%%%%%%%%%%%%%%%%%%%%%%%
%%%%%%%%%%%%%%%%%%%%%%%%%
 In comparison with a previous CDF measurement~\cite{bmix} that uses data
 collected in the $1992-1995$ collider run (Run I), the ratio $PP$/$BB$
 returned by the fit has increased  from 34\% to 76\%. In the present study,
 we do not remove $\Upsilon$ candidates ($13800 \pm 290$) which represent 
 32\% of the $PP$ contribution. After removing the $\Upsilon$ contribution,
 the $PP$ contribution in the present data sample is 50\% larger than in
 the Run I data. This is explained by a substantial increase in the rate of
 muons faked by prompt tracks. Therefore, in the next subsection, we evaluate
 the fraction of the $BB$ yield due to muons faked by hadronic tracks from
 heavy flavor decays (this contribution was found negligible in the Run I
 data~\cite{bmix,2mucdf}).
%%%%
\subsection{Fake muon contribution}\label{sec:ss-fake}
%%%%
 We use two methods to estimate the contribution of tracks arising from
 heavy flavor decays that mimic a CMUP signal. The first method is based 
 on a combination of data and simulation. We use the simulation to estimate 
 the relative yields, $R_K$ and $R_{\pi}$, of $\mu-K$ and $\mu-\pi$ 
 combinations with respect to that of real muon pairs in  the decay of 
 hadrons with heavy flavor (we select muons and tracks with 
 $p_T \geq 3 \; \gevc$ and $|\eta| \leq 0.7$, and we require the invariant
 mass of each pair to be larger than 5 $\gevcc$). These yields are listed in 
 Table~\ref{tab:tab_3}. The ratio of dimuons contributed by $\mu-$track 
 combinations to real dimuons from semileptonic decays is
 $F=(R_K \cdot P_f^{K} + R_{\pi} \cdot P_f^{\pi})/\epsilon_{\mu}$,
 where $P_f^{K}$ ($P_f^{\pi}$) is the probability that a pion (kaon) track
 mimics a muon signal. These probabilities are determined using
 a sample of $D^0 \rightarrow \pi K$ decays in App.~B. The corresponding
 efficiency for detecting a real muon is $\epsilon_{\mu}=0.5057$
 (see Table~\ref{tab:tab_9}). The errors
 in Table~\ref{tab:tab_3} are the sum in quadrature of statistical errors
 and the 10\% systematic uncertainty of the kaon and pion rates predicted
 by the simulation. This  systematic uncertainty is derived from a 
 comparison of  kaon and pion production rates measured at the
 $\Upsilon(4S)$ to the prediction of the {\sc evtgen} Monte Carlo
 program~\cite{babar1}. 
 %%%%%%%%%%%%%%%%%%%%%%%%%%%%%%%%%%
 \begin{table}[htp]
 \caption[]{Ratio of the numbers of $\mu-K(\pi)$ combinations to that of
            $\mu-\mu$ pairs, $R_{K(\pi)}$, in the simulation of different
            heavy flavor productions. The ratio $F$ of the number of 
            fake-real muon pairs to that of real dimuons is estimated using
            the fake muon probabilities derived in App.~B and the measured 
            detector efficiency $\epsilon_\mu=0.5057$ for a real muon.}
 \begin{center}
 \begin{ruledtabular}
 \begin{tabular}{lccc}
  Production   &  $R_K$ &  $R_\pi$ &  $F$ (\%)    \\
  $b\bar{b}$   & $~3.70 \pm 0.43 $  & $ ~7.58 \pm 0.82$  &  $~7.2 \pm 0.6$ \\ 
  $c\bar{c}$   & $21.73 \pm 2.51 $ & $23.47 \pm 2.68$  &  $32.4\pm 2.7 $   \\
  $bc$         & $16.78 \pm 2.71 $ & $10.83 \pm 2.03$  &  $21.5\pm 2.8 $   \\
 \end{tabular}
 \end{ruledtabular}
 \end{center}
 \label{tab:tab_3}
 \end{table}
 
\begin{figure}
 \begin{center}
  \leavevmode
  \includegraphics*[width=\textwidth]{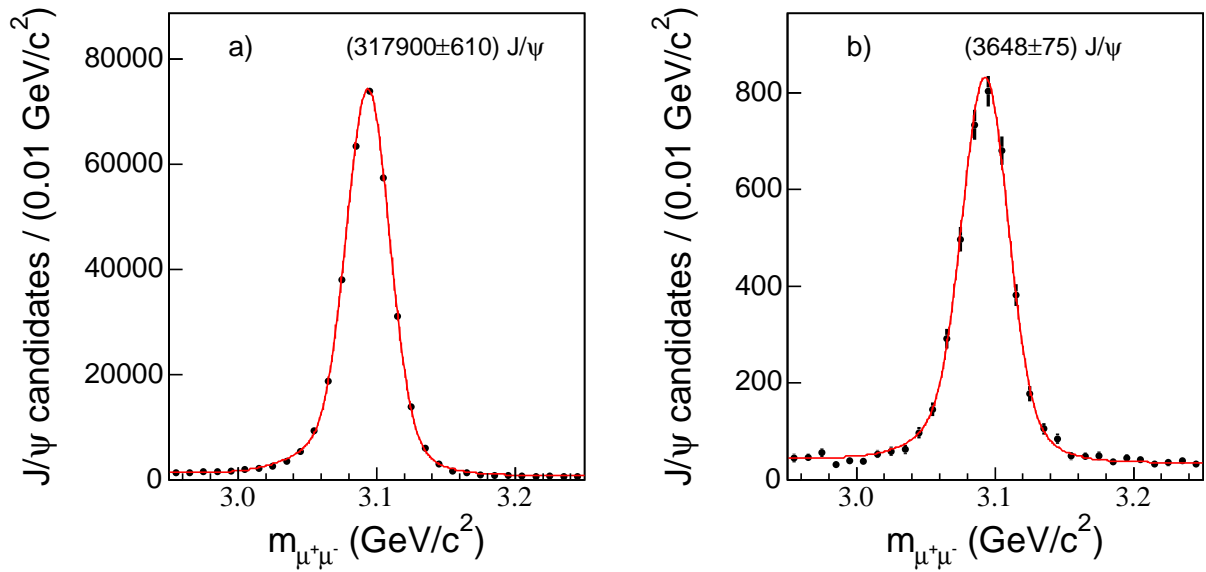}
  \caption[]{Invariant mass distribution of CMUP muon pairs with a randomly
             chosen muon that (a) satisfies or (b) fails the $\chi^2<9$
             requirement.}
  \label{fig:fig_7}
 \end{center}
\end{figure}

 We evaluate the purity $(1/(1+F))$ with a second method that is almost 
 independent of the simulation prediction. We make use of stricter muon 
 selection criteria by supplementing the $\Delta r \phi$ cut between the 
 muon track projection and the CMU and CMP stubs with the requirement, 
 referred to as $\chi^2$ cut, that the 
 extrapolated COT track and the CMU muon stub match within $3\;\sigma$ in 
 the $r-\phi$ plane, where $\sigma$ is a standard deviation that includes 
 the effect of multiple scattering and energy loss. The efficiency of the
 $\chi^2$ cut for real muons is measured using a sample of muons acquired
 with the $J/\psi$ trigger. We compare the invariant mass distributions 
 of CMUP pairs when a randomly chosen muon passes or  fails the
 $\chi^2 \leq 9$ cut. We fit the data with two Gaussian functions to model 
 the $J/\psi$ signal and a straight line to model the background.  
 The $\eta$ and $p_T$ distribution of CMUP muons from $J/\psi$ decays
 are weighted to model that of muons from $b$-hadron decays. As shown in
 Fig.~\ref{fig:fig_7}, the  $\chi^2$ cut reduces the efficiency for detecting
 a muon pair by $\epsilon_{\rm ineff}= 2.20\pm 0.04 $\%.
%%%%%%%%%%%%%%%%%%%%%
%%%%%%%%%%%%%%%%%%%%%%%%%%
%%%%%%%%%%%%%%%%%%%%%%%%%

  The corresponding fake muon probabilities are measured using
  $D^0 \rightarrow K \pi$ decays and are listed in Table~\ref{tab:tab_appb1}
  of App.~B.
  We select a sample of dimuons enriched in fake muons by requiring $\chi^2>9$
  for one muon and determine its heavy flavor composition by fitting the 
  impact parameter distribution. 
  The fit result is shown in Table~\ref{tab:tab_4}.
%%%%%%%%%%%%%%%%%%%%%%%%%%%%%%%%%%
 \begin{table}
 \caption[]{Number of events attributed to the different dimuon sources by
           the fit to the impact parameter distribution. We use events in
           which at least one muon  fails the $\chi^2 >9$ requirement 
           (see text).}
\begin{center}
\begin{ruledtabular}
 \begin{tabular}{lc}
 Component  &  No. of Events    \\
  $BB$      & $1103 \pm 102$    \\ 
  $CC$      & $1189 \pm 272$    \\
  $PP$      & $4249 \pm 131$    \\
  $BP$      & $1508 \pm 136$    \\
  $CP$      & $1218 \pm 194$    \\
  $BC$      & $~~51 \pm ~15$    \\
 \end{tabular}
 \end{ruledtabular}
 \end{center}
 \label{tab:tab_4}
 \end{table}
%%%%%%%%%%%%%%%%%%%%%%%%%%%%%%%%%%%%%%%%%%%
 For each heavy flavor component, we derive the fake muon contribution by
 solving the system of equations
 \begin{eqnarray}
  T &=& HF + P_f \cdot FK \\ \nonumber
  T(\chi^2 >9) &=& \epsilon_{\rm ineff} \cdot HF + P_f(\chi^2 >9) \cdot FK ,
 \end{eqnarray}
  where $T$ and $T(\chi^2 >9)$ are the size of the component determined by 
  the fits in Table~\ref{tab:tab_1} and~\ref{tab:tab_4}, respectively,
  $HF$ is the number of real muon pairs, and $FK$ is the number of
  dimuons one of which is faked by a track from heavy flavor decays.
  The fraction of real muon pairs reads
 \begin{eqnarray}
   1/(1+F)&=& \frac{ P_f(\chi^2 >9)\cdot T - P_f\cdot T(\chi^2 >9)  }
   {T\cdot (P_f(\chi^2 >9) - 0.022\cdot P_f)} \\ \nonumber
   & & \pm    \frac {P_f}{ T \cdot (  P_f(\chi^2 >9) - 0.022\cdot P_f) } 
              \sqrt{\delta T^2(\chi^2 >9)+
    T^2(\chi^2 >9)/T^2 \cdot \delta T^2 }.
 \end{eqnarray}
 This second method provides a determination of the fraction of real dimuons
 almost independent  of the pion and kaon rate predicted by the simulation.
 The fraction of real muon pairs determined with the two methods is
 shown in Table~\ref{tab:tab_5}. We use the average and take
 the maximum and minimum RMS deviation as systematic uncertainty 
 ($0.96\pm 0.04$ for $b\bar{b}$ and $0.81\pm 0.09$ for $c\bar{c}$ production).
 The contribution of pairs of muons that are both faked by tracks from 
 heavy flavor decays has been estimated to be less than 0.4\% in the 
 worst case, and it is ignored. 
%%%%%%%%%%%%%%%%%%%
 \begin{table}[htp]
 \caption[]{Fractions of real dimuons due to heavy flavor, $1/(1+F)$,
            determined with the simulation or by using the results returned
            by fits to the impact parameter distributions of all muon pairs 
            and of those pairs in which at least one muon fails the
            $\chi^2 \leq 9$ cut.}
 \begin{center}
 \begin{ruledtabular}
 \begin{tabular}{lcc}
  Production   & Simulation      &  $\chi^2 >9$     \\
  $b\bar{b}$   & $0.93 \pm 0.01$ & $1.01 \pm 0.01$  \\ 
  $c\bar{c}$   & $0.76 \pm 0.02$ & $0.86 \pm 0.06$  \\
 \end{tabular}
 \end{ruledtabular}
 \end{center}
 \label{tab:tab_5}
 \end{table}
%%%%%%%
\subsection{Results after  fake removal} 
\begin{figure}
 \begin{center}
  \leavevmode
 \includegraphics*[width=\textwidth]{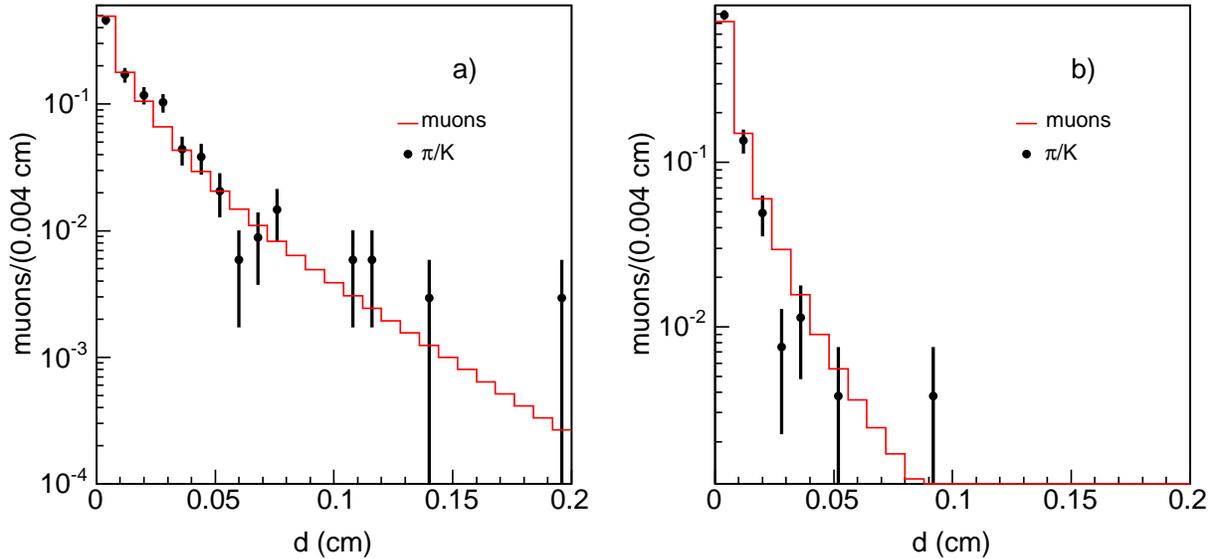}
  \caption[]{Simulated impact parameter distributions of pion and kaon tracks
             from (a) $b$- and (b) $c$-quark decays are compared to those of
             muons from semileptonic decays with the same kinematical
             requirements.}
  \label{fig:fig_7bis}
 \end{center}
\end{figure}

%%%%%%%
  Table~\ref{tab:tab_6} lists the various heavy flavor contributions
  to the dimuon sample after removing the contribution of tracks from
  heavy flavor decays that mimic a muon signal. As shown in 
  Table~\ref{tab:tab_5}, the contribution of muons faked by tracks from
  $c$-quark decays is not negligible. Therefore, we search the simulation
  for combinations of muons from $b$ semileptonic decays and pion 
  or kaon tracks from $b$- or $c$-quark decays (both with $p_T \geq 3 \; \gevc$
  and $|\eta| \leq 0.7$).
  Figure~\ref{fig:fig_7bis} compares impact parameter distributions of these 
  tracks to the standard muon templates. Distributions for muons and hadrons
  are quite similar. We have fitted the data with templates that include the 
  expected contribution of muons faked by tracks from heavy flavor decays as 
  listed in Table~\ref{tab:tab_5}. The result of this fit differs by less
  than 0.1\% from that of the standard fit in Table~\ref{tab:tab_1}.
%%%%%%%%%%%%%%%%%%%%%%%%%%%%%%%%%%
 \begin{table}[htp]
 \caption[]{Number of real muon pairs from heavy flavor sources after 
            removing the fake muon contributions. Errors include the 
            uncertainty of the fake removal.}
\begin{center}
\begin{ruledtabular}
 \begin{tabular}{lc}
 Component  &  No. of Events       \\
  $BB$      & $52400 \pm 2278$     \\ 
  $CC$      & $19811 \pm 2540$     \\
 \end{tabular}
 \end{ruledtabular}
 \end{center}
 \label{tab:tab_6}
 \end{table}
%%%%%%%%%%%%%%%%%%%%%%%%%%%%%%%%%%%%%%%%%%%
%%%%%%%%%%%%%%%%%%%%%%%%%%
%%%%%%%%%%%%%%%%%%%%%%%%%
 
  For completeness, we use the simulation to verify the ratio of the $BP$
  to $BB$ components returned by the fit performed in Sec.~\ref{sec:ss-comp}.
  The fit yields a ratio $BP/BB= 0.194 \pm 0.013$. We search the simulation
  for combinations of muons from $b$ semileptonic decays and prompt pion 
  or kaon tracks (both with $p_T \geq 3 \; \gevc$ and $|\eta| \leq 0.7$).
  The ratio of their number to that of dimuons from $b$ semileptonic
  decays is $32.8 \pm 0.6$ (stat.). Since the efficiency for detecting a 
  muon is $\epsilon_\mu=0.5$ and the probability that  prompt tracks fake
  a muon signal is 0.0032, the simulation predicts $BP/BB=0.21 \pm 0.01 $,
  in fair agreement with the fit result even without considering the 
  uncertainty of the rate of prompt pions and kaons predicted by the 
  simulation. In $c\bar{c}$ data, approximately 20\% of the muons are faked 
  by hadronic tracks from $c$-quark decays. This is in agreement with the
  result that $CP/BP$ is $ 0.95 \pm 0.14$ in the data and 0.83 in the heavy
  flavor simulation.
%%%%% 
\subsection{Dependence of the result on the muon {\boldmath $p_T$}
            distribution}\label{sec:ss-cross}
%%%%%
  The impact parameter of a track arising from heavy flavor decays depends
  on the proper decay time of the parent hadron and the decay angle between
  the daughter track and the parent hadron in its rest frame. Because we
  select muons above a given $p_T$ threshold, the range of accepted decay
  angles shrinks as the $p_T$ difference between the daughter track and 
  parent hadron decreases. Therefore, impact parameter templates have a 
  small dependence on the transverse momentum distribution of muons 
  (or parent heavy flavor) in the simulation.

\begin{figure}
 \begin{center}
 \leavevmode
 \includegraphics*[width=0.5\textwidth]{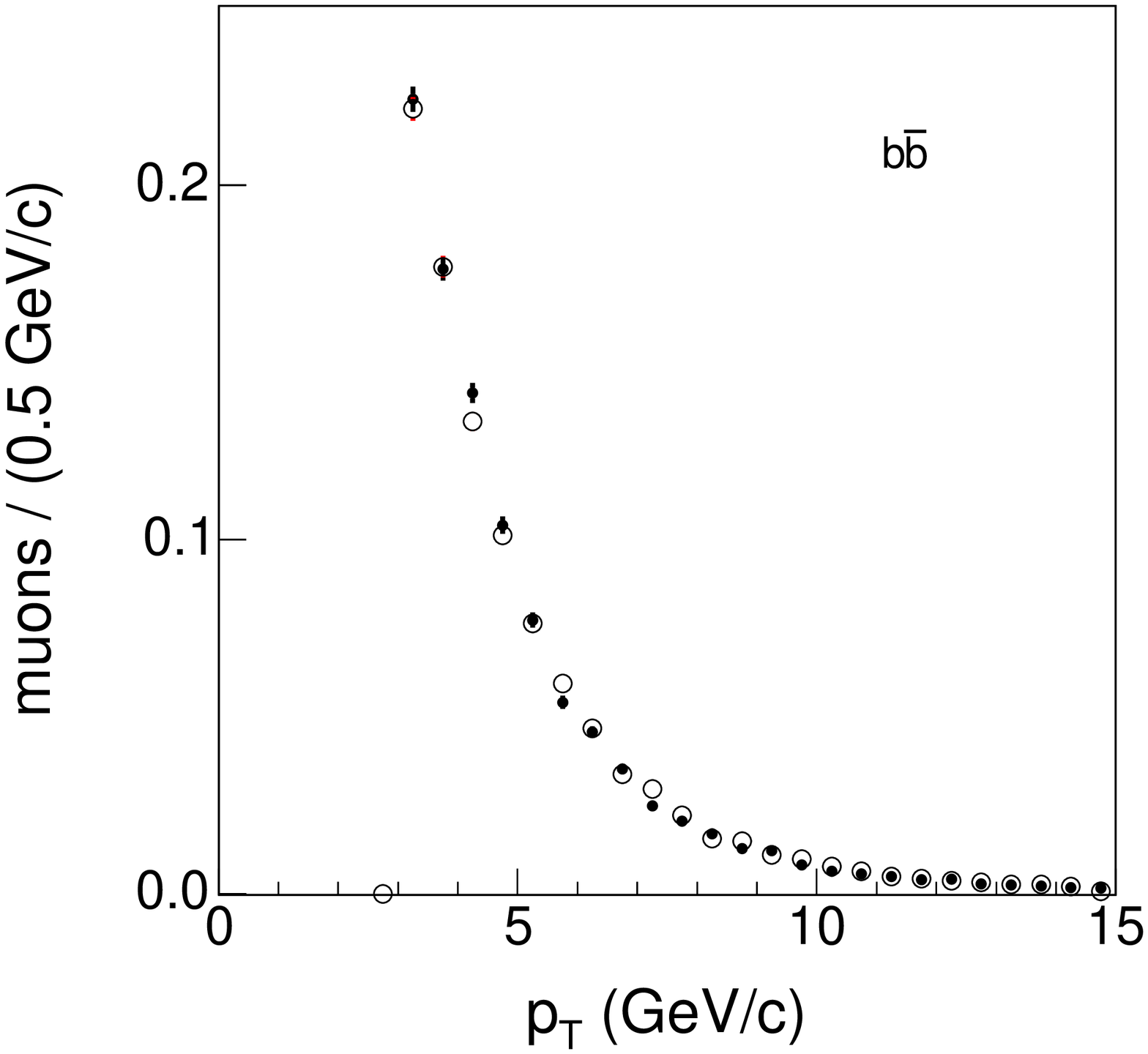}\includegraphics*[width=0.5\textwidth]{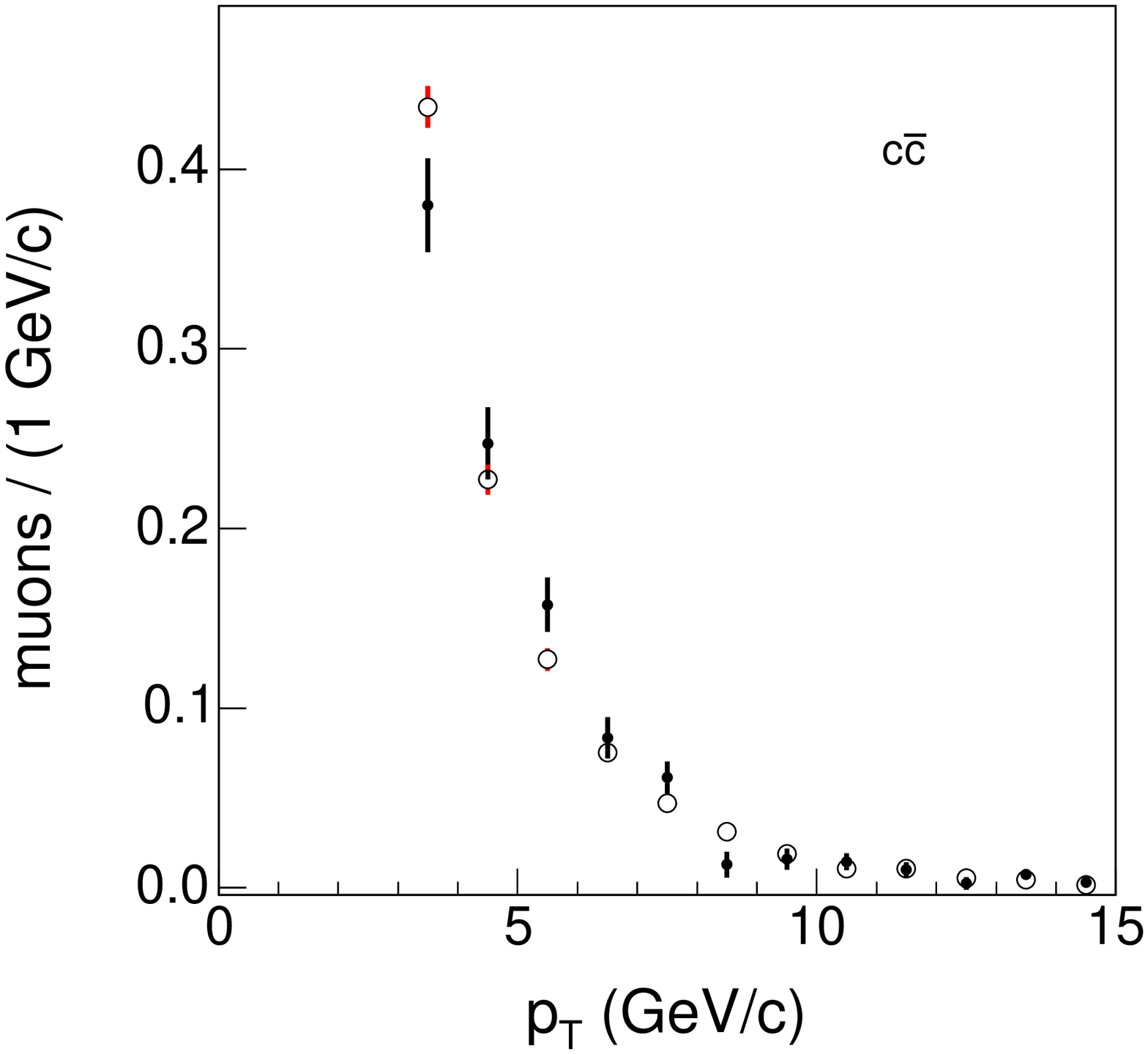}
 \caption[]{Transverse momentum distributions in the data ($\bullet$)
            and simulation ($\circ$) for muon pairs arising from 
            (left) $b \bar{b}$ and (right) $c\bar{c}$ production. 
            Distributions are normalized to unit area.}
  \label{fig:fig_8}
 \end{center}
\end{figure}

  In the data, we derive transverse momentum distributions for muons from
  $b$- and $c$-hadron decays by using the $_s{\cal P}lot$ statistical
  method~\cite{splot}. We call $f_n$ one of the six components used in the
  likelihood function $L$ in Eq.~(2), $N_n$ the number of events attributed
  to this component by the fit in Table~\ref{tab:tab_1},
  and $N$ the total number of events~\footnote{
  For example, $f_1 =  S_b \cdot S_b$ and  $N_1=BB$.}.
  Given an event $e$ in which  muons have impact parameters in the 
  $(i_e,j_e)$-th bin, the probability that the event belongs to the $n$-th
  component is
 \begin{eqnarray}
   P_n(i_e,j_e)= \frac {\sum_{l=1}^{6} V_{nl}\cdot f_l(i_e,j_e) }
  {\sum_{m=1}^{6} N_m \cdot f_m(i_e,j_e)     } \;,
 \end{eqnarray}
  where
 \begin{eqnarray}
  V_{nl}^{-1}= \frac{\sum_{e=1}^{N} f_n(i_e,j_e) \cdot f_l(i_e,j_e)}
  {(\sum_{m=1}^{6} N_m \cdot f_m(i_e,j_e))^2  } \;.
 \end{eqnarray}
  The transverse momentum distribution of muons from $b\bar{b}$ and
  $c\bar{c}$ production is obtained by weighting the muon transverse 
  momenta in the event $e$ by the corresponding probabilities 
  $ P_1(i_e,j_e)$ and $P_2(i_e,j_e)$, respectively.
  The corresponding errors have been evaluated with Monte Carlo
  pseudoexperiments. These distributions are compared to those in the 
  simulation in Fig.~\ref{fig:fig_8}.
 %%%%%%%%%%%%%%%%%%%%%%%%%%
%%%%%%%%%%%%%%%%%%%%%%%%%%
  The $b\bar{b}$ data are quite well modeled by the {\sc herwig} generator.
  We estimate the dependence of the fit result on the muon $p_T$ spectrum in
  the heavy flavor simulation by fitting the function $A \cdot p_T^\alpha$ to
  the ratio of these distributions in the simulation and in the data.
  The fit returns $\alpha=-0.029 \pm 0.015$. We have constructed templates 
  by reweighting the simulated  muon transverse momentum distribution with
  the function $p_T^ {\pm 0.044}$. When these templates are used,
  the $BB$ and $CC$ yields returned by the fit change by $\pm 1.5$\%, and
  $\mp 4$\%, respectively. We do not correct our result for this effect, 
  but we add this variation to other systematic effects evaluated in
  Sec.~\ref{sec:ss-cross1}.

\begin{figure}
 \begin{center}
 \leavevmode
 \includegraphics*[width=0.5\textwidth]{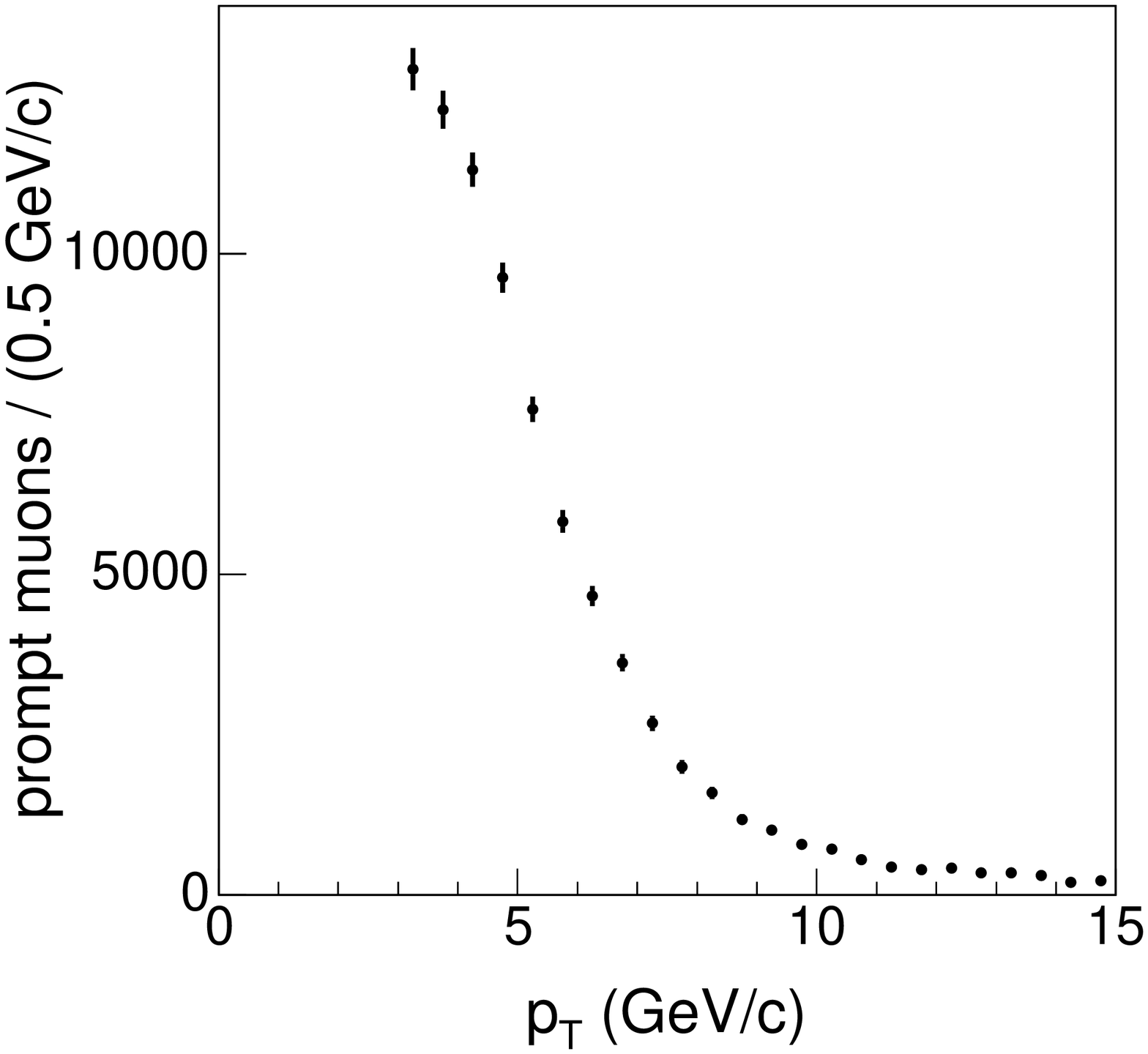}\includegraphics*[width=0.5\textwidth]{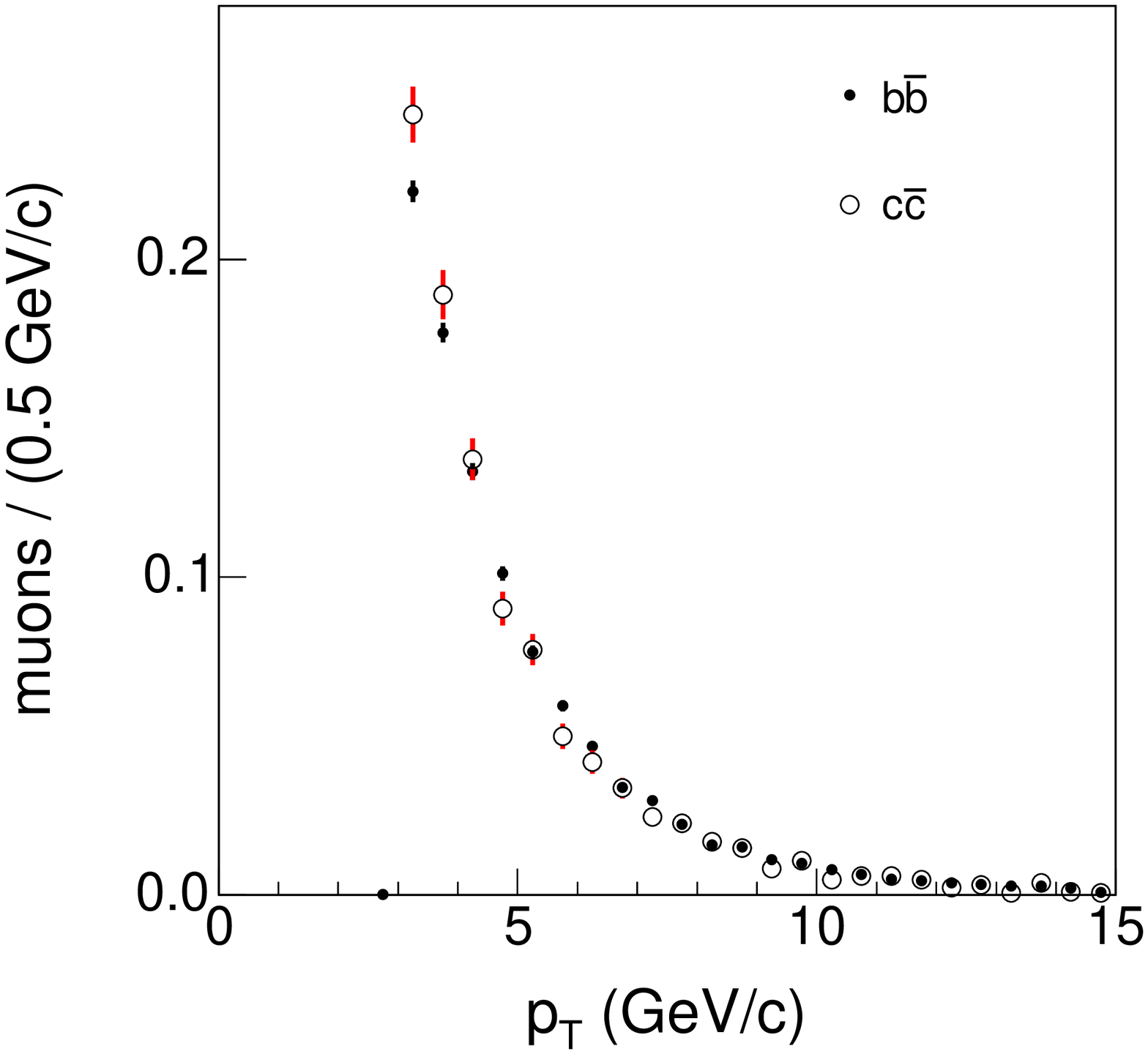}
 \caption[]{Transverse momentum distribution of (left) prompt muons in the
            data and (right) muons from $b$- or $c$-hadron decays in the
            simulation. Simulated distributions are normalized to unit area.}
 \label{fig:fig_8bis}
 \end{center}
 \end{figure}

\begin{figure}
 \begin{center}
 \leavevmode
 \includegraphics*[width=0.5\textwidth]{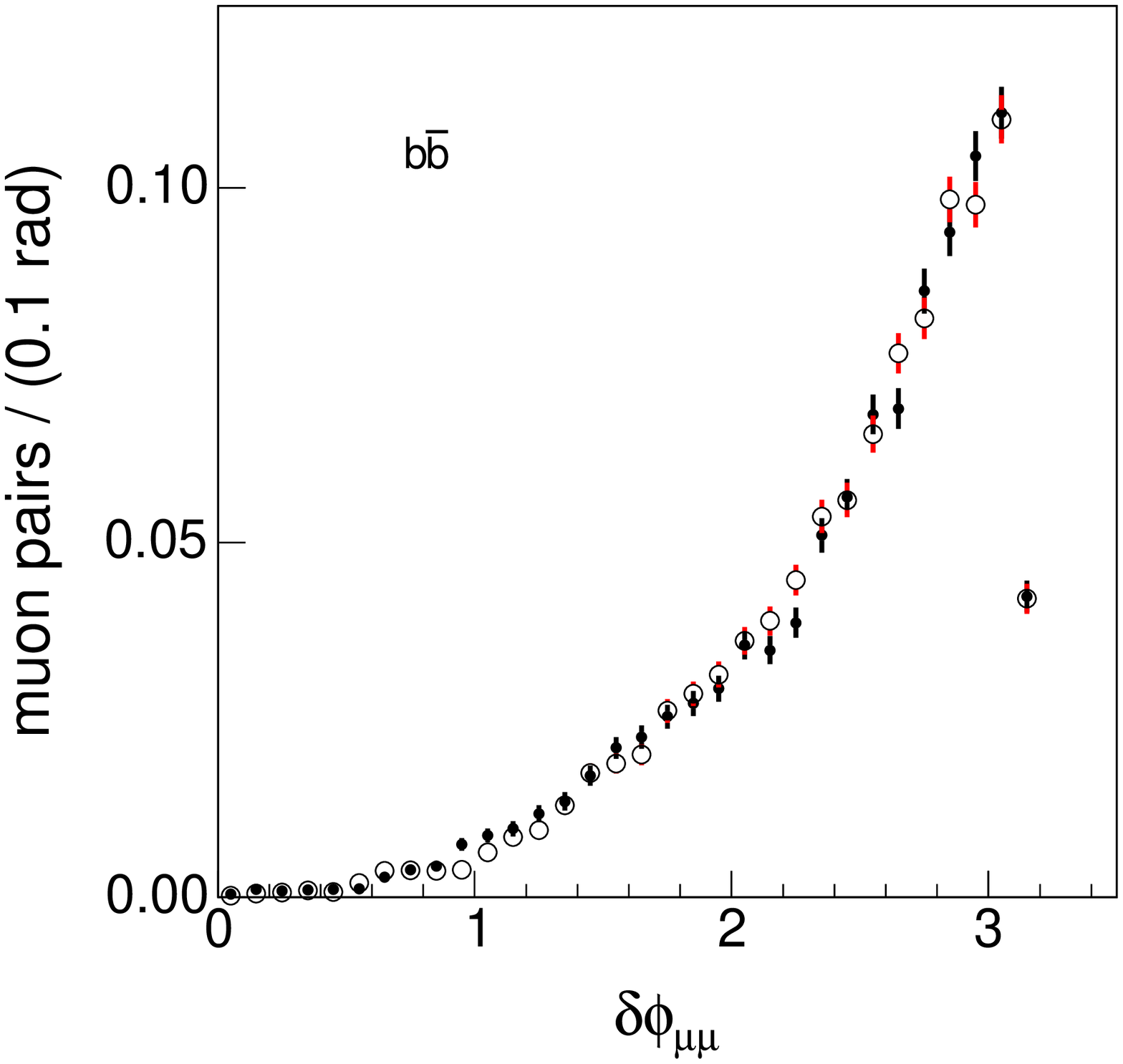}\includegraphics*[width=0.5\textwidth]{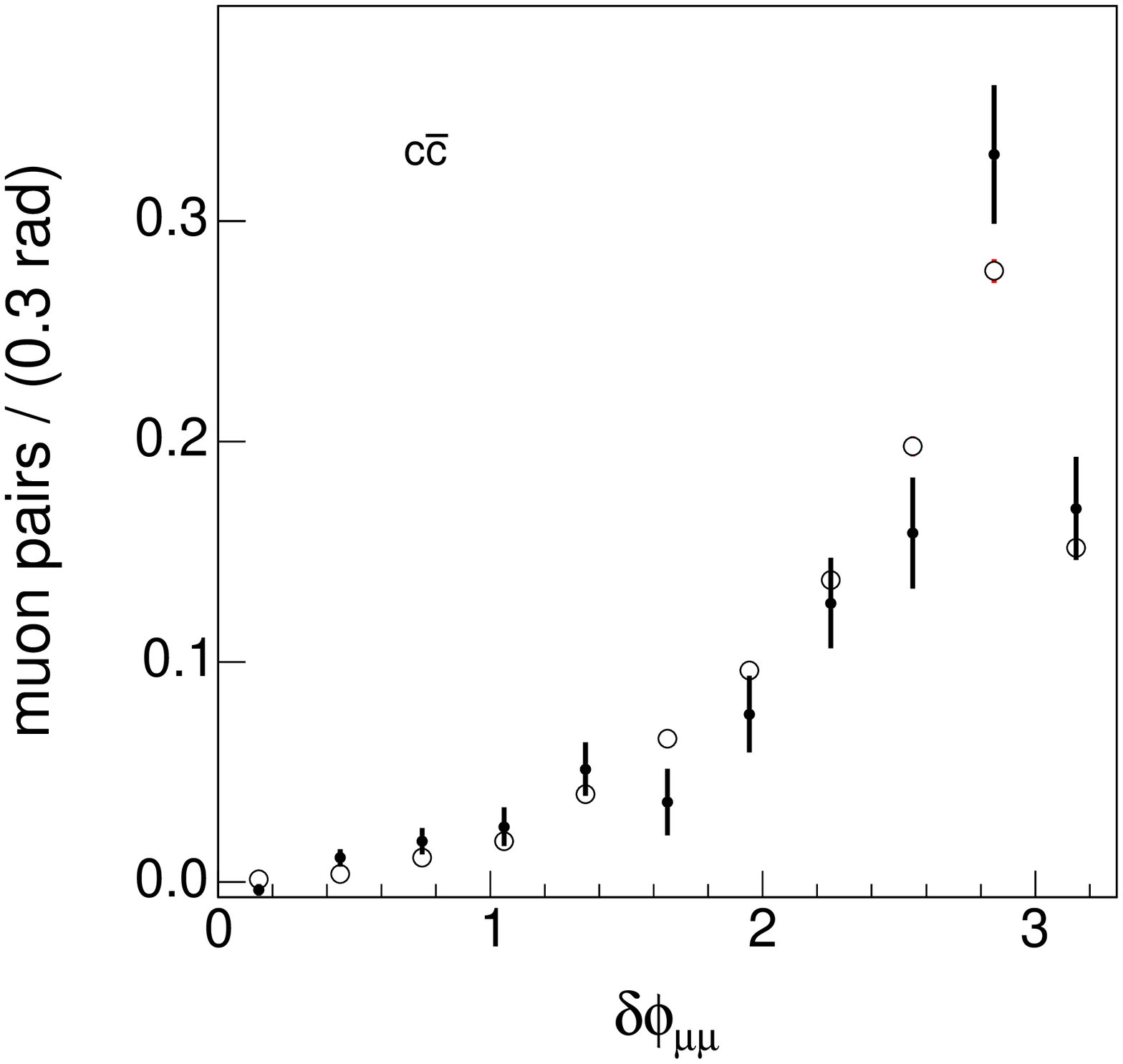}
 \caption[]{Distributions of the opening azimuthal angle between two muons
            from (left) $b \bar{b}$  and (right) $c\bar{c}$ production
            in the data ($\bullet$) and simulation ($\circ$). Distributions 
            are normalized to unit area.}
 \label{fig:fig_9}
 \end{center}
 \end{figure}

\begin{figure}
 \begin{center}
 \leavevmode
 \includegraphics*[width=0.5\textwidth]{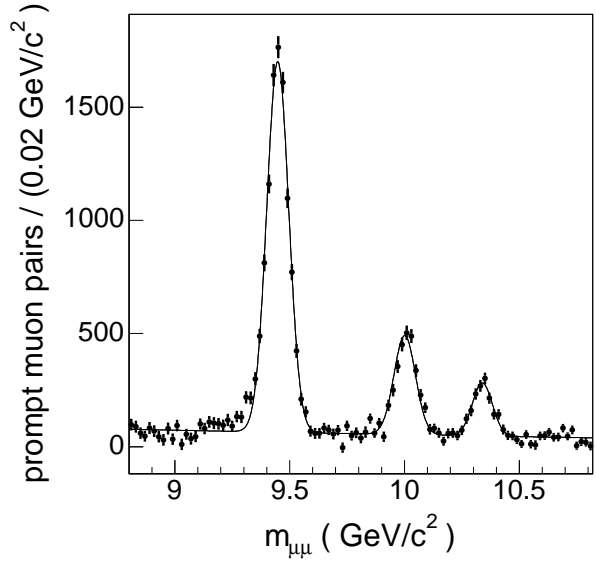}
 \caption[]{Distribution of the invariant mass of the prompt dimuon
             component as determined by the fit to the muon impact parameter.
             The solid line represents a fit to the distribution that returns 
             $9899 \pm 142$  $\Upsilon(1S)$ candidates, whereas the same fit 
             to the data in Fig.~\ref{fig:fig_2} yields $9952 \pm 122$
             candidates.  }
 \label{fig:fig_9bis}
 \end{center}
 \end{figure}

\begin{figure}
 \begin{center}
 \leavevmode
 \includegraphics*[width=0.5\textwidth]{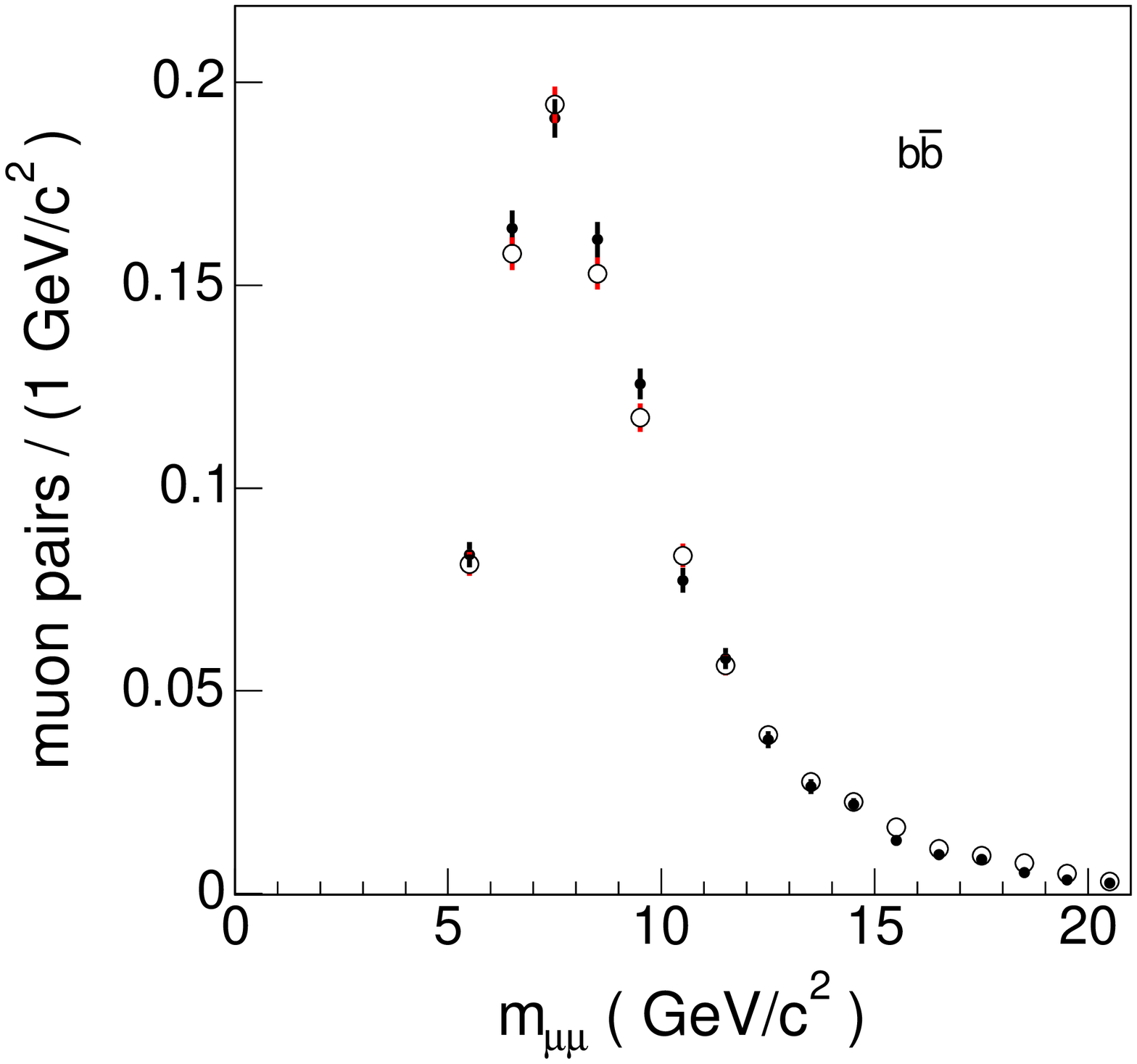}\includegraphics*[width=0.5\textwidth]{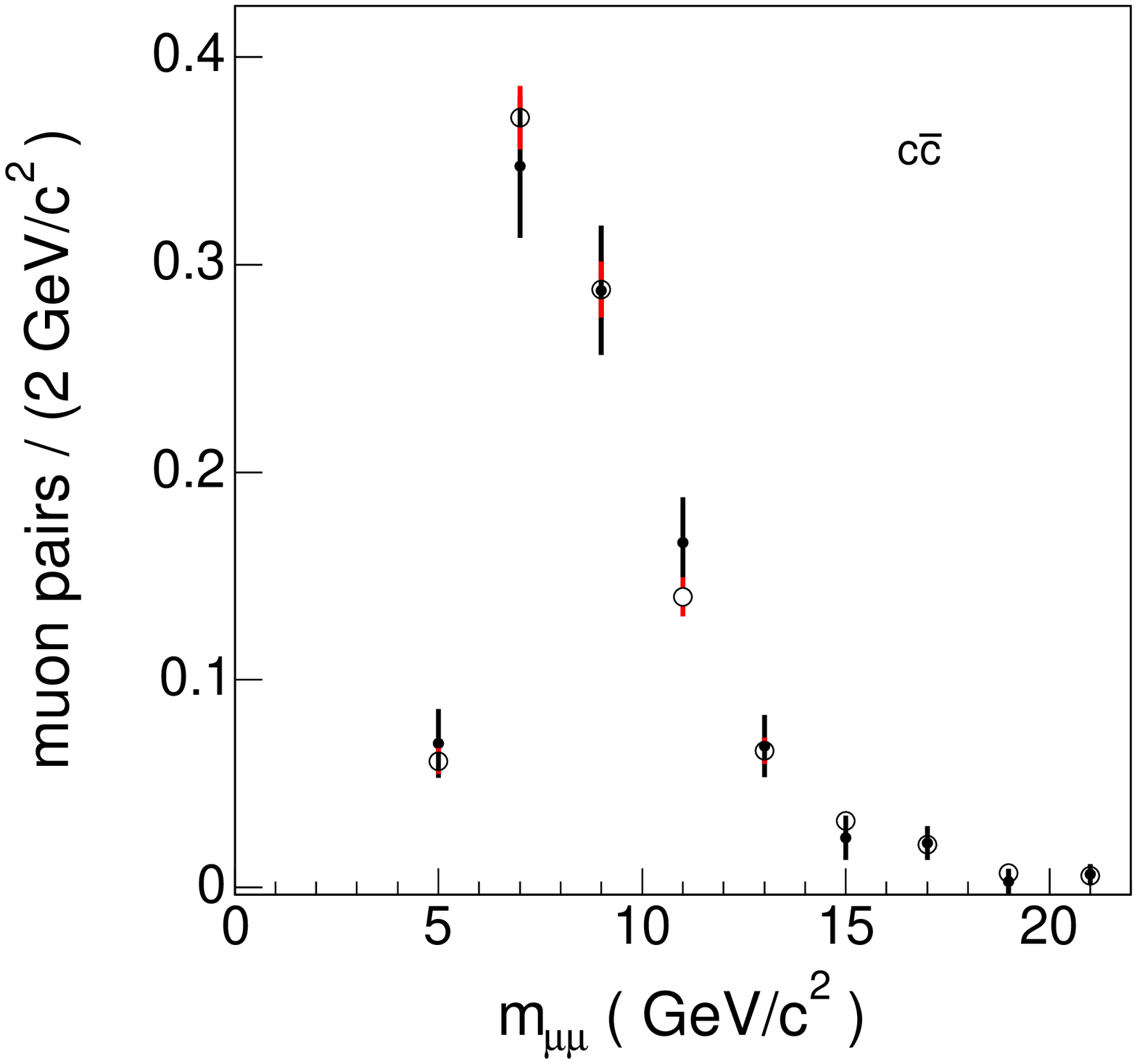}
 \caption[]{Distributions of the invariant mass of muon pairs from (left)
            $b \bar{b}$  and (right) $c\bar{c}$ production in the data
           ($\bullet$) and simulation ($\circ$). Distributions are 
           normalized to unit area.}
 \label{fig:fig_9tris}
 \end{center}
 \end{figure}

  For completeness, we use the probability defined in Eq.~(6) to show:
  a comparison of transverse momentum distributions of prompt muons and muons
  from simulated heavy flavor decays in Fig.~\ref{fig:fig_8bis}; distributions
  of $\delta \phi$, the azimuthal opening angle between two muons, in the data
  and the simulation in  Fig.~\ref{fig:fig_9}; 
  the invariant mass spectrum of the $PP$ component in the $\Upsilon$ mass
  region in  Fig.~\ref{fig:fig_9bis}; and a data to simulation comparison of
  the invariant mass spectrum of dimuon pairs from heavy flavor decays 
  in Fig.~\ref{fig:fig_9tris}.
 %%%%%%%%%%%%%%%%%%%%%%%%%%
%%%%%%%%%%%%%%%%%%%%%%%%% 
 %%%%%%%%%%%%%%%%%%%%%%%%%%
%%%%%%%%%%%%%%%%%%%%%%%%%
 %%%%%%%%%%%%%%%%%%%%%%%%%
%%%%%%%%%%%%%%%%%%%%%%%%%%

 %%%%%%%%%%%%%%%%%%%%%%%%%%
%%%%%%%%%%%%%%%%%%%%%%%%%%
  \subsection{Dependence of the result on the {\boldmath $b$}- 
              and {\boldmath $c$}-quark lifetime}\label{sec:ss-cross1}
%%%%%%%%%%%%%%%%%%%%%%%%%
  The lifetime of the $b$ hadron mixture with semileptonic decays produced
  at the Tevatron has a $0.6$\% uncertainty~\cite{pdg}. Impact parameter 
  templates, constructed by varying the lifetime by this uncertainty, 
  change the $BB$ size returned by the fit by $\pm 0.4$\% and the $CC$ 
  size by $\pm 1$\%. The lifetime of the $c$-hadron mixture has a $\pm 3.2$\%
  uncertainty, mostly due to the uncertainty of the relative fractions of
  produced hadrons, listed in Table~\ref{tab:tab_appa2}, and to the
  uncertainty of the semileptonic branching fractions of different
  $c$ hadrons~\cite{pdg}. When using simulated templates constructed by
  changing the average lifetime by  $\pm 3.2$\%, the $BB$ size returned 
  by the fit changes by $\pm 1$\% and the $CC$ size varies by $\pm 3$\%.
  By adding linearly these systematic uncertainties to that due to the muon
  $p_T$ spectrum in the simulation, we derive a $\pm 2.9$\% systematic 
  error for the $BB$ component and $\pm 8$\% systematic error for the 
  $CC$ component.
%%%%%%%%%%%%%%%%%%%%%%%%
\section{Acceptance and efficiencies}\label{sec:ss-acc}
%%%%%%%%%%%%%%%%%%%%%%%%%
  The kinematic and detector acceptance is calculated with the Monte Carlo
  simulation described at the beginning of Sec.~\ref{sec:ss-meth} and in
  App.~A. The detector response to muons produced by $b$- and $c$-hadron 
  decays is modeled with the CDF~II detector simulation that also models
  the L1 and L2 trigger responses. Simulated events are processed and
  selected with the same analysis code used for the data. The acceptance
  ($\cal A$) is the fraction of generated muon pairs that are identified 
  in the detector and pass all selection requirements. At generator level, 
  we select pairs of muons with invariant mass 
  $5 \leq m_{\mu\mu} \leq 80 \; \gevcc$, each having  $p_T \geq 3\; \gevc$ 
  and $|\eta| \leq 0.7$. Acceptances derived from the simulation are listed
  in Table~\ref{tab:tab_8}.
%%%%%%%
 \begin{table}
 \caption{Detector and kinematic acceptances, $\cal A$, for dimuon pairs
          arising from $b\bar{b}$ and  $c\bar{c}$ production. The 
          acceptance $\cal A_{\rm corr}$ includes corrections evaluated
          using the data.}
 \begin{center}
 \begin{ruledtabular}
 \begin{tabular}{lcc}
  Production      & $\cal A$ (\%)   & $\cal A_{\rm corr}$ (\%) \\
     $b\bar{b}$   & $4.21\pm 0.04$  & $4.56 \pm 0.15$ \\
     $c\bar{c}$   & $3.95\pm 0.10$  & $4.28 \pm 0.17$ \\
 \end{tabular}
 \end{ruledtabular}
 \end{center}
 \label{tab:tab_8}
 \end{table}
%%%%%%%%%%%%%%%%%%%%%%%%%%
  We use the data to verify the detector acceptance and efficiencies
  evaluated using the CDF~II detector simulation. We adjust the simulation
  to match measurements in the data of: (1) the offline COT track 
  reconstruction efficiency; (2) the CMUP detector acceptance and
  efficiency; (3) the efficiency for finding L1 CMU primitives;
  (4) the SVXII acceptance and efficiency; and (5) the efficiency
  of the L1, L2, and L3 triggers. 

 In the simulation, the offline COT track reconstruction efficiency 
 ($0.998 \pm 0.002$) is the fraction of tracks, which at generator level
 satisfy the  $p_T$ and $\eta$ selection cuts, that survives after
 selecting fully simulated events as the data. In the data, this efficiency
 has been measured to be 0.996 with a $\simeq 0.006$ systematic accuracy by
 embedding  COT hits generated from simulated tracks into $J/\psi$ 
 data~\cite{bishai}~\footnote{
 The efficiency measurement was performed in a subset of the data used for 
 this analysis. Studies of independent data samples collected in the data
 taking period used for this analysis show that changes of the track 
 reconstruction efficiency are appreciably smaller than the quoted systematic
 uncertainty~\cite{matt}.}.
 As in a previous study~\cite{bjk}, we conclude that the efficiencies
 for reconstructing muon pairs in the data and the simulation are equal
 within a 1.3\% systematic uncertainty.

\begin{figure}
 \begin{center}
 \leavevmode
 \includegraphics*[width=\textwidth]{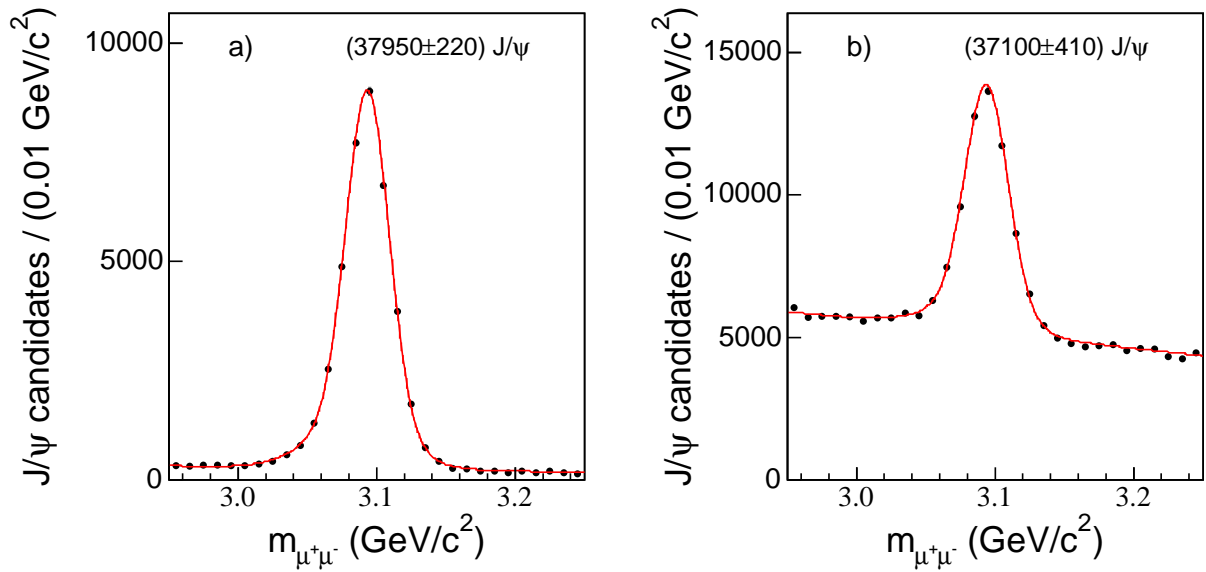}
 \caption[]{Invariant mass distribution of  CMUP muons paired with a 
            displaced track with (a) or without (b) a CMUP stub. Lines
            represent the fits described in the text.}
 \label{fig:fig_10}
 \end{center}
 \end{figure}

 In the simulation, the fraction of CMUP stubs generated by muon tracks with
 $p_T \geq 3\; \gevc$ and $|\eta| \leq 0.7$ is $0.5235 \pm  0.0022$. In the
 data, this efficiency is measured by using $J/\psi \rightarrow \mu^+\;\mu^-$
 decays acquired with the $\mu$-SVT trigger. 
 We evaluate the invariant mass of all pairs of a CMUP track and a track 
 with displaced impact parameter, $p_T\geq 3 \; \gevc$, and $|\eta| \leq 0.7$.
 We fit the invariant mass distribution with a first order polynomial plus
 two Gaussian functions to extract the $J/\psi$ signal. From the number
 of $J/\psi$ mesons reconstructed using displaced tracks with or without
 a CMUP stub (Fig.~\ref{fig:fig_10}(a) and~(b), respectively), we derive
 an efficiency of $0.5057 \pm 0.0032$. The integrated efficiency is 
 evaluated after having weighted the $p_T$ and $\eta$ distributions of 
 displaced tracks in the data to be equal to those of muons from heavy
 flavor decays in the simulation.

\begin{figure}
 \begin{center}
 \leavevmode
 \includegraphics*[width=\textwidth]{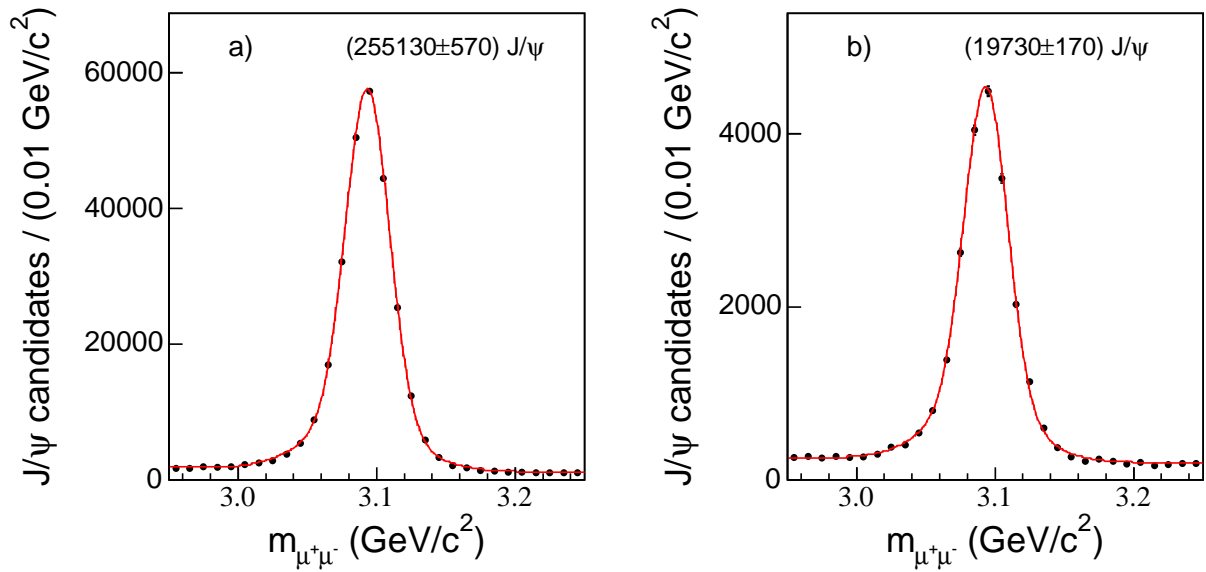}
 \caption[]{Invariant mass distribution of  CMUP muons paired with other
            CMUP muons in the event with (a) or without (b) a L1 CMU
            primitive. In order to derive the efficiency from the numbers of
            $J/\psi$ candidates in plots (a) and (b), the histogram in (a) 
            has an entry for each CMUP leg. Solid lines represent the fits 
            described in the text.}
 \label{fig:fig_11}
 \end{center}
 \end{figure}

 In the simulation, the efficiency for finding a L1 CMU primitive 
 (CMU stub matched by a XFT track) is $0.8489 \pm 0.0026$. This efficiency
 is measured in the data by using events acquired with the CMUP$p_T$4 trigger.
 We combine the CMUP muon with all other CMUP muons found in the event with
 and without a L1 CMU primitive. We extract the number of
 $J/\psi \rightarrow \mu^+ \mu^-$ mesons  by fitting the invariant mass
 distributions of all candidates with a first order polynomial plus two 
 Gaussian functions. By comparing the fitted numbers of $J/\psi$ candidates
 with and without a L1 CMU primitive (Fig.~\ref{fig:fig_11}(a) and~(b),
 respectively) we derive an efficiency of $0.92822 \pm 0.00006$.
 The integrated efficiency is evaluated after having weighted the $p_T$ and
 $\eta$ distributions of the additional CMU muons to be equal to that
 of muons from heavy flavor decays in the simulation.
%%%%%%%%%%%%%%%%%%%%%%%%%%
%%%%%%%%%%%%%%%%%%%%%%%%%
%%%%%%%%%%%%%%%%%%%%%%%%%%
%%%%%%%%%%%%%%%%%%%%%%%%%   

\begin{figure}
  \begin{center}
  \leavevmode
  \includegraphics*[width=\textwidth]{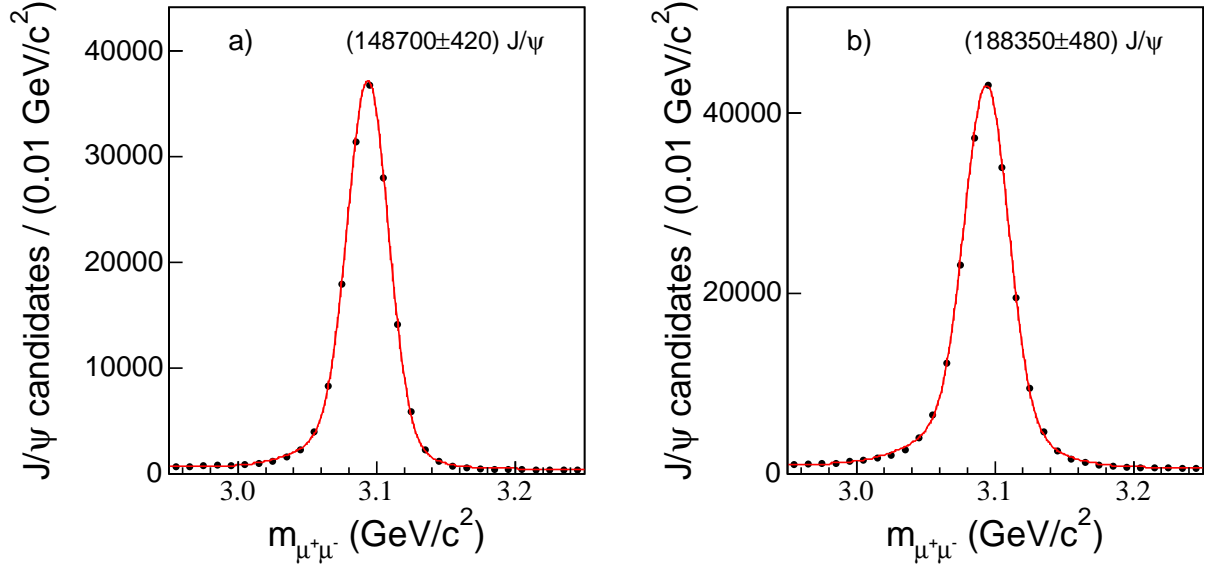}
  \caption[]{Invariant mass distribution of CMUP muon pairs in which  a
            first randomly chosen muon (a) passes or (b) fails the SVXII 
            requirements. Solid lines represent the fits described in the
            text.}
  \label{fig:fig_12}
  \end{center}
  \end{figure}

\begin{figure}
  \begin{center}
  \leavevmode
  \includegraphics*[width=\textwidth]{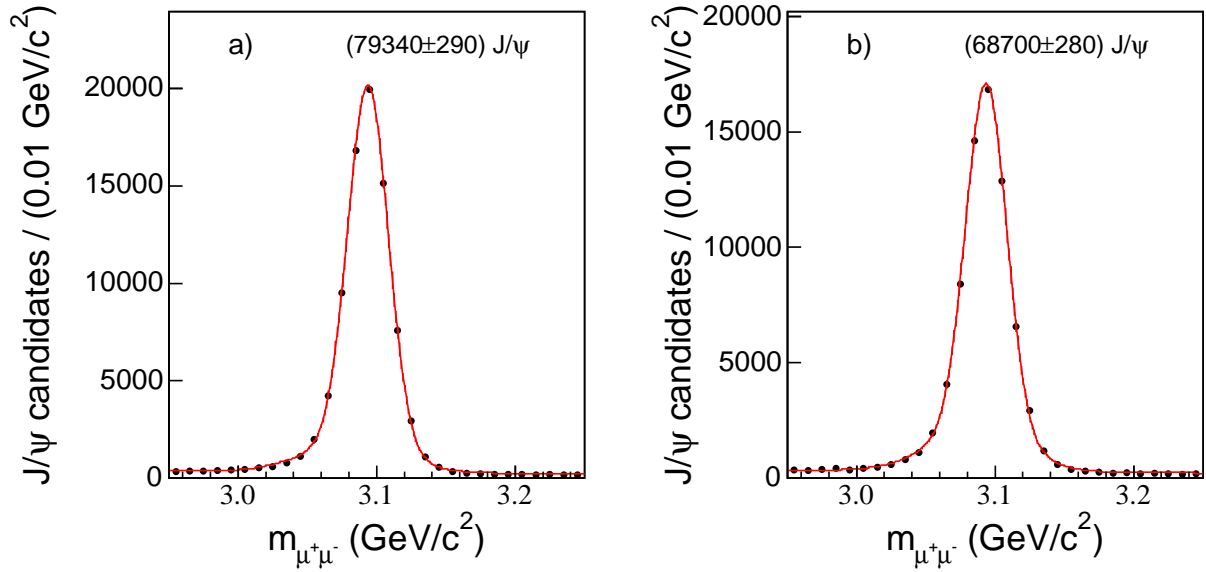}
  \caption[]{Invariant mass distribution of CMUP muon pairs in which a first
             randomly chosen muon track satisfies the SVXII requirements and
             the second muon track (a) passes or (b) fails the SVXII 
             requirements. Solid lines represent the fits described in the
             text. }
  \label{fig:fig_13}
  \end{center}
  \end{figure}

\begin{figure}
  \begin{center}
  \leavevmode
  \includegraphics*[width=\textwidth]{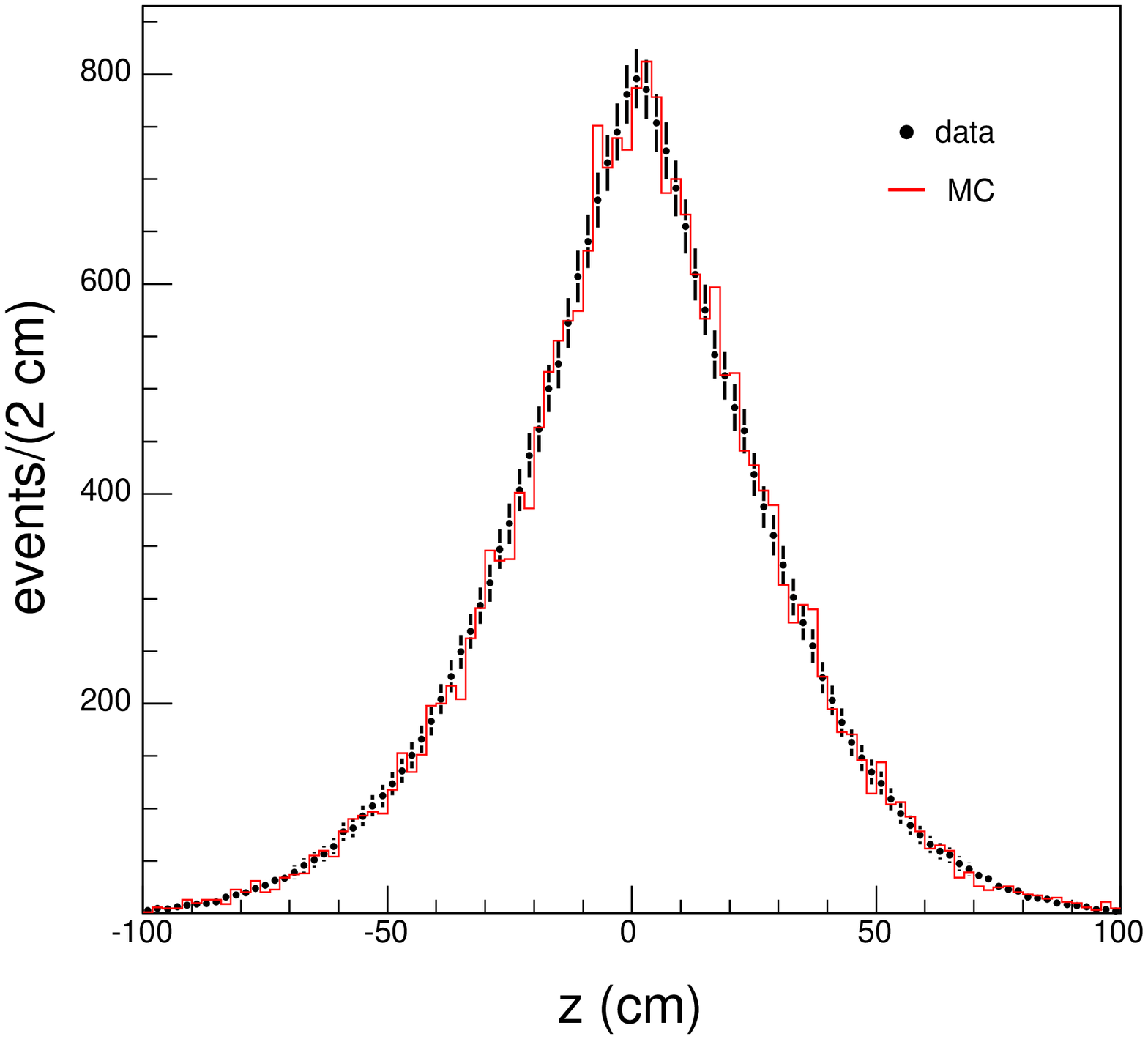}
  \caption[]{Distribution of the event vertex along the beam line in the data
             ($\bullet$) and in the heavy flavor simulation (histogram).}
  \label{fig:fig_14}
  \end{center}
  \end{figure}

  In the simulation, the probability  that a CMUP pair passes the SVXII
  requirements described in Sec.~\ref{sec:ss-anal} is $0.2206 \pm 0.0047$.
  This efficiency is measured in the data using muon pairs acquired with
  the $J/\psi$ trigger. We use CMUP muons with $p_T \geq  3 \; \gevc$ and
  $|\eta| \leq 0.7$. The efficiency is evaluated in two steps. 
  For each event, we first randomly choose a CMUP muon. After weighting 
  the $z_0$ and $\eta$ distributions of these muons to be equal to those
  of CMUP muons from simulated heavy flavor decays, we derive the SVXII
  efficiency $\epsilon_1$  from the number of CMUP muons that pass or fail
  the SVXII requirements by fitting the dimuon invariant mass distribution
  with a straight line plus two Gaussian functions
  (see Fig.~\ref{fig:fig_12}). For events in which the first randomly
  chosen muon passes the SVXII requirements, we derive the SVXII efficiency
  $\epsilon_2$ from the numbers of second muons that pass or fail the SVXII
  requirements. After weighting $z_0$ and $\eta$ distributions of the 
  second muons to be equal to those of CMUP muons from simulated heavy 
  flavor decays, we fit again the dimuon invariant mass distributions with
  a straight line plus two Gaussian functions (see Fig.~\ref{fig:fig_13}).
  The probability that a muon pair passes the SVXII requirements is 
  $\epsilon_1 \cdot \epsilon_2= 0.2365 \pm0.0013$. This measurement of 
  the SVXII efficiency rests on the verified assumption that the event
  vertex $z$-distribution is the same in the data and the simulation
 (see Fig.~\ref{fig:fig_14}). 
%%%%%%%%%%%%%%%%%%%%%%%%%%
%%%%%%%%%%%%%%%%%%%%%%%%%   
%%%%%%%%%%%%%%%%%%%%%%%%%%
%%%%%%%%%%%%%%%%%%%%%%%%%   
%%%%%%%%%%%%%%%%%%%%%%%%%%
%%%%%%%%%%%%%%%%%%%%%%%%%   

  In the simulation, the efficiencies of the L1 and L2 triggers are 1 and
  0.9976, respectively. By using muon pairs with CMU primitives acquired
  with the {\sc charm} trigger, we measure the L1 and L2 trigger efficiency
  to be $1 \pm 0.001$ and  $0.99943 \pm 0.00045$, respectively.
  The L3 trigger is not simulated.  The L3 trigger efficiency is dominated
  by differences between the online and offline reconstruction code
  efficiency~\footnote{
  Online algorithms are faster but less accurate than the offline 
  reconstruction code.}.
  The relative L3 efficiency for reconstructing a single muon identified
  by the offline code has been measured to be 
  $0.997 \pm 0.002$~\cite{bishai,bjk}. However, in a large fraction of 
  the data, the L3 trigger has selected muons with the requirement that
  the distance between the track projection to the CMP chambers and CMP 
  stub be $\Delta r\phi \leq 25$ cm, whereas the offline analysis
  requires $\Delta r\phi \leq 40$ cm. We have measured the efficiency
  of this L3 cut by using $J/\psi$ candidates acquired with the $J/\psi$ 
  trigger that has no $\Delta r\phi$ requirement. After weighting the
  $p_T$ distribution of muons from  $J/\psi$ candidates to model that 
  of muons from $b$ decays in the simulation, we measure the efficiency
  to be $0.948 \pm 0.005$ for a single muon. The reconstruction efficiencies
  in the data and in the simulation are summarized in Table~\ref{tab:tab_9}.
%%%%%%%%%%%%%%%%%%%%%%%%%%%%%%%%%%%%%%%%%
  \begin{table}
  \caption{Summary of efficiencies for reconstructing muon pairs from
           heavy flavor decays in the data and in the simulation. The 
           last column indicates the corrections applied to the simulated
           efficiencies and used to derive $\cal A_{\rm corr}$ in 
           Table~\ref{tab:tab_8}.} 
 \begin{center}
 \begin{ruledtabular}
 \begin{tabular}{lccc}
  Source   & Data & Simulation & Corr. \\
  COT tracking      & $(0.996\pm0.006)^2$   & $(0.998\pm0.002)^2$ & $1\pm0.013$    \\
 CMUP acc. and eff. & $(0.5057\pm0.0032)^2$ & $(0.5235\pm0.0022)^2$& $0.933\pm0.014$        \\
 L1 CMU primitives  & $(0.92822\pm0.0006)^2$ & $(0.8489\pm0.0026)^2$ & $1.196\pm0.007$      \\
 Sili acc. and eff. & $ 0.2365\pm0.0013$   & $ 0.2206\pm0.0047 $   & $1.072\pm0.024$        \\
 L1 eff.            & $ 1\pm0.001$ & $  1\pm0.001  $ & $1\pm0.0014$   \\
 L2 eff.            & $ 0.99943\pm0.00045$ & $ 0.9976\pm0.001 $ & $1.002\pm0.001$      \\
 L3 eff.            & $ 0.90 \pm0.01 $     & $  1 $    & $0.90 \pm 0.01$  \\
 Total        & $0.0471\pm0.001$      & $0.0435\pm0.001$ & $1.084\pm0.035$ \\
 \end{tabular}
 \end{ruledtabular}
 \end{center}
 \label{tab:tab_9}
 \end{table}
%%%%%%%%%%%%%%%%%%%%%%%%%%%%%%%%%%%%%%%%%%%%%
 \section{Dimuon cross section and comparison with previous results}
 \label{sec:ss-disc}
%%%%%%%%%%%%%%%%%%%%%%%%
  We have selected pairs of muons, each  with $p_T \geq 3\; \gevc$ and
  $|\eta| \leq 0.7$, with invariant mass $5\leq m_{\mu\mu}\leq 80\; \gevcc$
  and produced by double semileptonic decays of heavy flavors.
  The production cross section is given by  
 \begin{equation}
    \sigma = \frac{ N}{ {\cal L} \times {\cal A_{\rm corr}} },
 \end{equation}
  where $N=BB=52400\pm 2747$ for $b\bar{b}$ production 
 ($N=CC=19811 \pm 2994$ for $c\bar{c}$ production).
  The geometric and kinematic acceptance, ${\cal A_{\rm corr}}$, that 
  includes trigger and tracking efficiencies measured with the data is
  listed in Table~\ref{tab:tab_8}. The integrated luminosity of the
  data sample is $ {\cal L}= 742 \pm 44 $ pb$^{-1}$.

  We derive $\sigma_{b\rightarrow\mu,\bar{b}\rightarrow \mu}=1549 \pm 133$ pb,
  where the 8.6\% error is the sum in quadrature of the 1.2\% statistical
  error, the 2.9\% systematic uncertainty due to the fit likelihood function, 
  the 4.2\% systematic uncertainty in the removal of the fake muon 
  contribution, the 6\% uncertainty of the luminosity, and the 3.2\%
  uncertainty of the acceptance calculation.

  We also derive $\sigma_{c\rightarrow\mu,\bar{c}\rightarrow\mu}=624\pm 104$
  pb. In this case, the statistical error is 6.4\%, the uncertainty due to
  the fit likelihood function is $\pm 8$\%, and the uncertainty in the
  removal of the fake muon contribution is 11.1\%.

  We evaluate the exact NLO prediction of 
  $\sigma_{b\rightarrow\mu,\bar{b}\rightarrow \mu}$ and 
  $\sigma_{c\rightarrow\mu,\bar{c}\rightarrow \mu}$ by complementing
  the {\sc mnr} generator with the {\sc evtgen} Monte Carlo program.
  We use the Peterson fragmentation function with $\epsilon=0.006 \;(0.06)$
  for $b$ ($c$) quarks and the measured fragmentation fractions listed in
  Table~\ref{tab:tab_appa2}. The NLO prediction is estimated  using
  $m=4.75 \;(1.5) \; \gevcc$, the factorization and normalization scale
  $\mu_R=\mu_F=\sqrt{p_T^2 +m^2}$, where $m$ is the $b\; (c)$ quark mass,
  and the MRST PDF fits~\cite{mrst} (we use the five flavor scheme and 
  $\Lambda_5=0.22 \;\gevcc$). In the following, we refer to it as the standard 
  NLO calculation. We generate heavy flavor quarks with $p_T \geq 2\; \gevc$
  and $|y|\leq 1.3$.

   The values of $\sigma_{b\rightarrow\mu,\bar{b}\rightarrow \mu}$ and
  $\sigma_{c\rightarrow\mu,\bar{c}\rightarrow \mu}$ predicted by the 
  standard NLO calculation have a 2\% uncertainty, estimated by using
  different but reasonable procedures to sum the positive and negative 
  weights, due to real and virtual soft gluon emission, returned by
  the {\sc mnr} computation. The theoretical prediction also carries the 
  uncertainty of the semileptonic branching fractions
  $b \rightarrow \mu =10.71 \pm 0.22 $, 
  $b \rightarrow c \rightarrow \mu =9.63 \pm 0.44 $, and
  $c \rightarrow \mu = 9.69 \pm 0.31$\%~\cite{lepewg}.
  In the simulation, 79.4\% of the muon pairs are due to $b \rightarrow \mu$
  decays, 1.3\% to $b \rightarrow c \rightarrow \mu$ decays and the rest to 
  a mix of these decays. Therefore, the rate of predicted dimuon pairs due
  to $b\bar{b}$ ($c\bar{c}$) production has a 3.7\% (6.4\%) uncertainty.
  For muon pairs selected with the same kinematic cuts of the data, the
  standard NLO prediction is  
  $\sigma_{b\rightarrow\mu,\bar{b}\rightarrow \mu}= 1293 \pm 55$ pb.
  For $c\bar{c}$ production, the standard NLO prediction is
  $\sigma_{c\rightarrow\mu,\bar{c}\rightarrow \mu}= 230 \pm 16$ pb~\footnote{
  For comparison, the prediction of the {\sc herwig} generator is
  $\sigma_{b\rightarrow\mu,\bar{b}\rightarrow \mu} =904 \pm 33$ pb and 
  $\sigma_{c  \rightarrow\mu,\bar{c}\rightarrow \mu} =173 \pm 11$ pb.}. 
  The ratio of the data to the standard NLO prediction with the above
  mentioned uncertainties ($R1= 1.20 \pm 0.11$ for $b\bar{b}$ production 
  and $R1=2.71 \pm 0.49$ for $c\bar{c}$ production) can be used to extract 
  the value of $\sigma_{b\bar{b}}$ and $\sigma_{c\bar{c}}$ in the data.
  In addition, as discussed in Sec.~\ref{sec:ss-intro}, the theoretical
  prediction has a 15\% uncertainty due to the choice of the heavy quark
  pole-mass~\footnote{
  Following tradition, we vary the pole mass of $b$ quarks by 
  $\pm 0.25\;\gevcc$ and that of $c$ quarks by $\pm 0.2\;\gevcc$.}, 
  PDF fits, and renormalization and factorization scales. After including
  the latter uncertainty, the ratio of the data to the standard NLO prediction
  is $1.20 \pm 0.21$ for $b\bar{b}$ production and $ 2.71 \pm 0.64$ for 
  $c\bar{c}$ production. 

  The $c\bar{c}$ correlation measurement has no previous result to compare
  with. However, the CDF study in Ref.~\cite{bmix} has measured the ratio
  of dimuon pairs due to $c\bar{c}$ to those due to $b\bar{b}$ production.
  That study uses muon pairs selected in the same kinematic region as our
  measurement, and finds a ratio $CC/BB =0.15 \pm 0.02$~\footnote{ 
  The error is statistical. Systematic effects due to the fit likelihood 
  functions or $c$-hadron lifetime were not investigated.},
  whereas our fit to the impact parameter distributions yields 
  $CC/BB =0.38\pm 0.07$. In the simulation, this ratio is $0.17\; (0.16)$ 
  when using the {\sc herwig} ({\sc mnr}) generator.
 
  The extraction of  $\sigma_{b\bar{b}}$ from the dimuon production cross
  section and the comparison to other measurements is not a trivial issue.
  Muons with $p_T \geq 3\; \gevc$ and $|\eta| \leq 0.7$ are mostly 
  contributed by $b$ quarks with $p_T \geq 6.5 \; \gevc$ and $|y|\leq 1$.
  However, there are  tails contributed from $b$ quarks with $p_T$ as small
  as 2 $\gevc$ and  $|y|$ as large as 1.3. If these contributions are 
  included, the resulting value of $\sigma_{b\bar{b}}$ is dominated by the
  production of $b$ quarks with the smallest $p_T$ that, unfortunately, has
  a large statistical error because of the small kinematic acceptance.
  The measurement of $\sigma_{b\bar{b}}$ in Ref.~\cite{2mucdf} is based 
  upon muon pairs selected with the same kinematic cuts as this study. 
  That study does not report the value of 
  $\sigma_{b\rightarrow\mu,\bar{b}\rightarrow \mu}$ but, in the assumption
  that these muon pairs are produced by $b$ quarks with 
  $p_T \geq 6.5 \; \gevc$ and $|y|\leq 1$, quotes 
  $\sigma_{b\bar{b}}(p_T \geq 6.5 \; \gevc, |y| \leq 1)=2.42\pm 0.45$ $\mu$b.
  It seems more appropriate to derive this cross section assuming that the
  ratio of data to theory is the same as that of the measured to predicted 
  dimuon cross section. Using this method, the ratio $R1$ yields
  $\sigma_{b\bar{b}}(p_T \geq 6.5 \; \gevc, |y| \leq 1) =1324 \pm 121$ nb
  (the standard NLO prediction is $1103\pm 169$ nb).
  In the simulation, only 75\% of the muon pairs arise from $b$ and 
  $\bar{b}$ quarks with $p_T \geq 6.5 \; \gevc$ and $|y|\leq 1$.
  The result of  Ref.~\cite{2mucdf}, rescaled by 75\%, becomes 
  $\sigma_{b\bar{b}}(p_T \geq 6.5 \; \gevc, |y|\leq 1)=1.80 \pm 0.34$ $\mu$b.

  The D${\not\! {\rm O}}$ collaboration~\cite{d0b2} has measured 
  $\sigma_{b\rightarrow\mu,\bar{b}\rightarrow \mu} =1027 \pm 260$ pb
  using muon pairs with $6 \leq m_{\mu\mu} \leq 35\; \gevcc$. That study
  selects muons with $4 \leq p_T \leq 25 \; \gevc$, $|\eta|\leq 0.8$,
  and contained in a jet with transverse energy $E_T \geq 12$ GeV.
  Reference~\cite{d0b2} compares data to the exact NLO prediction that, 
  evaluated with the {\sc hvqjet} Monte Carlo program~\cite{hvqjet}, is 357 pb.
  For this kinematical selection, except the request that muons are embedded
  in jets, the standard NLO prediction is
  $\sigma_{b\rightarrow\mu,\bar{b}\rightarrow \mu} = 550 $ pb.
  When applying these kinematical cuts to our data, and before asking
  that muons are contained in jets with  $E_T \geq 12$ GeV, we measure
  $\sigma_{b\rightarrow\mu,\bar{b}\rightarrow \mu}  = 658 \pm 55$ pb.

  Using the ratio $R1$, the data yield 
  $\sigma_{b\bar{b}}(p_T \geq 6\; \gevc, |y|\leq 1)= 1618\pm 148$ nb. 
  The standard NLO prediction is
  $\sigma_{b\bar{b}}(p_T \geq 6\; \gevc, |y|\leq 1)= 1348\pm 209$ nb~\footnote{
  For charmed quarks, the standard NLO prediction is
  $\sigma_{c\bar{c}}(p_T \geq 6 \; \gevc, |y| \leq 1) = 2133 \pm 323$ nb.}.
  For comparison, the {\sc herwig} parton-level prediction is
  $\sigma_{b\bar{b}}(p_T \geq 6 \; \gevc, |y| \leq 1) = 1327 $ nb,
  and the cross section returned by the {\sc mc@nlo} generator~\footnote{
  We input the same $b$-quark mass, scales and PDF fits used in the
  standard NLO calculation.}
  is  $\sigma_{b\bar{b}}(p_T \geq 6 \; \gevc, |y|\leq 1) = 1704$ nb,
  27\%  larger than the {\sc mnr} result~\footnote{
  The total $b\bar{b}$ cross section predicted by both {\sc mc@nlo} and 
  {\sc mnr} generators is 56.6 $\mu$b and compares well with the result 
  of the {\sc nde} calculation (57.6 $\mu$b). However, the inclusive 
  single $b$ cross section for $p_T \geq 6 \; \gevc$ and $|y| \leq 1$ 
  predicted by the {\sc nde} and {\sc mnr} programs are 5.5 and 5.6 $\mu$b, 
  respectively, whereas {\sc mc@nlo} generator predicts 11.8  $\mu$b.
  In contrast, for both $b$ and $\bar{b}$ quarks with $p_T \geq 25\; \gevc$
  and $|y| \leq 1.2$ the {\sc mc@nlo} prediction is approximately 12\% 
  smaller than that of the {\sc mnr} generator.}.

  The value of $\sigma_{b\bar{b}}$ has been extracted from the data using
  a fragmentation model based on the Peterson function. As previously noted,
  the {\sc mnr} and {\sc herwig} generators predict the same parton-level
  cross section  $\sigma_{b\bar{b}}(p_T \geq 6 \; \gevc, |y| \leq 1)$.
  The {\sc herwig} generator models the $b$-quark fragmentation differently, 
  and this difference results in a prediction of  
  $\sigma_{b\rightarrow\mu,\bar{b}\rightarrow \mu}$ which is 30\% smaller
  than that of the {\sc mnr} generator implemented with the Peterson
  fragmentation model. In the transverse momentum range of this study,
  the FONLL prediction for the single $b$-quark cross section is fairly well
  reproduced by the {\sc nde} calculation when using the Peterson
  fragmentation function with $\epsilon=0.002$~\cite{cana,matteo2}.
  When using this fragmentation function, the exact NLO prediction becomes
  $\sigma_{b\rightarrow\mu,\bar{b}\rightarrow \mu}= 1543 $ pb,
  which is 20\% higher than the standard exact NLO prediction. 

  As argued in Ref.~\cite{cacc-greco}, the charmed quark production in
  $e^+e^-$ data can be described at NLO accuracy using the Peterson
  fragmentation model with $\epsilon=0.02$. In this case, the NLO prediction
  becomes $\sigma_{c\rightarrow\mu,\bar{c}\rightarrow \mu}= 383 $ pb,
  66\% larger  than the standard NLO prediction. When using the smaller
  values of the $\epsilon$ parameter, the ratio of data to theory becomes
  $1.0 \pm 0.2$ for  $\sigma_{b\rightarrow\mu,\bar{b}\rightarrow \mu}$
  and $1.6 \pm 0.4$ for $\sigma_{c\rightarrow\mu,\bar{c}\rightarrow \mu}$. 
%%%%%%%%%%%%%%%%%%%%%%%%%%%%%%%%%%%
\section{Conclusions}
  We have measured the production cross section of muon pairs from double
  semileptonic decays of $b$ and $\bar{b}$ quarks produced at the Tevatron
  Fermilab collider operating at $\sqrt{s}=1.96$ TeV. We select muons with
  $p_T \geq 3 \; \gevc$ and $|\eta| \leq 0.7$. We select dimuons with
  $5 \leq m_{\mu\mu} \leq 80\; \gevcc$ to reject the contribution of 
  sequential decays of single $b$ quarks and $Z^0$ decays. The main sources
  of these muon pairs are semileptonic decays of $b$ and $c$ quarks, prompt
  decays of quarkonia, and Drell-Yan production. We determine the $b\bar{b}$
  content of the data by fitting the impact parameter distribution of muon
  tracks with the templates expected for the various sources. Previous 
  measurements of the $b\bar{b}$ correlations at the Tevatron yield
  contradictory results. The ratio of the data to exact NLO prediction
  is approximately $1.15 \pm 0.21 $ when  $b$ quarks are selected via
  secondary vertex identifications, whereas this ratio is found to be
  significantly larger than two when identifying $b$ quarks through their
  semileptonic decays.

  We measure $\sigma_{b\rightarrow\mu,\bar{b}\rightarrow \mu}=1549\pm 133$ pb.
  The exact NLO prediction is evaluated using the {\sc mnr} calculation
  complemented with the {\sc evtgen} generator. In the calculation, we use
  $m_b=4.75 \;\gevcc$, the factorization and normalization scale 
  $\mu_R=\mu_F=\sqrt{p_T^2 +m_b^2}$, and the MRST PDF fits (we use the five
  flavor scheme and $\Lambda_5=0.22 \;\gevcc$). We use the Peterson
  fragmentation function with $\epsilon=0.006$, and the PDG values for the
  fragmentation fractions. The NLO prediction is 
  $\sigma_{b\rightarrow\mu,\bar{b}\rightarrow \mu}= 1293 \pm 201$ pb.
  The ratio of the data to the NLO prediction is $1.20 \pm 0.21$.

  From this measurement, we also derive
  $\sigma_{b\bar{b}}(p_T \geq 6 \; \gevc, |y| \leq 1) = 1618 \pm 148$ nb
  (the exact  NLO prediction is  $1348 \pm 209$ nb). The extraction of 
  $\sigma_{b\bar{b}}$ from the data depends on the choice of the fragmentation
  functions that connect a muon to the parent $b$ quark. No fragmentation 
  functions are available that match the accuracy of the NLO calculation. 
  Reasonable changes of the fragmentation model indicate that the value of
  $\sigma_{b\bar{b}}$ extracted from the data has an additional uncertainty
  of approximately 25\%.
\label{sec:ss-concl} 
%%%%%%%%%%%%%%%%%%%%%%%%%%%%%%%%%%%%%%%%%%%
\section{Acknowledgments}
  We thank the Fermilab staff and the technical staffs of the participating
  institutions for their vital contributions. This work was supported by the
  U.S. Department of Energy and National Science Foundation; 
  the Italian Istituto Nazionale di Fisica Nucleare; the Ministry of Education,
  Culture, Sports, Science and Technology of Japan; the Natural Sciences and 
  Engineering Research Council of Canada; the National Science Council of the
  Republic of China; the Swiss National Science Foundation; the A.P. Sloan 
  Foundation; the Bundesministerium f\"ur Bildung und Forschung, Germany; 
  the Korean Science and Engineering Foundation and the Korean Research 
  Foundation; the Particle Physics and Astronomy Research Council and the 
  Royal Society, UK; the Institut National de Physique Nucleaire et Physique
  des Particules/CNRS; the Russian Foundation for Basic Research; 
  the Comisi\'on Interministerial de Ciencia y Tecnolog\'{\i}a, Spain;
  the European Community's Human Potential Programme; the Slovak R\&D Agency;
  and the Academy of Finland.
%%%%%%%%%%%%%%%%%%%%%%%%%%%%%
\appendix
\section{Settings of the HERWIG  Monte Carlo program}
\begin{figure}
 \begin{center}
 \leavevmode
 \includegraphics*[width=\textwidth]{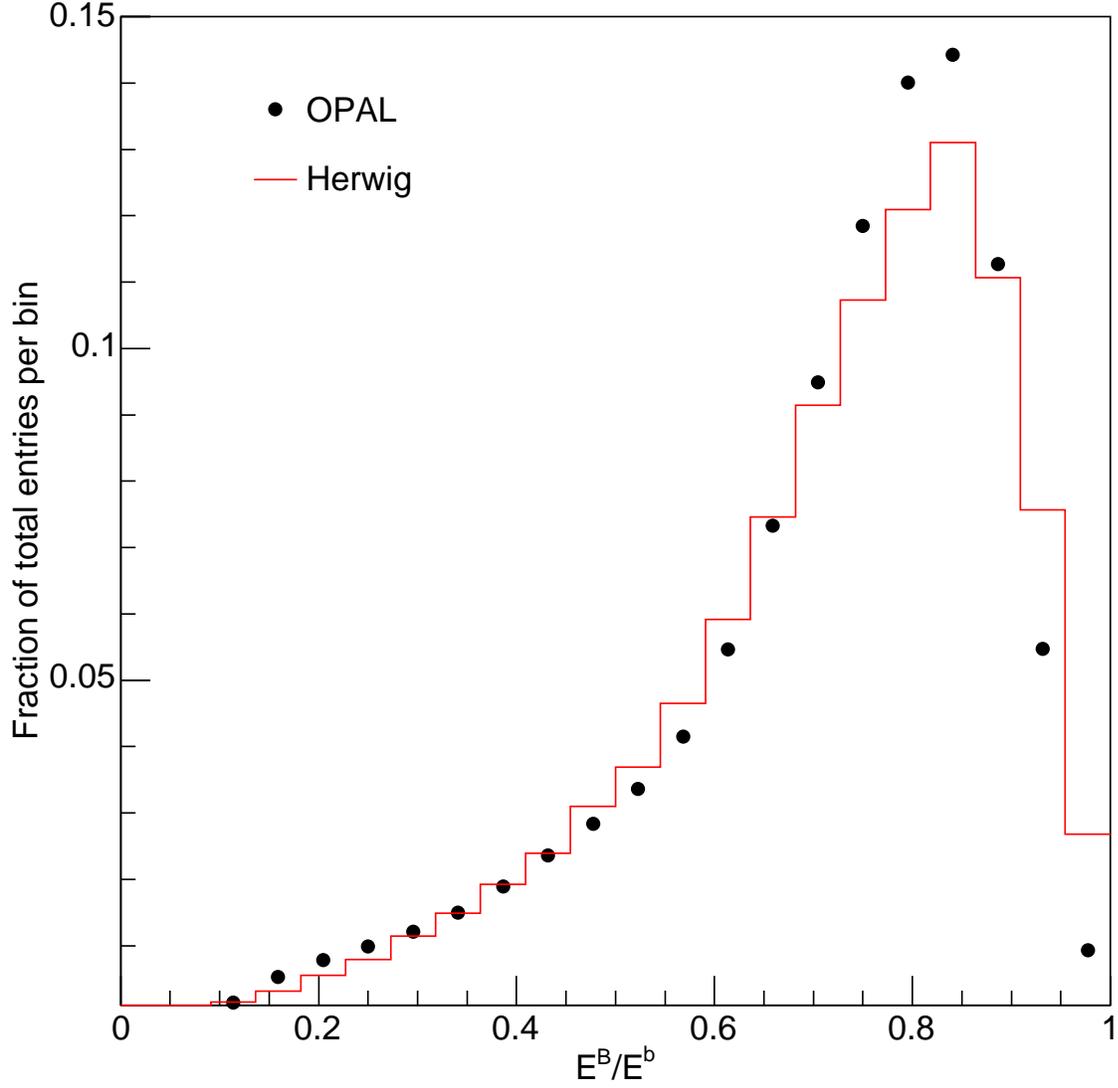}
 \caption[]{Distribution of ratio of  the energy carried by all $B$ hadrons
            to that of the parent $b$ quarks. The data are OPAL measurements
            at the $Z$ pole~\cite{op}, while the {\sc herwig} distribution
            is obtained with the parameter tuning listed in
            Table~\ref{tab:tab_appa1}.} 
 \label{fig:fig_appa1}
 \end{center}
 \end{figure}

\begin{figure}
 \begin{center}
 \leavevmode
\includegraphics*[width=\textwidth]{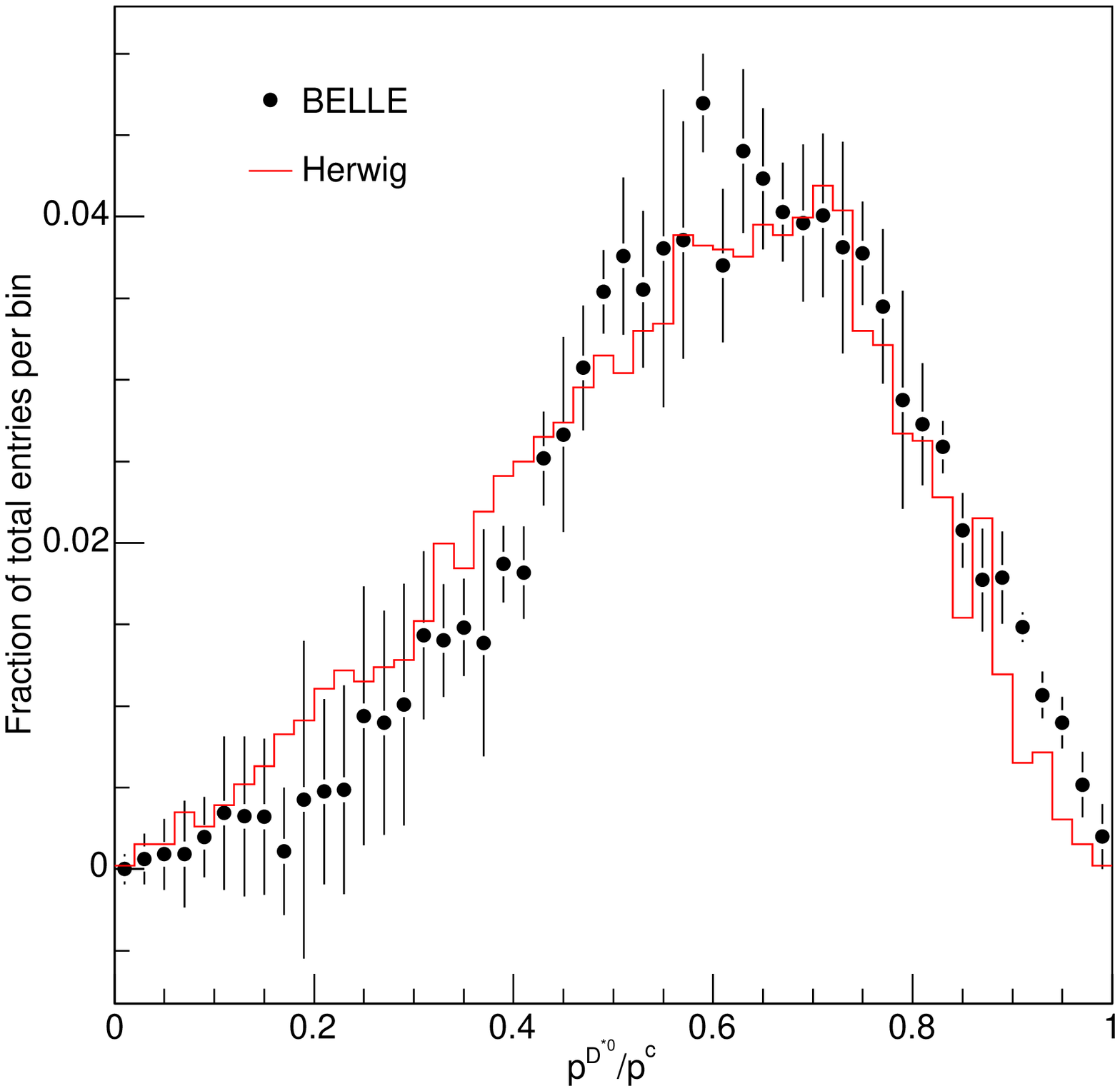}
 \caption[]{Distribution of ratio of  the momentum carried by $D^*$ mesons 
            to that of the parent $c$ quarks. The data are BELLE
            measurements~\cite{belle}, while the {\sc herwig} distribution
            is obtained with the parameter tuning listed in
            Table~\ref{tab:tab_appa1}.} 
 \label{fig:fig_appa2}
 \end{center}
 \end{figure}

  We generate  generic $2 \rightarrow 2$ hard scattering, process 1500, 
  using version 6.5 of the {\sc herwig} Monte Carlo program.
  In the generic hard parton scattering, $b\bar{b}$ and $c\bar{c}$ pairs 
  are generated by {\sc herwig} through processes of order $\alpha_s^{2}$ 
  (LO) such as $gg \rightarrow b\bar{b}$ (direct production). Processes
  of order $\alpha_s^{3}$ are implemented in {\sc herwig} through flavor
  excitation processes, such as $gb \rightarrow g b$, or gluon splitting, 
  in which the process $gg \rightarrow gg$ is followed by 
  $g \rightarrow b\bar{b}$. We generate final state partons with
  $p_T \geq 5\; \gevcc$ and $|y|\leq 1.7$. The hard scattering cross section
  is evaluated using the MRST fits to the parton distribution
  functions~\cite{mrst}. Hadrons with heavy flavor, produced by the 
  {\sc herwig} generator, are decayed with the {\sc evtgen} Monte Carlo 
  tuned by the BABAR collaboration~\cite{evtgen}. We retain simulated events
  that contains a pair of muons, each of them with  $p_T \geq \; 2.8\;\gevcc$
  and $|\eta| \leq 0.8$~\footnote{
  We also produced simulated samples requiring the presence of only one muon or
  no muons at all.}.
  We find one good event in approximately $10^8$ generated events.
  These events are used to determine the kinematical and
  detector acceptance as well as the impact parameter templates used to
  extract the heavy flavor composition of the data. Since different 
  $b$ hadrons, and especially different $c$ hadrons, have quite different
  lifetimes and semileptonic branching fractions, it is important that the
  generator models correctly the known fragmentation fractions and functions
  of $b$ and $c$ quarks. The {\sc herwig} generator makes use of a large 
  number of parameters that can be adjusted to this purpose. Unfortunately,
  their default setting~\cite{herw_sett} does not yield a satisfactory 
  modeling of the heavy quark fragmentation that we have studied by comparing 
  simulated $Z^0$ decays (process 2160 of {\sc herwig}) to $e^+e^-$ data. 
%%%%%%%%%%%%%%%%%%%%%%%%
 \begin{table}
 \caption[]{Parameter settings used in our simulation are compared to the
            {\sc herwig} default values.}
 \begin{center}
 \begin{ruledtabular}
 \begin{tabular}{lcc}
 Parameter             &   Default &  This Study  \\
 QCDLAM                &    0.180  &    0.18  \\
 RMASS(4) ($c$ quark)  &           &    1.50  \\
 RMASS(5) ($b$ quark)  &           &    4.75  \\
 RMASS(13)             &    0.75   &    0.75  \\
 CLMAX        	       &    3.35   &    3.75  \\
 CLPOW                 &    2.00   &    1.06  \\
 LCLPW                 &           &    2.20  \\
 DCLPW                 &           &    1.30  \\
 PSPLT(1)              &    1.00   &    0.50  \\
 PSPLT(2)              &    1.00   & 	1.10  \\
 CLSMR(1)              &    0.00   &    0.00  \\
 CLSMR(2)              &    0.00   &    0.40  \\
 PWT(3)                &    1.00   &    0.70  \\
 PWT(7)                &    1.00   &    0.45  \\
 SNGWT  	       &    1.00   &    1.00  \\
 DECWT  	       &    1.00   &    1.00  \\ 
 REPWT(0,1,0)          &    1.00   &   10.00  \\
 \end{tabular}
 \end{ruledtabular}
 \end{center}
 \label{tab:tab_appa1}
 \end{table}
%%%%%%%%%%%%%%%%%%%%%%
  The available parameters do not allow to tune the ratio of baryon to mesons
  simultaneously for bottom and charmed flavors, nor to reproduce the
  measured ratio of vector to pseudoscalar resonances produced in the heavy
  quark hadronization. The first deficiency becomes a problem when the 
  simulation of a QCD process, such as ours, is extremely time consuming.
  The second deficiency impacts the evaluation of the kinematical efficiency
  and lifetime templates for $c$ quarks because $D^*$ mesons mostly decay to
  $D^0$ mesons, the lifetime and semileptonic branching fractions of which
  differ by a factor of three from that of $D^+$ mesons. We have solved 
  these issues by adding two additional parameters, analogous of CLPOW.
  In the {\sc hwuinc.f} routine of the {\sc herwig} program, 
  the parameter CLPOW tunes the invariant mass distribution of cluster
  generated in the heavy quark hadronization. We use this parameter for 
  $b$ quarks only. For $c$ quarks, the parameter CLPOW is replaced with 
  two parameters, LCLPW and DCLPW, that separately control the yield of
  $c$-quark (mesons) and $c$-diquark (baryons) clusters, respectively.
  Table~\ref{tab:tab_appa1} lists the {\sc herwig} parameter settings used
  in our simulation.
  Table~\ref{tab:tab_appa2} compares fragmentation fractions in the tuned
  simulation at the $Z$-pole to the data. The fragmentation fractions for
  $b$ quarks are taken from Ref.~\cite{pdg}. For $c$ quarks, the 
  fragmentation fractions are taken from Refs.~\cite{belle, cleo, hera, 
  glad, glad1}.
%%%%%%%%%%%%%%%%%%%%%%%%%%%%%%%%%%%%%%%%%
 \begin{table}
 \caption{Fragmentation fractions in the tuned {\sc herwig} simulation are
          compared to data. The fragmentation fractions of $b$ quarks (first
          three rows) are defined according to the PDG notation~\cite{pdg}. } 
 \begin{center}
 \begin{ruledtabular}
 \begin{tabular}{rccc}
 \multicolumn{3}{c}{Data} & {\sc herwig} \\
  $f_u$ = $f_d$                       &=& (39.7$\pm$1.0)\%                        & 39.6\%  \\
  $f_s$                              &=&(10.7$\pm$1.1)\%                          & 11.2\% \\
  $f_{\rm baryon}$                   &=&(9.9$\pm$1.7)\%                  & 9.6\%  \\ 
  $f (c \rightarrow D^+)$ = $f (c \rightarrow D^0)$  &= &(16.4$\pm$2.3)\%  & 16.9\%\\
 $f (c \rightarrow D^{*+})$ = $f (c \rightarrow D^{*0})$ &= &(22.8$\pm$2.5)\% &  22.5\% \\
   $f (c \rightarrow D_s+D^*_s)$        &= &(12.1$\pm$2.5)\%                       & 11.5\% \\
   $f (c \rightarrow {\rm baryons}) $    &=&(9.5$\pm$4.0)\%                       & 9.7\% \\
 \end{tabular}
 \end{ruledtabular}
 \end{center}
 \label{tab:tab_appa2}
 \end{table}
%%%%%%%%%%%%%%%%%%%%%%%%%%
  The fragmentation functions are tuned in the {\sc herwig} simulation by
  adjusting the parameters PSPLT and CLSMR to the values listed in
  Table~\ref{tab:tab_appa1}. Figures~\ref{fig:fig_appa1} 
  and~\ref{fig:fig_appa2} compare some fragmentation functions predicted
  by  the tuned {\sc herwig} simulation to data. Simulated fragmentation 
  functions are derived using $Z^0$ decays generated with process 2160. 
  Figure~\ref{fig:fig_appa1} compares  the distribution of the fraction of
  energy of parent $b$ quarks carried by all $B$ hadrons resulting from
  the heavy quark fragmentation to OPAL data~\cite{op} that in turn are
  consistent with Aleph and SLD measurements~\cite{al,sld}. 
  Figure~\ref{fig:fig_appa2} compares the fraction of momentum of the parent
  $c$ quarks carried by $D^*$ mesons to BELLE data~\cite{belle} that in turn
  are consistent with the CLEO result~\cite{cleo} and the Aleph measurement
  at the $Z$-pole~\cite{alcha}.
%%%%%%%%%%%%%%%%%%%%%%%%%%%%%%%%%%%%%%%%%%%%%%%%%%%
%%%%%%%%%%%%%%%%%%%%%%%%%%%%%%%%%%%%%%%%%%%%%%%%%%
%%%%%%%%%%%%%%%%%%%%%%%%%%%%%%%%%%%%%%%%%%%%%%%%%%%
%%%%%%%%%%%%%%%%%%%%%%%%%%%%%%%%%%%%%%%%%%%%%%%%%%
  The data are fairly well modeled by the {\sc herwig} generator with the
  parameter settings listed in Table~\ref{tab:tab_appa1}. A similar agreement
  for the fragmentation functions, but not the fragmentation fractions, can
  be achieved using the tuning proposed by some of the {\sc herwig} 
  authors~\cite{corcella}.  
%%%%%%%%%%%%%%%%%%%%%%%%%%%%%%%
 \section{Rate of fake muons}
\begin{figure}
  \begin{center}
  \leavevmode
  \includegraphics*[width=0.5\textwidth]{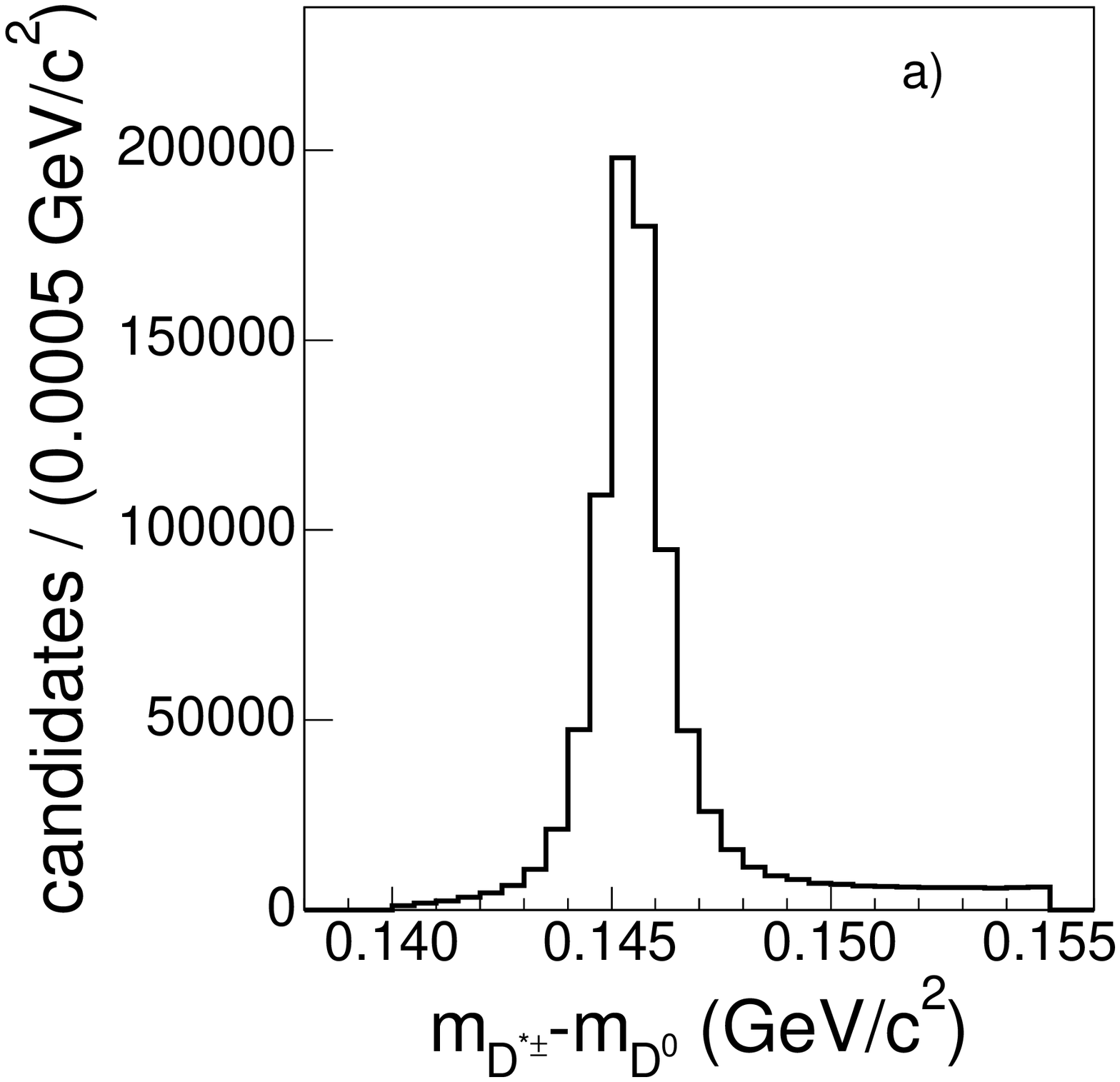}\includegraphics*[width=0.5\textwidth]{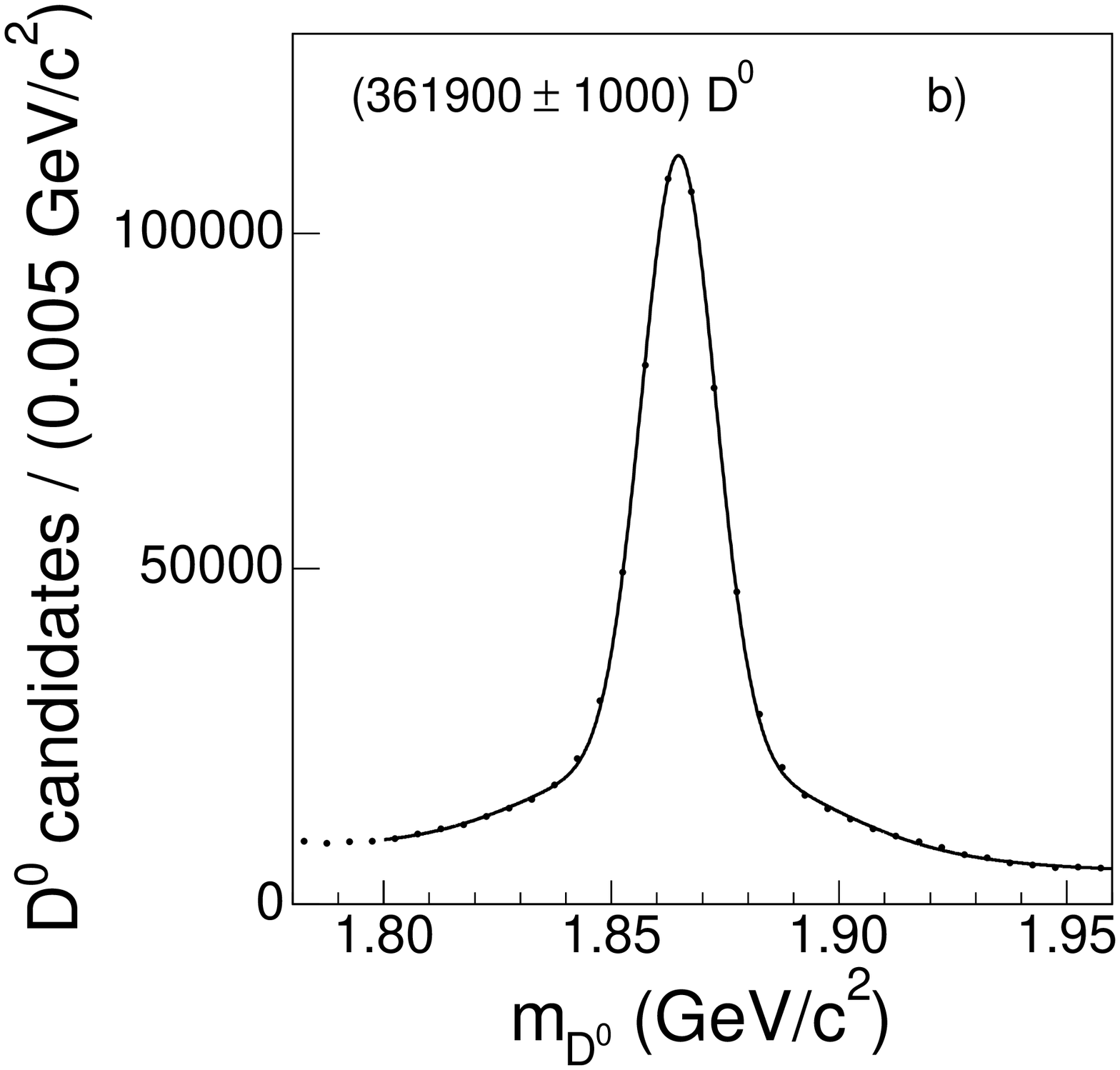}
  \caption[]{Distributions of (a) $m_{D^{*\pm}}-m_{D^0}$ and 
             (b) the invariant mass of $D^0$ candidates
             (the solid line represents the fit described in the text).}
  \label{fig:fig_appb1}
  \end{center}
  \end{figure}

\begin{figure}
  \begin{center}
  \vspace{-0.5cm}
  \leavevmode
  \includegraphics*[width=0.5\textwidth]{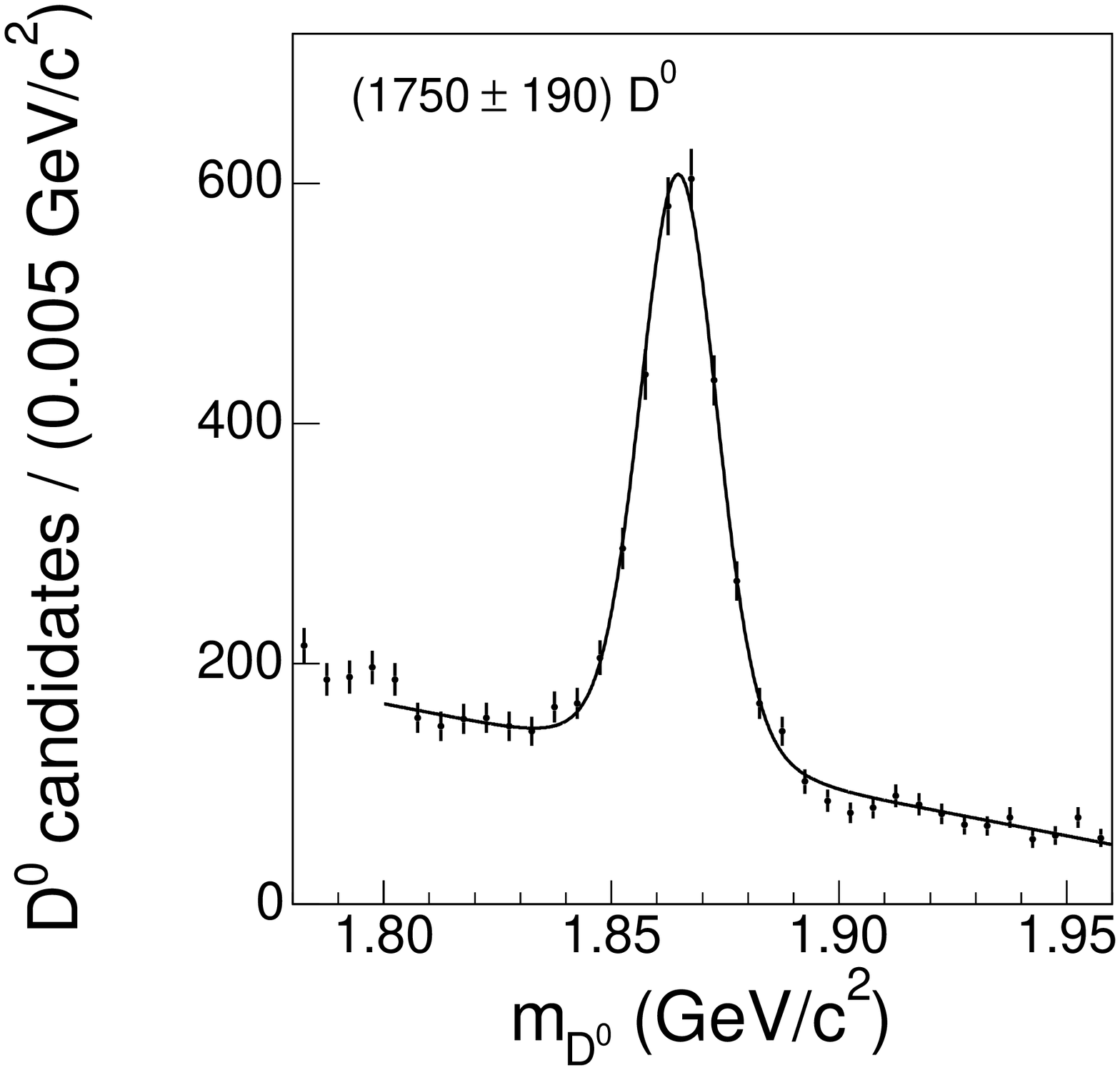}\includegraphics*[width=0.5\textwidth]{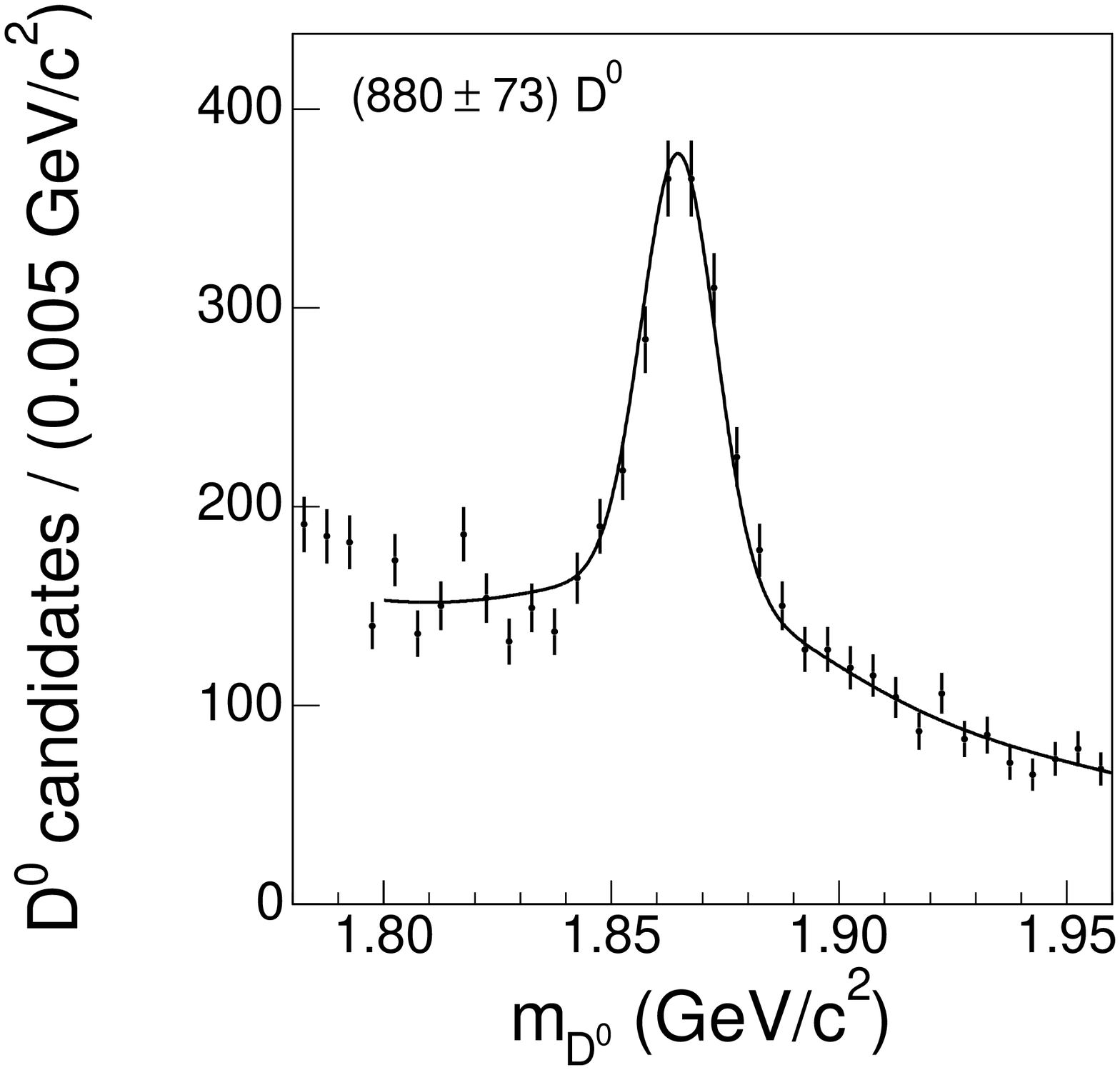}
  \caption[]{Invariant mass distribution of $D^0$ candidates with (left)
             a kaon  or (right) a pion leg identified as a CMUP muon.
             Solid lines represent the fits described in the text.}
  \label{fig:fig_appb3}
  \vspace{-0.5cm}
  \end{center}
  \end{figure}

\begin{figure}
  \begin{center}
  \leavevmode
  \includegraphics*[width=0.5\textwidth]{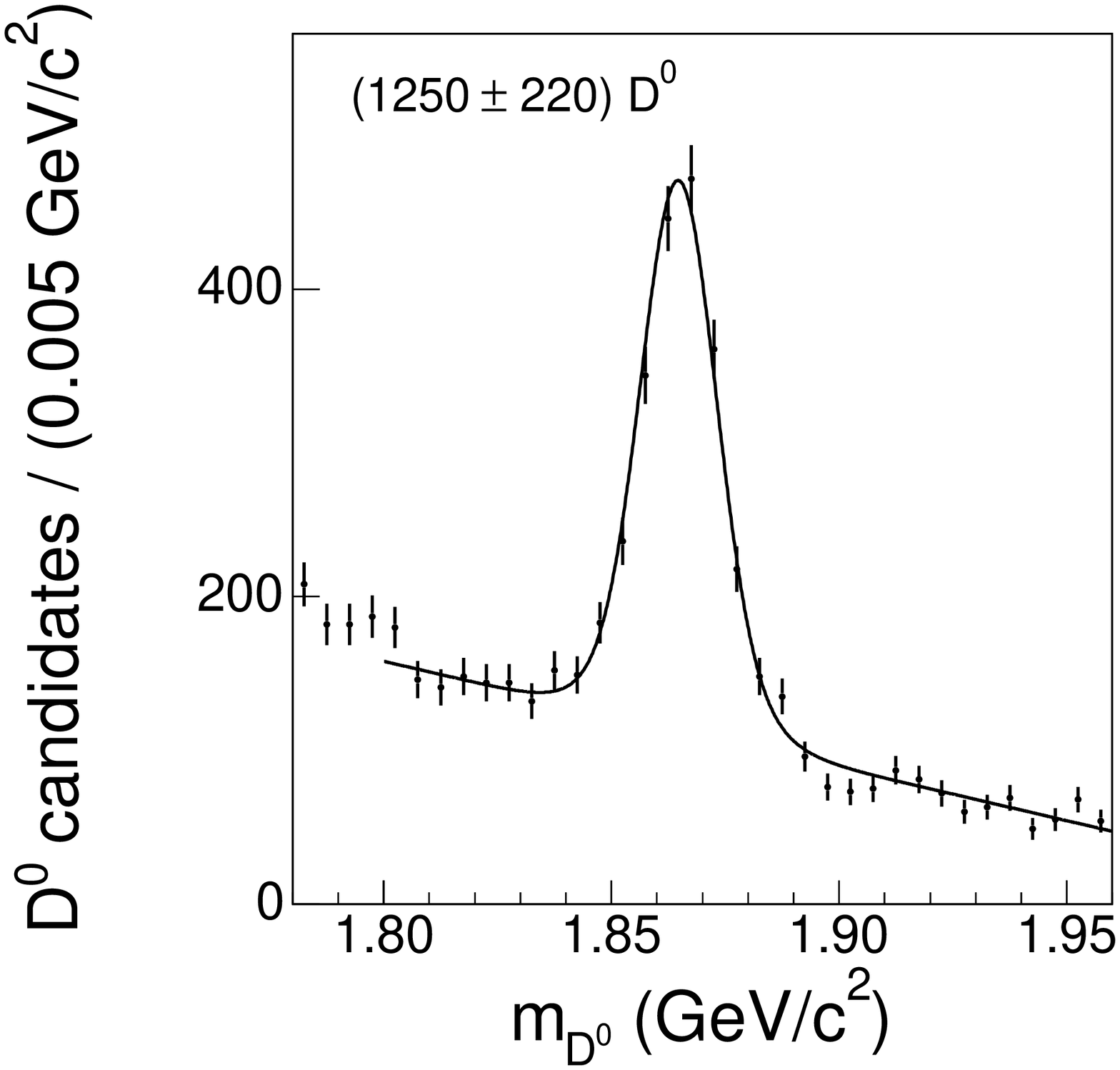}\includegraphics*[width=0.5\textwidth]{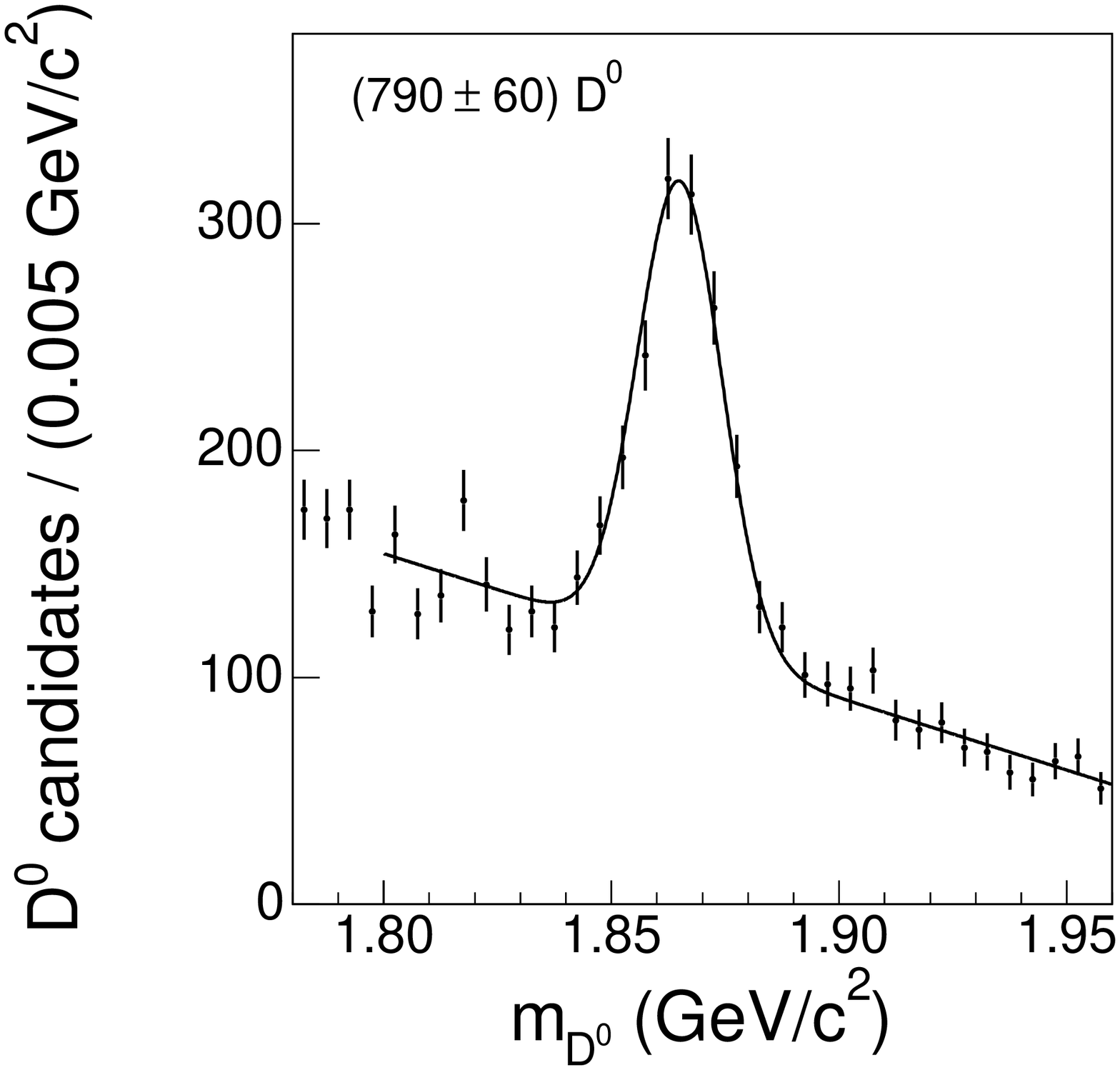}
  \includegraphics*[width=0.5\textwidth]{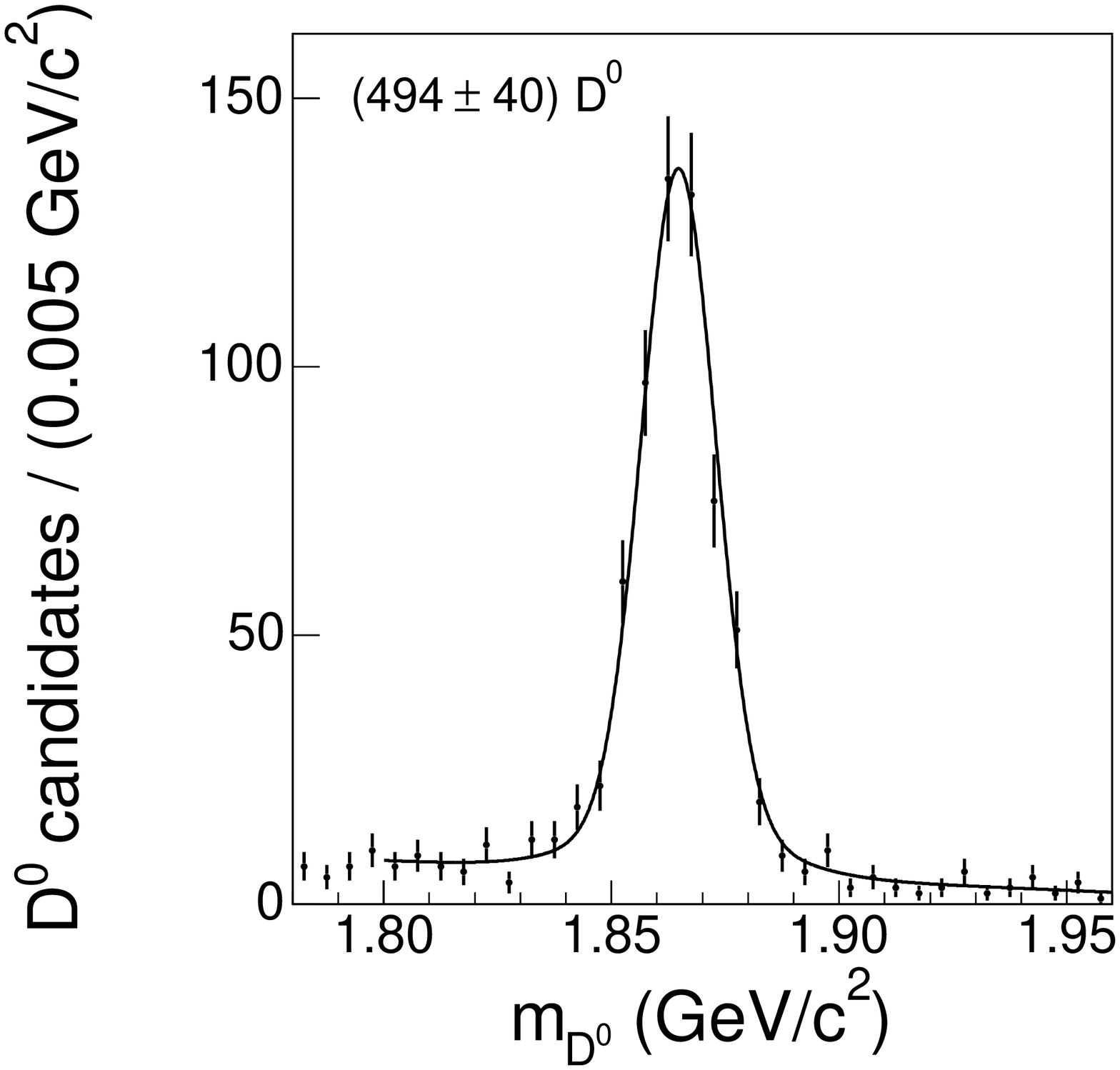}\includegraphics*[width=0.5\textwidth]{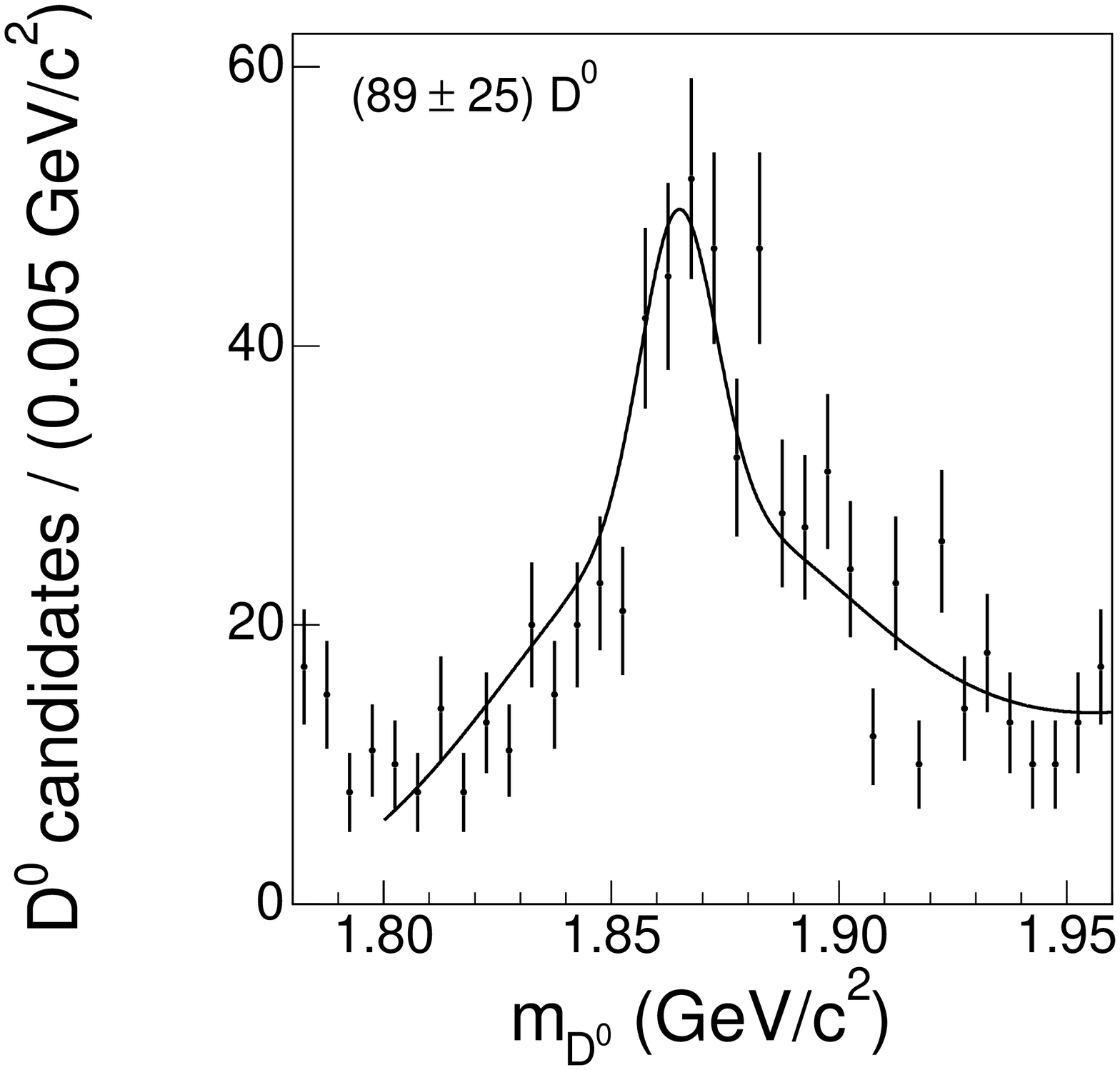}
  \caption[]{Invariant mass distribution of $D^0$ candidates with (left) 
             a kaon or (right) a pion leg identified as a CMUP muon.
             Top (bottom) plots require muons passing (failing) the $\chi^2<9$
             cut. Solid lines represent the fits described in the text.}
  \label{fig:fig_appb4}
  \end{center}
  \end{figure}

\begin{figure}
 \begin{center}
 \leavevmode
 \includegraphics*[width=0.5\textwidth]{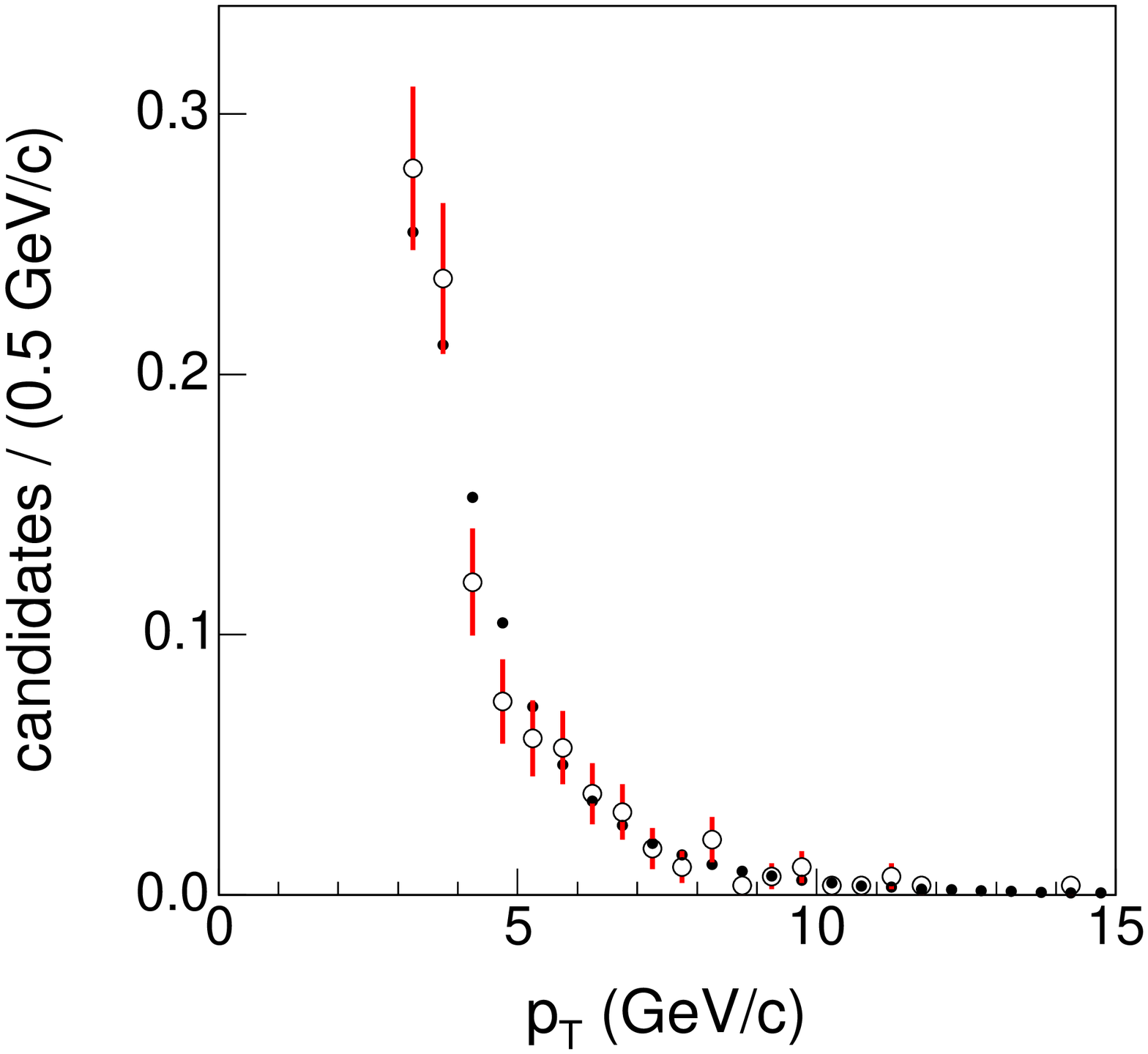}\includegraphics*[width=0.5\textwidth]{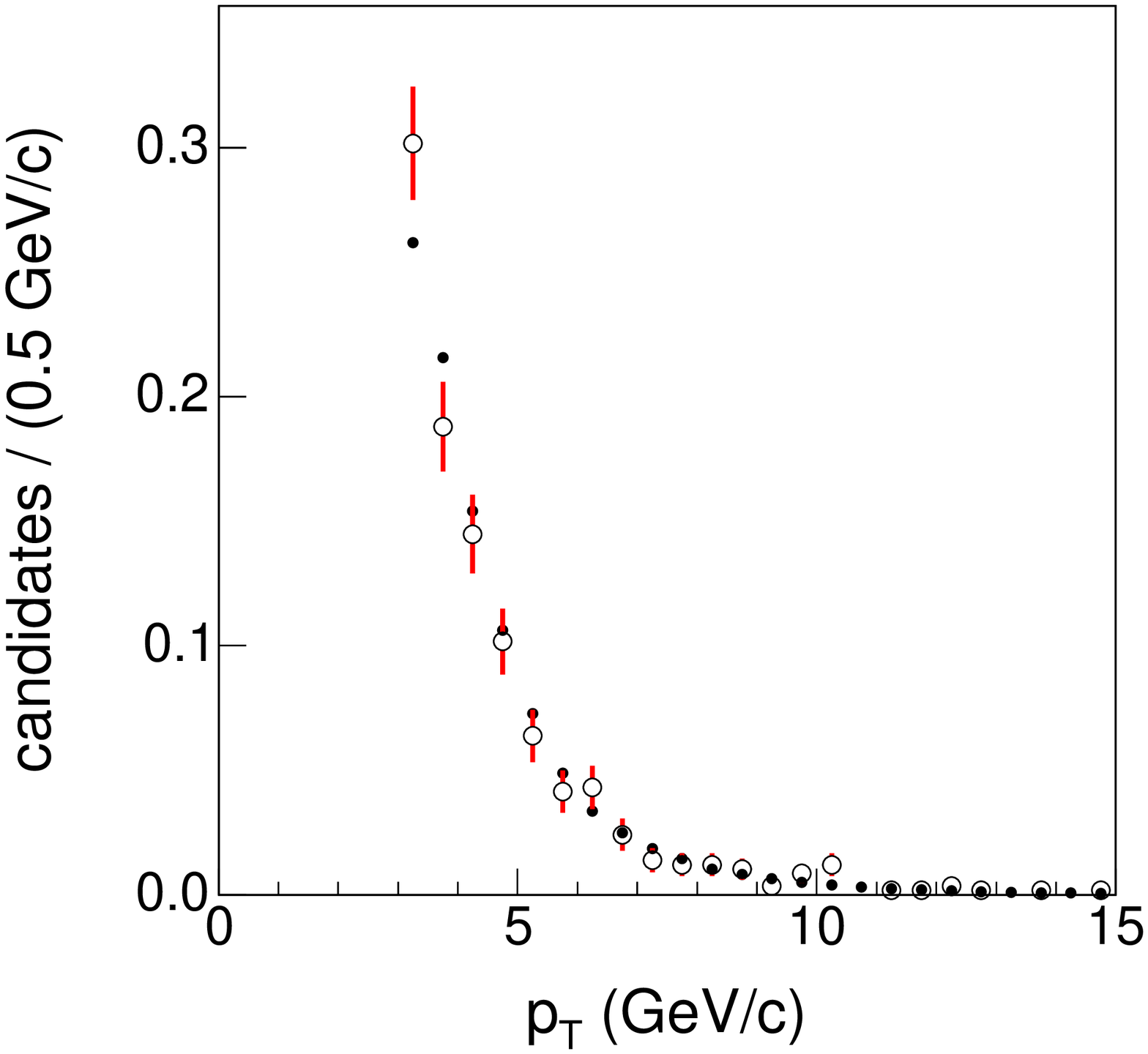}
 \caption[]{Transverse momentum distributions of (left) kaons and (right)
            pions from $D^0$ decays ($\circ$) are compared to those of
            tracks ($\bullet$) arising from simulated $b$-hadron decays.}
  \label{fig:fig_appb5}
 \end{center}
 \end{figure}

\begin{figure}
 \begin{center}
 \leavevmode
 \includegraphics*[width=0.5\textwidth]{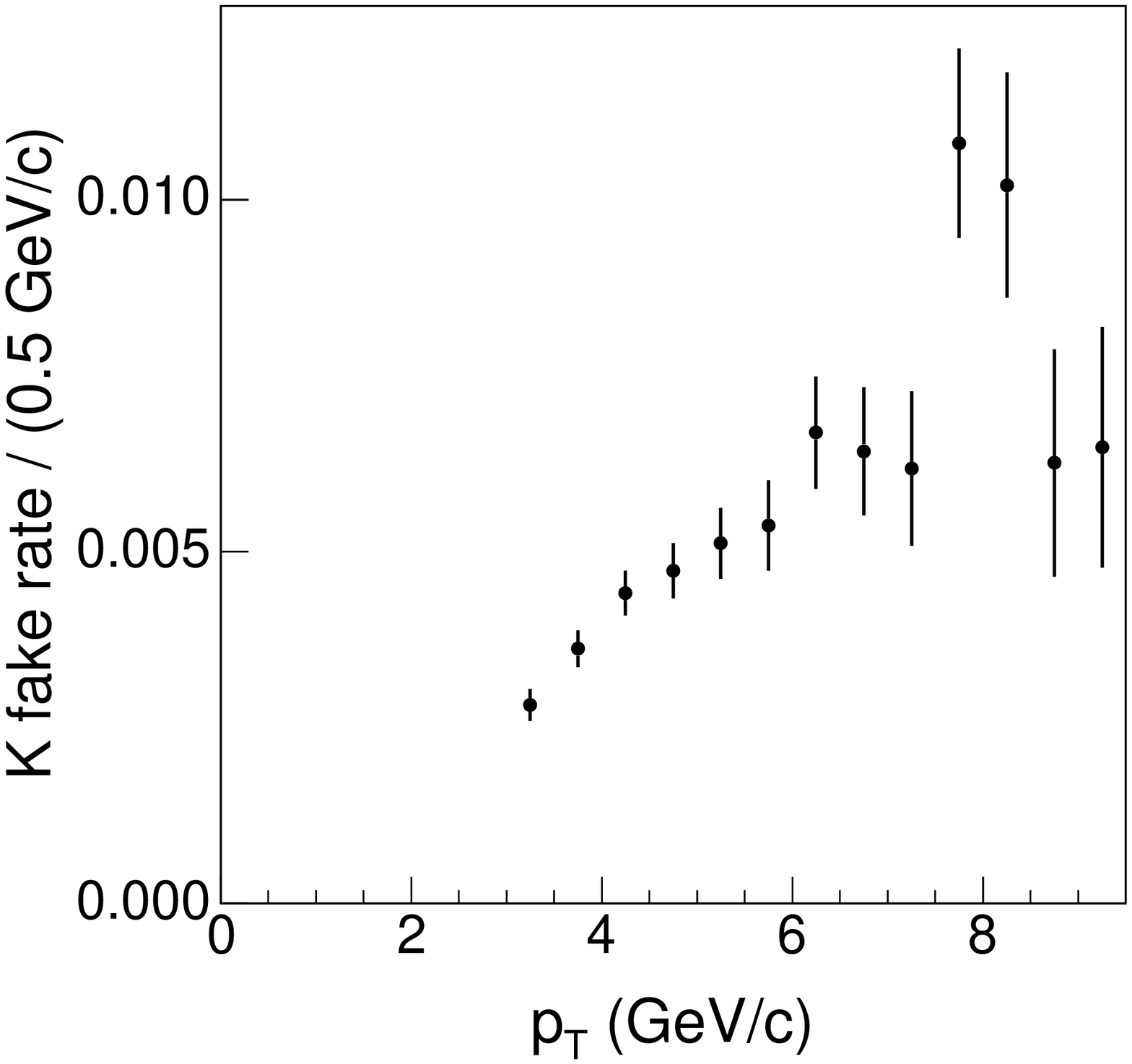}\includegraphics*[width=0.5\textwidth]{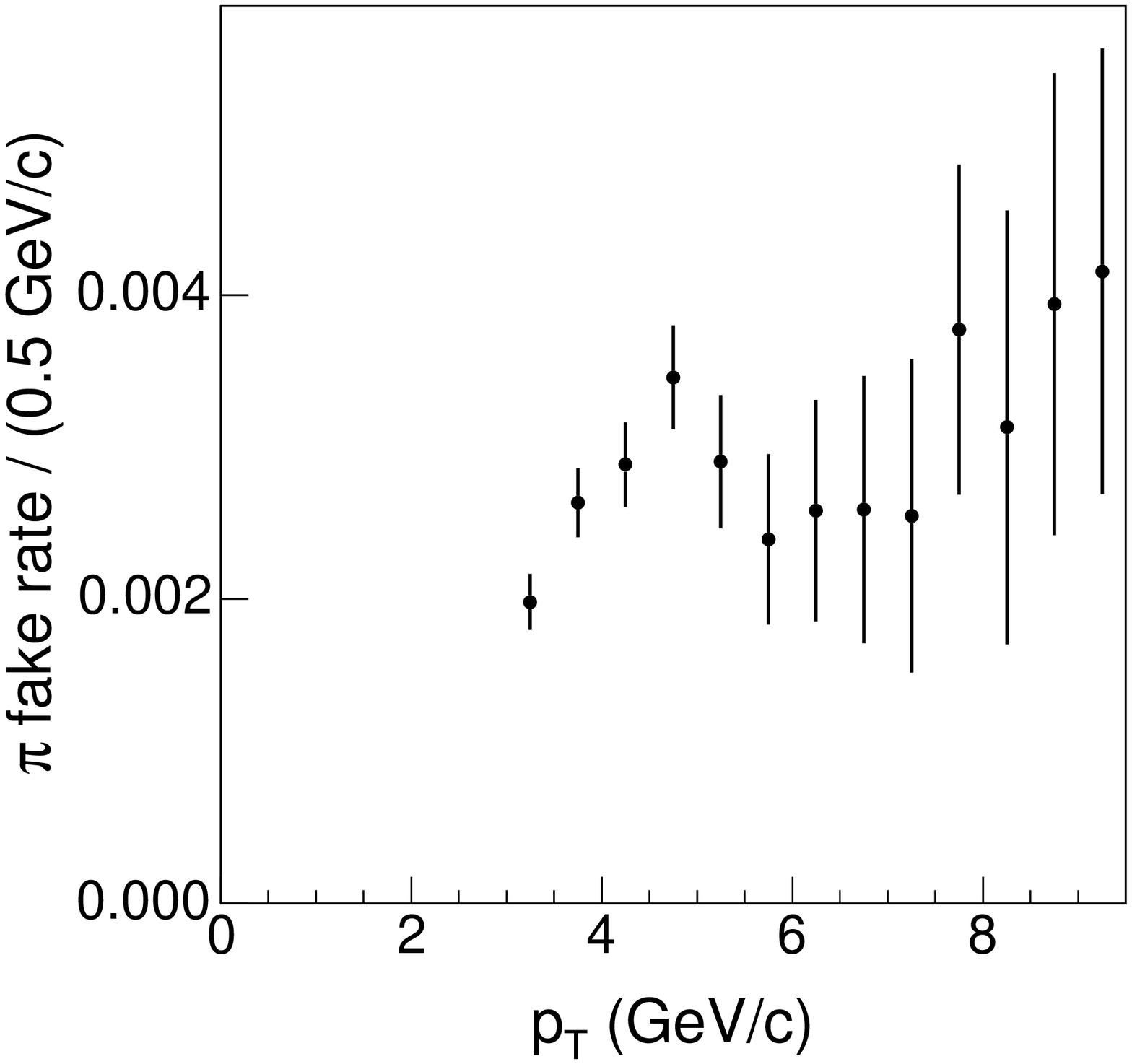}
 \caption[]{Fake muon probability as a function of the (left) kaon and
           (right) pion transverse momentum.}
 \label{fig:fig_appb6}
 \end{center}
\end{figure}

   Muons reconstructed in the CMUP detector are divided in this study into
   real and fake muons. Real muons originate from semileptonic decays of 
   hadrons with heavy flavor, the Drell-Yan process, and $\Upsilon$ decays. 
   Fake muons include muons from $\pi$ or K decays and hadronic punchthroughs
   that mimic a muon signal. The probability that a $\pi$ or K track is 
   misidentified as a muon is evaluated using $D^0 \rightarrow K \pi$ decays
   reconstructed in data collected with the {\sc charm} trigger.
   We select oppositely charged particles, each with $p_T \geq 3 \; \gevc$
   and $|\eta| \leq 0.7$, with $|\delta z_0| \leq 0.5$ cm.
   We require that each track is reconstructed in the microvertex detector
   with hits in at least four of the eight silicon layers. We evaluate
   the pair invariant mass for all pion-kaon mass assignments. The invariant
   mass is evaluated by constraining the two tracks to originate from a
   common point in the three-dimensional space (vertex constraint). We reject
   pairs if the probability of originating from a common vertex is smaller
   than 0.0002 or their invariant mass is outside the interval
   $1.77-1.97 \; \gevcc$. We also require that the displacement of the
   $D^0$-candidate vertex from the primary event vertex, projected onto 
   the $D^0$ transverse momentum vector, be larger than 0.02 cm. To further
   reduce the combinatorial background, we also require the $D^0$ candidate
   to originate from a $D^{*\pm}$ decay. We reconstruct $D^{*\pm}$ decays 
   by combining $D^0$ candidates with all additional COT tracks with 
   a distance $|\delta z_0| \leq 0.5$ cm with respect to the $D^0$ vertex.
   Additional tracks are assumed to be pions and the $D^*$ invariant mass 
   is evaluated by vertex constraining  pion and $D^0$ candidates and 
   rejecting combinations with probability smaller than 0.0002.
   The observed $m_{D^{*\pm}}-m_{D^0}$ distribution is shown in
   Fig.~\ref{fig:fig_appb1}~(a). We retain $D^0$ candidates with 
   $0.144 \leq m_{D^{*\pm}}-m_{D^0}\leq 0.147$ (their invariant mass 
   distribution is plotted in  Fig.~\ref{fig:fig_appb1}~(b)).
   The fake muon probability is derived using the invariant mass spectrum of 
   $D^0 \rightarrow \pi K$ decays in which one of the decay products is 
   matched to a CMUP stub (see Fig.~\ref{fig:fig_appb3}). We fit the data
   with two Gaussian functions to model the $D^0$ signal and a polynomial
   function to model the underlying background. The Gaussian functions model
   separately the right and wrong sign $D^0$ decays. In the fits to the data 
   in Fig.~\ref{fig:fig_appb3}, the width and peak of the first Gaussian 
   function and the peak of the second one are constrained to the value
   returned by the best fit to the data in Fig.~\ref{fig:fig_appb1}~(b)
   (peak at $1.865 \; \gevcc$  and $\sigma=0.008 \; \gevcc$). Using the 
   same method, we also evaluate the  rate of fake CMUP muons that pass 
   or fail the stricter $\chi^2 \leq 9$ selection cut described in 
   Sec.~\ref{sec:ss-fake} (see Fig.~\ref{fig:fig_appb4}). By using 
   $361902 $ $D^0 \rightarrow K \pi$ candidates we measure the fake muon 
   probabilities listed in Table~\ref{tab:tab_appb1}. The fake muon
   probabilities have been evaluated after weighting the transverse momentum
   distributions of kaons (pions) from $D^0$ decays to model that of kaons
   (pions) produced by simulated $b$-hadron decays (unweighted distributions
   are shown in Fig.~\ref{fig:fig_appb5}). Since the fake muon probability is
   not a strong function of the transverse momentum 
   (see Fig.~\ref{fig:fig_appb6}), we ignore the statistical uncertainty of 
   the simulated distributions because its effect is negligible compared to 
   the 10\% uncertainty of the kaon and pion rates predicted by the simulation.
%%%%%%%%%%%%%%%%%%%%%%%%%
%%%%%%%%%%%%%%%%%%%%%%%%%%%%%%%%%%%%%%%%%%%%
%%%%%%%%%%%%%%%%%%%%%%%%%%
%%%%%%%%%%%%%%%%%%%%%%%%%%
%%%%%%%%%%%%%%%%%%%%%%%%%
%%%%%%%%%%%%%%%%%%%%%%%%%%
%%%%%%%%%%%%%%%%%%%%%%%%%
%%%%%%%%%%%%%%%%%%%%%%%%%%
%%%%%%%%%%%%%%%%%%%%%%%%%%
%%%%%%%%%%%%%%%%%%%%%%%%%
%%%%%%%%%%%%%%%%%%%%%%%%%%%%%%%%%
 \begin{table}[htb]
 \caption[]{Probabilities, $P_f^K$ and $P_f^\pi$, that pions and kaons, 
            respectively, mimic a CMUP signal for different selection
            criteria.}
 \begin{center}
 \begin{ruledtabular}
 \begin{tabular}{lcc}
  CMUP selection      &  $P_f^K$ (\%)      &  $P_f^\pi$ (\%)   \\
   standard           & $0.483 \pm  0.003$ & $0.243\pm 0.004$  \\ 
  $ \chi^{2} \leq 9$  & $0.347 \pm  0.003$ & $0.219\pm 0.003$  \\
  $ \chi^{2} > 9$     & $0.136 \pm  0.001$ & $0.025\pm 0.002$  \\
 \end{tabular}
 \end{ruledtabular}
 \end{center}
 \label{tab:tab_appb1}
 \end{table}
%%%%%%%%%%%%%%%%%%%%%%%%%%%%%%%%%%%%%%%%%%%%%%%%%%%%%%%%%%%%%%%%%%%%%%%%%%%%%

%%%%%%%%%%%%%%%%%%%%%%%%%%%%%%%

\clearpage
  \begin{thebibliography}{99}
 \label{bibliography}
%%%%%%%%%%%%%%%%%%%%%%%%%%%%%%%%%%%%%%%
% Introduction
%%%%%%%%%%%%%%%%%%%%%%%%%%%%%%%%%%%%%%%
\bibitem{mnr}     M.~L.~Mangano, P.~Nason, and G.~Ridolfi, 
                  Nucl.~Phys.~{\bf B373}, 295 (1992).
                  The {\sc fortran} code, also referred to as {\sc hvqmnr},
                  is made available by the authors and can be downloaded
                  from $\tt http://www.ge.infn.it/\sim ridolfi$.
\bibitem{nde}     P.~Nason, S.~Dawson, and R.~K.~Ellis, Nucl.~Phys.~{\bf B327},
                  49 (1989); {\bf B335}, 260 (1990). 
\bibitem{mlmri}   S.~Frixione {\it et al.}
                  Adv.~Ser.~Direct.~High~Energy~Phys.~{\bf 15}, 609 (1998)
		  [arXiv:hep-ph/9702287].
\bibitem{qcdan}   M.~Cacciari {\it et al.,} J.~High~Energy~Phys.~0407, 033
                  (2004).
\bibitem{ajets}   D.~Acosta {\it et al.,} Phys.~Rev.~D~{\bf 69}, 072004 (2004).
\bibitem{bstatus} F.~Happacher {\it et al.,} Phys.~Rev.~D~{\bf 73}, 014026
                  (2006); 
                  F.~Happacher, {\it Status of the Observed and Predicted
                  $b \bar{b}$ Cross Section at the Tevatron},
                  $\tt www-conf.kek.jp/dis06/doc/WG5/hfl20-happacher.ps$,
                  to appear in the Proceedings of DIS 2006, Tsukuba, Japan.
\bibitem{mrst}    A.~D.~Martin {\it et al.,}  Eur.~Phys.~J.~C~{\bf 4}, 463 
                  (1998).
\bibitem{cteq}    J.~Pumplin  {\it et al.,}  J.~High~Energy~Phys.~0207, 012
                  (2002).
\bibitem{pet}     C.~Peterson  {\it et al.,} Phys.~Rev.~D~{\bf 27}, 105 (1983).
\bibitem{chrin}   J.~Chrin, Z.~Phys.~C~{\bf 36}, 163 (1987).
\bibitem{cana}    M.~Cacciari and P.~Nason, Phys.~Rev.~Lett.~{\bf 89}, 122003
                  (2002).
\bibitem{fonll}   M.~Cacciari {\it et al.,}  J.~High~Energy~Phys.~9805, 007
                  (1998).
\bibitem{f1}      P.~Nason and C.~Oleari, Nucl.~Phys.~{\bf B565}, 245 (2000);
                  B.~Mele and P.~Nason, Nucl.~Phys.~{\bf B361}, 626 (1991);
                  G.~Colangelo and P.~Nason, Phys.~Lett.~B~{\bf 285}, 167
                  (1992).
\bibitem{f2}      H.~Heister {\it et al.,} Phys.~Lett.~B~{\bf 512}, 30 (2001);
                  K.~Abe {\it et al.,} Phys.~Rev.~D~{\bf 65}, 092006 (2002).
 \bibitem{herwig} G.~Marchesini and B.~R.~Webber, Nucl.~Phys.~B~{\bf 310}, 
                  461 (1988); G.~Marchesini {\it et al.,} 
                  Comput.~Phys.~Commun. {\bf 67}, 465 (1992).
\bibitem{pythia}  T.~Sj\"{o}strand and M.~Bengtsson, 
                  Comp.~Phys.~Commun.~{\bf 43}, 367 (1987);
                  H.~Bengtsson and T.~Sj\"{o}strand,
                  Comp.~Phys.~Commun.~{\bf 46}, 43 (1987).       
\bibitem{mcnlo}   S.~Frixione {\it et al.,} J.~High~Energy~Phys.~0308, 007
                  (2003);
                  S.~Frixione and B.~R.~Webber, J.~High~Energy~Phys.~0206, 029
                  (2002).
                  The code is made available by the authors
                  and can be downloaded from 
                  $\tt http://www.hep.phy.cam.ac.uk/theory/webber/MCatNLO/$.
\bibitem{pnason}  P.~Nason {\it et al.,} hep-ph/0003142.
\bibitem{bdis}    S.~Alekhin {\it et al.,} hep-ph/0204316.
 \bibitem{shears} T.~Shears, ``Charm and Beauty Production at the Tevatron'',
                  Proceedings of the Int.~Europhys.~Conf. on High Energy
                  Phys., PoS (HEP2005), 072 (2005).
\bibitem{derw}    F.~Abe {\it et al.,} Phys.~Rev.~D~{\bf 53}, 1051 (1996).
\bibitem{2mucdf}  F.~Abe {\it et al.,} Phys.~Rev.~D~{\bf 55}, 2546 (1997).
\bibitem{bjk}  A.~Abulencia {\it et al.,} Phys.~Rev.~D.~{\bf 75}, 012010
                  (2007).
\bibitem{d0b2}    B.~Abbott {\it et al.,} Phys.~Lett.~B~{\bf 487}, 264 (2000).                
\bibitem{bmix}    D.~Acosta {\it et al.,} Phys.~Rev.~D~{\bf 69}, 012002 (2004).
\bibitem{det1}    F.~Abe {\it et al.,} Nucl.~Instrum.~Methods~Phys.~Res.,
                  Sect.~A~{\bf 271}, 387 (1988).
\bibitem{det2}    R.~Blair {\it et al.,} Fermilab Report No.
                  FERMILAB-Pub-96/390-E (1996).
\bibitem{det3_0}  C.~S.~Hill {\it et al.,} Nucl.~Instrum.~Methods~Phys.~Res.,
                  Sect.~A~{\bf 530}, 1 (2004).
\bibitem{det3}    A.~Sill {\it et al.,} Nucl.~Instrum.~Methods~Phys.~Res.,
                  Sect.~A~{\bf 447}, 1 (2000).
\bibitem{det4_0}  T.~Affolder {\it et al.,} Nucl.~Instrum.~Methods.~Phys.~Res.,
                  Sect.~A~{\bf 453}, 84 (2000).
\bibitem{det4}    T.~Affolder {\it et al.,} Nucl.~Instrum.~Methods~Phys.~Res.,
                  Sect.~A~{\bf 526}, 249 (2004).
\bibitem{det5}    G.~Ascoli {\it et al.,} Nucl.~Instrum.~Methods~Phys.~Res., 
                  Sect.~A~{\bf 268}, 33 (1988).
\bibitem{det6}    J.~Elias {\it et al.,} Nucl.~Instrum.~Methods.~Phys.~Res.,
                  Sect.~A~{\bf 441}, 366 (2000).
\bibitem{det7}    D.~Acosta {\it et al.,} Nucl.~Instrum.~Methods~Phys.~Res., 
                  Sect.~A~{\bf 461}, 540 (2001).
\bibitem{det8}     R.~Downing {\it et al.,} Nucl.~Instrum.~Methods~Phys.~Res., 
                  Sect.~A~{\bf 570}, 36 (2007).
\bibitem{sigmatot} M.~M.~Block and R.~N.~Cahn, Rev.~Mod.~Phys.~{\bf 57}, 563
                  (1985).
\bibitem{klimen}  S.~Klimenko {\it et al.,} Fermilab Report No.
                  FERMILAB-FN-0741 (2003).
\bibitem{svt}     B.~Ashmanskas {\it et al.,} 
                  Nucl.~Instrum.~Methods~Phys.~Res., Sect.~A~{\bf 518}, 532 
                 (2004).
\bibitem{evtgen}  D.~J.~Lange, Nucl.~Instrum.~Meth.~A~{\bf 462}, 152 (2001).
                  We use version {\sc V00-14-05}\\ downloaded from
           $\tt http://www.slac.stanford.edu/BFROOT/dist/packages/EvtGen/$.
\bibitem{geant}   R.~Brun  {\it et al.,} CERN Report No. CERN-DD-78-2-REV;
                  R.~Brun  {\it et al.,} CERN Programming Library Long 
                  Write-up W5013 (1993).
\bibitem{minuit}  F.~James and M.~Roos, Comput.~Phys.~Commun.~{\bf 10}, 343 
                  (1975).
\bibitem{pdf}     H.~Plothow-Besch, {\it PDFLIB: Nucleon, Pion and Photon 
                  Parton Density Function and $\alpha_s$ Calculations}, 
                  User's Manual - Version~6.06, W5051 PDFLIB, 1995.03.15,
                  CERN-PPE.
\bibitem{topxsec} T.~Affolder {\it et al.,} Phys.~Rev.~D~{\bf 64}, 032002
                  (2001).
\bibitem{babar1}  The Babar collaboration,
                  $\tt https://oraweb.slac.stanford.edu/pls/slacquery/$
                  $\tt BABAR\_DOCUMENTS.DetailedIndex?P\_BP\_ID=3553$,
                  unpublished.
\bibitem{splot}   M.~Pivk and F.~R.~Le~Diberder, 
                  Nucl.~Instrum.~Methods~A~{\bf 555}, 356 (2005).
\bibitem{herw_sett} $\tt http://hepwww.rl.ac.uk/theory/seymour/herwig/HWtune.html$.
\bibitem{pdg}     W.-M.~Yao {\it et al.,} J.~Phys.~G~{\bf 33}, 1 (2006).
\bibitem{belle}   R.~Seuster {\it et al.,} Phys.~Rev.~D~{\bf 73}, 032002 
                 (2005).
\bibitem{cleo}    M.~Artuso {\it et al.,} Phys.~Rev.~D~{\bf 70}, 112001 (2004).
\bibitem{hera}    A.~Atkas {\it et al.,}  Eur.~Phys.~J.~C~{\bf 38}, 447 (2005).
\bibitem{glad}    L.~Gladilin, hep-ex/9912064.
\bibitem{glad1}   L.~K.~Gladilin, hep-ex/0607036.
\bibitem{op}      G.~Abbiendi {\it et al.,} Eur.~Phys.~J.~C~{\bf 29}, 463
                  (2003).
\bibitem{al}      A.~Heister {\it et al.,} Phys.~Lett.~B~{\bf 512}, 30 (2001).
\bibitem{sld}     K.~Abe {\it et al.,} Phys.~Rev.~D~{\bf 65}, 092006 (2002).
\bibitem{alcha}   R.~Barate {\it et al.,} Eur.~Phys.~J.~C~{\bf 16}, 597 (2000).
\bibitem{corcella} G.~Corcella {\it et al.,} hep-ph/0602191.
\bibitem{bishai}  D.~Acosta {\it et al.,} Phys.~Rev.~D~{\bf 71}, 032001 (2005).
\bibitem{matt}    A.~Abulencia {\it et al.,} Phys.~Rev.~Lett.~{\bf 95}, 221805
                 (2005).
\bibitem{lepewg}  The ALEPH Collaboration {\it et al.,} Phys.~Rept.~{\bf 427},
                  257 (2006).
\bibitem{hvqjet}  M.~Baarmand and F.~Page, private communication.
\bibitem{matteo2} M.~Cacciari, private communication.
\bibitem{cacc-greco} M.~Cacciari {\it et al.,} Phys.~Rev.~D~{\bf 55}, 7134 
                     (1997).
%%%%%%%%%%%%%%%%%%%%%%%%%%%%%%%%%%%

%\bibitem{psi-runii} D.~Acosta {\it et al.,} Phys.~Rev.~D~{\bf 71}, 032001
%                   (2005).
%\bibitem{fonll}   M.~Cacciari {\it et al.,}  J.~High~Energy~Phys.~{\bf 9805}, 007 (1998).
%\bibitem{cacc}    M.~Cacciari {\it et al.,} JHEP~{\bf 0407}, 033 (2004).
%\bibitem{mlm}     M.~L.~Mangano, hep-ph/0411020.
%\bibitem{mc}      M.~Cacciari, hep-ph/0407187.
%\bibitem{herwig}  G.~Marchesini and B.~R.~Webber, Nucl.~Phys.~B~{\bf 310}, 
%                 461 (1988); G.~Marchesini {\it et al.,} 
%                 Comput.~Phys.~Commun. {\bf 67}, 465 (1992).
%\bibitem{pythia}  T.~Sj\"{o}strand and M.~Bengtsson, 
%                  Comp.~Phys.~Commun.~{\bf 43}, 367 (1987);
%                  H.~Bengtsson and T.~Sj\"{o}strand,
%                  Comp.~Phys.~Commun.~{\bf 46}, 43 (1987);
%\bibitem{ri}      S.~Frixione {\it et al.}
%                   Adv.~Ser.~Direct.~High~Energy~Phys.~{\bf 15}, 609 (1998)
%		  [arXiv:hep-ph/9702287].
%\bibitem{mrsd0}   A.~D.~Martin, W.~J.~Stirling and R.~G.~Roberts,
%	          Phys.~Rev.~D~{\bf 47}, 867 (1993).
%\bibitem{mrsa}    A.~D.~Martin, W.~J.~Stirling and R.~G.~Roberts, 
%                  Phys.~Lett.~B~{\bf 354}, 155 (1995).
%\bibitem{cdf1}    F.~Abe {\it et al.,} Phys.~Rev.~Lett.~{\bf 68}, 3403 (1992).
%\bibitem{d01}     S.~Abachi {\it et al.,} Phys.~Lett.~B~{\bf 370}, 239 (1996).
%\bibitem{cdf2}    F.~Abe {\it et al.,} Phys.~Rev.~Lett.~{\bf 69}, 3704 (1992).
%\bibitem{pet}     C.~Peterson  {\it et al.,} Phys.~Rev.~D~{\bf 27}, 105 (1983).
%\bibitem{chrin}   J.~Chrin, Z.~Phys.~{\bf C36}, 163 (1987).
%\bibitem{qq}      P.~Avery, K.~Read, G.~Trahern, Cornell Internal Note CSN-212,
%	          March 25, 1985 (unpublished).
%\bibitem{babar}  B.~Aubert{\it et al.,} Phys.~Rev.~D~{\bf 67}, 032002
%                  (2003). 
%\bibitem{cdfb1}   D.~Acosta {\it et al.,} Phys.~Rev.~D~{\bf 65}, 052005 (2002).
%\bibitem{mrst-n}  A.~D.~Martin {\it et al.,} Eur.~Phys.~J.~C{\bf 4}, 463 
%                  (1998).
%\bibitem{pdg02}   K.~Hagiwara {\it et al.,} Phys.~Rev.~D~{\bf 66}, 010001
%                  (2002).
%\bibitem{cdfb2}   F.~Abe {\it et al.,} Phys.~Rev.~Lett.~{\bf 75}, 1451 (1995).
%\bibitem{cdfb3}   F.~Abe {\it et al.,} Phys.~Rev.~Lett.~{\bf 71}, 2396 (1993).
%\bibitem{dflm}    M.~Diemoz {\it et al.,} Z.~Phys.~{\bf C39}, 21 (1988).
%\bibitem{cdfb4}   F.~Abe {\it et al.,} Phys.~Rev.~Lett.~{\bf 71}, 500 (1993).
%\bibitem{cdfb5}   F.~Abe {\it et al.,} Phys.~Rev.~Lett.~{\bf 79}, 572 (1997).
%\bibitem{d0b1}    S.~Abachi {\it et al.,} Phys.~Rev.~Lett.{\bf 74}, 3548
%		  (1995).
%\bibitem{d0b2}    B.~Abbott {\it et al.,} Phys.~Lett.~B~{\bf 487}, 264 (2000).
%\bibitem{mrsr2}   A.~D.~Martin, R.~G.~Roberts and W.~J.~Stirling,
%                  Phys.~Lett.~B~{\bf 387}, 419 (1996).
%\bibitem{d0b3}    B.~Abbott {\it et al.,} Phys.~Rev.~Lett.~{\bf 85}, 5068
%                  (2000).
%\bibitem{cana}    M.~Cacciari and P.~Nason, Phys.~Rev.~Lett.~{\bf 89}, 122003
%                  (2002).
%\bibitem{monica}  M.~D'Onofrio, hep-ex/0505036.
%\bibitem{pe1}     P.~Nason, S.~Dawson and R.~K.~Ellis, Nucl.~Phys.~B~{\bf 303},
%		  607 (1988).
%\bibitem{pe2}     J.~C.~Collins and R.~K.~Ellis, Nucl.~Phys.~B~{\bf 360}, 3 
%		  (1991).
%\bibitem{pe3}     S.~Catani, M.~Ciafaloni and F.~Hautmann, 
%	          Nucl.~Phys.~B~{\bf 366}, 135 (1991).
%\bibitem{pe4}     M.~Cacciari and M.~Greco, Nucl.~Phys.~B~{\bf 421}, 530
%                  (1994).
%\bibitem{f1}      P.~Nason and C.~Oleari, Nucl.~Phys.~B~{\bf 565}, 245 (2000);
%                  B.~Mele and P.~Nason, Nucl.~Phys.~B~{\bf 361}, 626 (1991);
%                  G.~Colangelo and P.~Nason, Phys.~Lett.~B~{\bf 285}, 167
%                  (1992).
%\bibitem{f2}      H.~Heister {\it et al.,} Phys.~Lett.~B~{\bf 512}, 30 (2001);
%                  K.~Abe {\it et al.,} Phys.~Rev.~D~{\bf 65}, 092006 (2002).
%\bibitem{mrst2001} A.~D.~Martin {\it et al.,} Eur.~Phys.~J.~C {\bf 23}, 73 (2002).
%\bibitem{cteq}    H.~L.~Lai {\it et al.,} JHEP {\bf 0207}, 012 (2002);
% Eur.~Phys.~J.~C{\bf 12}, 375 (2000).
%\bibitem{seymour} M.~H.~Seymour, Nucl.~Phys.~B~{\bf 436}, 163 (1995); 
%	          M.~L.~Mangano, Nucl.~Phys.~B~{\bf 405}, 536 (1993).
%\bibitem{pdf}     H.~Plothow-Besch, ``PDFLIB: Nucleon, Pion and Photon Parton
%                  Density Functions and $\alpha_s$ Calculations'',
%                  User's manual-Version 6.06, W5051 PDFLIB, 1995.03.15, 
%	          CERN-PPE.
%\bibitem{ajets}   D.~Acosta {\it et al.,} Phys.~Rev.~D~{\bf 69}, 072004 (2004).
%\bibitem{super}   D.~Acosta {\it et al.,} Phys.~Rev.~D~{\bf 65}, 052007 (2002).
%\bibitem{xsec}    T.~Affolder {\it et al.,} Phys.~Rev.~D~{\bf 64}, 032002
%                  (2001); Erratum-ibid.~D~{\bf 67}, 119901 (2003).
%\bibitem{field}   R.~D.~Field, Phys.~Rev.~D~{\bf 65}, 094006 (2002). 
%\bibitem{derw}    F.~Abe {\it et al.,} Phys.~Rev.~D~{\bf 53}, 1051 (1996).
%\bibitem{2mucdf}  F.~Abe {\it et al.,} Phys.~Rev.~D~{\bf 55}, 2546 (1997).
%\bibitem{shears}  T.~Shears, ``Charm and Beauty Production at the Tevatron'',
%                  talk presented at the Int.~Europhys.~Conf. on High Energy
%                  Phys., Lisboa, Portugal (2005);\\
% $\tt http://www.lip.pt/events/2005/hep2005/talks/hep2005\_talk\_TaraShears.ppt$. 
%\bibitem{webber}  S.~Frixione {\it et al.,} JHEP~{\bf 0308}, 007 (2003);
%                   S.~Frixione and B.~R.~Webber, JHEP~{\bf 0206}, 029 (2002).
%\bibitem{pdgold} L.~Montanet{\it et al.,} Phys.~Rev.~D~{\bf 50}, 1173 (1994).
%\bibitem{isajet}  F.~Paige and S.~Protopopescu, BNL report BNL38034, 1986 (unpublished).
%                   The {\sc qq} decay table is implemented starting with version V7.22.
%
\end{thebibliography}
 \end{document}